\documentclass[aps,pra,twocolumn, showpacs,groupedaddress]{revtex4}
\usepackage{epsfig,amssymb,amsmath, mathrsfs,color} 
\usepackage[hypertex,linkcolor=red]{hyperref}
\usepackage[matrix,frame,arrow]{xypic}\input{Qcircuit}
\newtheorem{lemma}{Lemma} 
\newtheorem{corollary}{Corollary} \newtheorem{theorem}{Theorem}
\newtheorem{postulate}{Postulate} \newtheorem{definition}{Definition}
\def\spc #1{{\mathscr #1}}
\def\Proof{\medskip\par\noindent{\bf Proof. }}
\def\qed{$\,\blacksquare$\par} 
\def\ugualonelike#1{\quad\begin{matrix} \\ \\
    #1 \\\end{matrix}\quad} \def\>{\rangle}
\def\<{\langle} \def\trnsfrm#1{\mathscr
  #1} \def\rA{{\rm A}}\def\rB{{\rm
    B}}\def\rC{{\rm C}}\def\rD{{\rm D}} \def\rE{{\rm E}} \def\rF{{\rm
    F}} \def\rG{{\rm G}} \def\rI{{\rm I}} \def\rR{{\rm
    R}} \def\rS{{\rm S}} \def\rX{{\rm X}} \def\rY{{\rm Y}} \def\rZ{{\rm Z}}
\def\tA{\trnsfrm A}\def\tB{\trnsfrm B}\def\tC{\trnsfrm
  C}\def\tD{\trnsfrm D} \def\tE{\trnsfrm E} \def\tF{\trnsfrm F}  \def\tI{\trnsfrm
  I}\def\tT{\trnsfrm T}\def\tU{\trnsfrm U} \def\tV{\trnsfrm V}
\def\tW{\trnsfrm W}
\def\tZ{\trnsfrm Z}\def\tF{\trnsfrm
  F}\def\tE{\trnsfrm E} \def\tR{\trnsfrm R}
\def\tT{\trnsfrm T} \def\Cntset{{\mathfrak E}}
\def\grp#1{{\mathbf{#1}}} \def\Span{\mathsf{Span}}
\def\Stset{{\mathfrak S}} \def\Trnset{{\mathfrak T}}
\def\K#1{\left|#1\right)\!}  \def\B#1{\left(#1\right|}
\def\SC#1#2{\left(#1\right|\left.\!#2\right)\!}  \def\Tr{{\rm Tr}}

\def\Reals{{\mathbb R}}
\newcommand{\multiprepareC}[2]{*+<1em,.9em>{\hphantom{#2}}\save[0,0].[#1,0];p\save
  !C *{#2},p+RU+<0em,0em>;+LU+<+.8em,0em> **\dir{-}\restore\save
  +RD;+RU **\dir{-}\restore\save +RD;+LD+<.8em,0em> **\dir{-}
  \restore\save +LD+<0em,.8em>;+LU-<0em,.8em> **\dir{-} \restore \POS
  !UL*!UL{\cir<.9em>{u_r}};!DL*!DL{\cir<.9em>{l_u}}\restore}
\newcommand{\prepareC}[1]{*{\xy*+=+<.5em>{\vphantom{#1\rule{0em}{.1em}}}*\cir{l^r};p\save*!L{#1} \restore\save+UC;+UC+<.5em,0em>*!L{\hphantom{#1}}+R **\dir{-} \restore\save+DC;+DC+<.5em,0em>*!L{\hphantom{#1}}+R **\dir{-} \restore\POS+UC+<.5em,0em>*!L{\hphantom{#1}}+R;+DC+<.5em,0em>*!L{\hphantom{#1}}+R **\dir{-} \endxy}}

\begin{document}
\title{Probabilistic theories with purification}
\author{Giulio Chiribella}\email{gchiribella@perimeterinstitute.ca} 
\affiliation{Perimeter Institute for Theoretical Physics, 31 Caroline Street North, Ontario,  Ontario N2L 2Y5, Canada.}
\homepage{http://www.perimeterinstitute.ca}
\author{Giacomo Mauro D'Ariano}\email{dariano@unipv.it}
\affiliation{{\em QUIT Group}, Dipartimento di Fisica  ``A. Volta'' and  INFN Sezione di Pavia, via Bassi 6, 27100 Pavia, Italy}
\homepage{http://www.qubit.it}
\author{Paolo Perinotti}\email{paolo.perinotti@unipv.it} 
\affiliation{{\em QUIT Group}, Dipartimento di Fisica  ``A. Volta'' and  INFN Sezione di Pavia, via Bassi 6, 27100 Pavia, Italy}
\homepage{http://www.qubit.it}
\date{\today}
\begin{abstract}
  We investigate general probabilistic theories in which every mixed
  state has a purification, unique up to reversible channels on the
  purifying system. We show that the purification principle is
  equivalent to the existence of a reversible realization of every
  physical process, that is, to the fact that every physical process can be regarded
  as arising from a reversible interaction of the system with an
  environment, which is eventually discarded.  From the purification
  principle we also construct an isomorphism between transformations
  and bipartite states that possesses all structural properties of the
  Choi-Jamio\l kowski isomorphism in quantum theory.  Such an
  isomorphism allows one to prove most of the basic features of
  quantum theory, like e.g.  existence of  pure bipartite states
  giving perfect correlations in independent
  experiments, no information without disturbance, no joint
  discrimination of all pure states, no cloning, teleportation, no
  programming, no bit commitment, complementarity between correctable channels
  and deletion channels, characterization of entanglement-breaking
  channels as measure-and-prepare channels, and others, without
  resorting to the mathematical framework of Hilbert spaces.
\end{abstract} \maketitle

\newcommand{\poloFantasmaCn}[1]{{{}^{#1}_{\phantom{#1}}}}

\tableofcontents
\section{Introduction}

In the past two decades the field of quantum information theory has
brought to light an enormous amount of protocols and tasks
that originate from the structure of quantum theory and have
dramatic consequences in the way information can be processed.
Non-locality, no-cloning, teleportation, dense coding, quantum key
distribution, quantum algorithms, and quantum error correction are
only the most celebrated examples of a much longer list. An important
lesson from this experience is that the abstract formalism of quantum
mechanics has a huge number of operational consequences.

At the same time, the question whether quantum theory is the only
conceivable theory with such operational consequences has attracted
the attention of an increasing number of researchers. In a seminal
paper \cite{poprohr}, Popescu and Rohrlich showed that non-locality
is not an exclusive feature of quantum theory, and that there are
in fact possible theories that  exhibit stronger nonlocality than
quantum theory without violating relativistic no-signaling. An
intense work on non-locality in general non-signaling theories has
followed this observation, opening a very active line of research (see
e.g. \cite{barpironio,cryptobar,cryptoacin,wolf}).  On the other hand,
the authors of Refs.  \cite{noclon,nobroad} have analyzed tasks like
cloning and broadcasting of states, showing that the impossibility of
achieving them is a highly generic property, while Ref.
\cite{info-processingBarrett} thoroughly discussed theories with a
local discriminability property that share other features of quantum
mechanics, like the non-unique convex decomposition of a mixed state
or the non-existence of ideal non-disturbing measurements.
Entanglement swapping and teleportation protocols have been considered
in Refs \cite{swapping,tele}, where the authors noticed the remarkable
fact that the no-signaling boxes of Popescu and Rohrlich do not allow
for entanglement swapping, nor for teleportation. Very recently, the
authors of Ref. \cite{infocaus} have introduced the new physical
principle of information causality, showing that while the principle
holds for quantum theory, it is violated by Popescu-Rohrlich boxes.

Despite the numerous advancements in the understanding of general
probabilistic theories, the fundamental problem of deriving quantum
mechanics from basic physical principles is still completely open.  In
particular, no physical principle is known that can single out quantum
mechanics in the physically motivated set of \emph{causal theories
  with local discriminability}. With this expression we mean
probabilistic theories where \emph{i)} the probability of outcomes of
an experiment performed at a given time does not depend on the choice
of experiments that will be performed at later times, and \emph{ii)}
if two bipartite states are different, then one can discriminate
between them using only local devices with an error
probability that is smaller than 1/2, the random guess value.  
In the case of classical physics, finding a
description is relatively simple: among theories in the above family,
classical probability is the only one where all pure states are
perfectly distinguishable. On the contrary, every current description
of quantum theory is a description of its mathematical apparatus:
e.g. one can say that quantum theory is the theory where pure
states are unit vectors in complex Hilbert spaces and probabilities
are given by the Born rule, or, equivalently, that it is the theory
where observables form a C*-algebra of complex matrices.

In the past there have been many attempts to find a more basic
description of quantum theory, in particular by discussing it from
the point of view of logic \cite{birkvn,mackey,jaupir,ludwig} (see also
Ref. \cite{coeckerev} and references therein).  More recently, Hardy
\cite{hardy5} has approached the problem from a different perspective,
providing a characterization of quantum theory based on  principles
of mathematical simplicity in the interplay among dimension of the
state space, structure of subsystems and subspaces, number of
distinguishable states, and topology of the set of pure states.
On the other hand, in recent years one of the authors has tackled the
problem using physical principles related to tomography and
calibration of physical devices, experimental complexity, and to the
composition of elementary (atomic) transformations (see Ref.
\cite{maurolast} for the state of the art of this project).  In
particular, Ref. \cite{maurofirst} firstly introduced the concept of
\emph{dynamically and preparationally faithful state}, which will play
an important role in this paper.

In this paper we introduce the purification principle ``Every mixed
state has a purification, unique up to reversible channels on the
purifying system''.  The main message of our work is simple: most of
the characteristic features of quantum theory can be summarized in
the physical statement ``quantum theory is a causal theory with
purification and local discriminability''.  In particular, from the
purification principle we derive the following features: no
information without disturbance, no joint discriminability and no
cloning of pure states, existence of pure entangled states with
perfect correlations, probabilistic teleportation, one-to-one
correspondence between transformations and bipartite states, dilation
of physical processes to reversible interactions with an environment,
necessary and sufficient conditions for error correction in terms of
the reversible dilation, no bit commitment, no programming of
reversible channels without perfectly distinguishable program states,
and identification of causal channels with sequences of channels with
memory, and characterization of entanglement breaking channels as
measure-and-prepare channels.  Moreover, we also discuss a stronger
version of the purification principle: ``For every system $\rA$ there
exists a \emph{conjugate system} $\tilde \rA$ such that every state of
$\rA$ has a purification in $\rA \tilde \rA$.  The conjugate of
$\tilde \rA$ is $\rA$ (symmetry), and the conjugate of a composite
system $\rA \rB$ is the composite system $\tilde \rA \tilde \rB$
(regularity under composition)''. With this further property one can
prove deterministic teleportation and show that its structure is
unique: the resource state for deterministic teleportation
must be a purification of the unique mixed state that is invariant under all reversible channels.
  
As we will show, the purification principle is equivalent to the fact
that every irreversible process arises from a reversible interaction
with an environment that is eventually lost. This can be viewed as a
law of ``conservation of information'': information cannot be erased,
it can only be discarded. Moreover, we will see that the purification
principle has other remarkable consequences: From the structural point
of view, a theory with purification is completely identified by the
states of all possible systems in it. Once the states are given, all
possible measurements and evolutions are fixed.  Even more strongly,
the purification postulate implies the completeness property
``whatever transformation is mathematically admissible (in a sense that
will be made precise later) must be feasible''. Conversely, we can explicitly say that whatever limitation to the feasibility of a mathematically admissible map results in a limitation to the purifiability of some state.    The analogue of  this property in quantum information is that every trace-preserving completely positive map must be feasible.

It is important to stress that we are not claiming that we derived
quantum theory.  What we can say is that we ``zipped'' a large part
of it, by reducing a long list of features to a single physical
principle.  In the process of doing this, we found proofs that are
often simpler (or at least more intuitive) than the original quantum
proofs.

In order to minimize the notational burden due to the lack of
a commonly established formalism, in presenting these proofs we opted
for a graphical notation, which is equivalent to formulae and replaces
them in most of the paper.  Since this notation is exactly the same
notation used in quantum circuits, a reader with a background in
quantum information can easily read the general equations without
spending too much time in the introductory part of the paper.  On the
other hand, an extended discussion on graphical calculus can be found
in the work by Penrose \cite{penrose} and in the rigorous
formalization by Joyal and Street within the theory of symmetric
monoidal categories \cite{joyal} (we also suggest the beautiful
introductions in the topic by Selinger \cite{selinger} and Coecke
\cite{practicingcoecke}).  We anyway stress that in the
present paper the choice of graphical notation is just the choice of a more
user-friendly way of presenting formulae, and that no prerequisite on
e.g. category theory is needed from the reader.

\section{Operational-probabilistic theories}\label{sec:op-prob}

In this Section we introduce some basic notions that will be used in the paper. In particular, we introduce the notion of operational-probabilistic theory
as a theory that \emph{i)} describes a set of possible experiments
that can be done with physical devices and \emph{ii)} gives
predictions about the probabilities of the outcomes in these
experiments.

\subsection{Systems and tests}

\emph{Systems} and \emph{tests} are the primitive notions of an
operational theory.  Each test represents one use of a \emph{physical
  device}, like a Stern-Gerlach magnet, a beamsplitter, or a photon
counter.  Systems play the role of labels attached to physical
devices: any device has an input and an output port labeled by an \emph{output} and an \emph{input system}, respectively.
These labels establish a rule for connecting physical devices among
themselves: two devices can be connected in a sequence only if the
output of the first device is a system of the same type as the input
of the second.

All throughout the paper we will denote systems with capital letters,
like $\rA, \rB, \rC,$ and so on. We reserve the letter $\rI$ for
the \emph{trivial system}, which simply means ``nothing''.  A device with input (output) system $\rI$ is a device with no input (no output). 

Let us now make more precise the notion of test. We already mentioned
that a test represents one use of a physical device. When the physical
device is used, it produces an \emph{outcome} $i$ in some set
$\rX$, e.g. the outcome could be a sequence of digits appearing on a
display, a light, or a sound emitted by the device. The outcome
produced by the device heralds the fact that some \emph{event} has
occurred.  These intuitive features concur in the definition of test:
\begin{definition}[Test]
  A test with input system $\rA$ and output system $\rB$ is a
  collection of events $\{\tC_i\}_{i\in \rX}$ labeled by outcomes in
  some outcome set $\rX$.  Diagrammatically, the test $\{\tC_i\}_{i \in \rX}$ is represented as follows 
\begin{equation}
  \Qcircuit @C=1em @R=.7em @! R {& \qw \poloFantasmaCn \rA & \gate {\{\tC_i\}_{i \in \rX}} & \qw \poloFantasmaCn \rB &\qw}
\end{equation}
while the specific event $\tC_i$ is represented by
\begin{equation}
  \Qcircuit @C=1em @R=.7em @! R {& \qw \poloFantasmaCn \rA & \gate {\tC_i} & \qw \poloFantasmaCn \rB &\qw}
\end{equation}
\end{definition} 
We denote by $\Trnset (\rA, \rB)$ the set of all events appearing in
all tests from $\rA$ to $\rB$. When $\rB \equiv \rA$ we will write
$\Trnset (\rA)$.

Tests with trivial input will be called \emph{preparation-tests}, and
the corresponding events will be called \emph{preparation-events}.  In
quantum information, a preparation-test is what is called a ``random
source of quantum states''.  In analogy we will adopt for
preparation-events the usual notation as for states in quantum
circuits:
\begin{equation}
 \Qcircuit @C=1em @R=.7em @! R {\prepareC{\rho_i}& \qw \poloFantasmaCn \rB &\qw} ~:= ~   \Qcircuit @C=1em @R=.7em @! R {& \qw \poloFantasmaCn \rI & \gate {\tC_i} & \qw \poloFantasmaCn \rB &\qw}
\end{equation}
In formulae, we will often use the ``Dirac-like'' notation $\K{\rho_i}_\rB$ to denote a preparation event of system $\rB$.  We will denote by $\Stset (\rA)$ the set of preparation-events for system $\rA$, namely $\Stset (\rA) := \Trnset (\rI, \rA)$.

Similarly, we will call tests with trivial output
\emph{observation-tests}, and the corresponding events
\emph{observation-events}. In quantum theory, an observation-test is a quantum measurement, and is represented by \emph{positive operator valued measure (POVM)}, that is, by a collection of positive operators $\{P_i\}_{i \in \rX}$ satisfying $\sum_{i\in \rX} P_i =I_\rA$, where $I_\rA$ is the identity on the Hilbert space of system $\rA$. For observation-tests we will then adopt the usual notation
for measurements in quantum circuits:
\begin{equation}
 \Qcircuit @C=1em @R=.7em @! R {& \qw \poloFantasmaCn \rA &\measureD {a_j}} ~:= ~   \Qcircuit @C=1em @R=.7em @! R {& \qw \poloFantasmaCn \rA & \gate {\tC_j} & \qw \poloFantasmaCn \rI &\qw}
\end{equation}
In formulae, we will often denote observation-events with the notation $\B {a_j}_{\rA}$.   We will denote by $\Cntset (\rA)$ the set of observation-events for system $\rA$, namely $\Cntset (\rA) := \Trnset (\rA, \rI)$.

For tests from the trivial system to itself we will omit the box and the wires, as follows: 
\begin{equation}
p_k ~:=~  \Qcircuit @C=1em @R=.7em @! R {& \qw \poloFantasmaCn \rI &\gate {p_k} & \qw \poloFantasmaCn{\rI}&\qw } 
\end{equation}
In Subsect. \ref{subsect:probint} we will interpret events from the
trivial system to itself as \emph{probabilities}.

Another important case of tests is that of \emph{single-outcome}
tests, in which the outcome space $\rX$ consists of a single element:
$\rX =\{i_0\}$. Whenever a device represented by a single-outcome test
is used, the experimenter is sure that only one
event can take place.  This motivates the following definition:

\begin{definition}[Deterministic tests]\label{def:deterministic} A test is \emph{deterministic} if its outcome set has a single element, namely $|\rX|=1$.
\end{definition}



\subsection{Sequential composition of tests}
  
Physical devices can be used in sequences, as long as the output of each device coincides with the input of the next one.  When two tests are composed in a sequence we obtain a new test, as in the following
\begin{definition}[Sequential composition of tests]  
If $\{\tC_i\}_{i \in \rX}$ is a test from $\rA$ to $\rB$ and $\{\tD_j\}_{j \in \rY}$ is a test from
$\rB$ to $\rC$, then their sequential composition is  test from $\rA$ to $\rC$, with outcomes $(i,j)
\in \rX \times \rY$, and events  $\{\tD_j\circ\tC_i\}_{(i,j) \in \rX \times \rY}$. 
Diagrammatically, the events $\tD_j\circ \tC_i$ are represented as follows 
\begin{equation}
  \Qcircuit @C=1em @R=.7em @! R {& \qw \poloFantasmaCn \rA & \gate {\tC_i} & \qw \poloFantasmaCn \rB &\gate{\tD_j} & \qw\poloFantasmaCn\rC &\qw} ~:= ~ \Qcircuit @C=1em @R=.7em @! R {& \qw \poloFantasmaCn \rA & \gate {\tD_j \circ \tC_i} & \qw \poloFantasmaCn \rC &\qw} 
\end{equation}
\end{definition} 

We will say that test $\{\tD_j\}$ ``follows'' test $\{\tC_i\}$, or,
equivalently, $\{\tC_i\}$ ``precedes'' $\{\tD_j\}$. For the moment,
the order of composition is not necessarily temporal. The
interpretation of sequential composition as a sequence of time-steps will be
given  in Subsect. \ref{sec:causaltheo} within the framework of causal theories.

The sequential composition of tests brings immediately  the notion of \emph{identity test}. 
\begin{definition}[Identity test]  The identity test for system $\rA$ is a test with a single event $\tI_{\rA}$ such that for every system $\rB$
\begin{equation}
\begin{split}
  \Qcircuit @C=1em @R=.7em @! R {& \qw \poloFantasmaCn \rA & \gate {\tI} & \qw \poloFantasmaCn \rA &\gate{\tC_i} & \qw\poloFantasmaCn\rB &\qw} &=~   \Qcircuit @C=1em @R=.7em @! R {& \qw \poloFantasmaCn \rA &\gate{\tC_i} & \qw\poloFantasmaCn\rB &\qw} \qquad \forall \tC_i \in \Trnset (\rA, \rB) \\
&\\
  \Qcircuit @C=1em @R=.7em @! R {&\qw \poloFantasmaCn \rB &\gate{\tD_j} & \qw\poloFantasmaCn\rA  & \gate {\tI} &  \qw \poloFantasmaCn \rA &\qw} &=~   \Qcircuit @C=1em @R=.7em @! R {& \qw \poloFantasmaCn \rB &\gate{\tD_i} & \qw\poloFantasmaCn\rA &\qw} \qquad \forall \tD_j \in \Trnset (\rB, \rA)
\end{split} 
\end{equation}
\end{definition}
Performing the identity test on a system just means ``doing nothing'' on it. We can think of the
outcome of the identity test as a blank character, which provides no information.

In some protocols, such as teleportation, one wants to emphasize that
one is dealing with two different systems ``of the same type''. For
examples, in quantum theory one can have two electrons in different
(spatially separated) regions. Distinguishing two systems of the same
type is essentially a matter of bookkeeping. Moreover, we can have
different physical systems that are ``operationally equivalent'', {\em
  e.~g.} the polarization of a single photon and the spin of an
electron in quantum theory are both represented by a {\em qubit},
and can be (at least in principle) converted one to another in a reversible fashion. For this
reason we introduce a formal notion of operational equivalence between systems, based on their mutual convertibility:

\begin{definition}[Operationally equivalent
  systems]\label{def:opeq} Two systems $\rA$ and $\rA'$ are
  \emph{operationally equivalent}---denoted as $\rA'\simeq\rA$---
  if there exist a deterministic test $\{\tI_{\rA , \rA'}\}$ from
  $\rA$ to $\rA'$ and a deterministic test $\{\tI_{\rA', \rA}\}$ from
  $\rA'$ to $\rA$, respectively, such that
\begin{equation}
\begin{split}
  \Qcircuit @C=1em @R=.7em @! R {& \qw \poloFantasmaCn \rA & \gate  \tI & \qw \poloFantasmaCn {\rA'} & \gate \tI &\qw \poloFantasmaCn \rA &\qw} &=~   \Qcircuit @C=1em @R=.7em @! R {& \qw \poloFantasmaCn \rA &\gate \tI  & \qw \poloFantasmaCn\rA &\qw}  \\
  &\\
  \Qcircuit @C=1em @R=.7em @! R {&\qw \poloFantasmaCn {\rA'}
    &\gate{\tI} & \qw\poloFantasmaCn\rA & \gate {\tI} & \qw
    \poloFantasmaCn {\rA'} &\qw} &=~ \Qcircuit @C=1em @R=.7em @! R {&
    \qw \poloFantasmaCn {\rA'} &\gate{\tI} & \qw\poloFantasmaCn{\rA'}
    &\qw}
\end{split} 
\end{equation}
\end{definition}   
Accordingly, if $\{\tC_i\}_{i \in \rX}$ is a test for system $\rA$,
performing the ``same test'' on system $\rA'$ means performing the
test $\{\tC_i'\}_{i \in \rX}$ defined by
\begin{equation}
  \Qcircuit @C=1em @R=.7em @! R {& \qw \poloFantasmaCn {\rA'} &\gate {\tC'_i}  & \qw \poloFantasmaCn{\rA'} &\qw} ~=~  \Qcircuit @C=1em @R=.7em @! R {& \qw \poloFantasmaCn {\rA'} & \gate \tI & \qw \poloFantasmaCn {\rA} & \gate {\tC_i} & \qw \poloFantasmaCn{\rA} & \gate \tI &\qw \poloFantasmaCn {\rA'} &\qw}   
\end{equation}

Clearly, the above notion of ``same test on a different system'' depends on the choice of the privileged test $\{\tI_{\rA, \rA'}\}$ used to set up the operational equivalence between $\rA$ and $\rA'$.  We will often drop the primes and write $\tC_i$ instead of $\tC_i'$.


\subsection{Composite systems and parallel composition of tests}

Given two systems $\rA$ and $\rB$, one can consider them together, thus forming the corresponding
\emph{composite system}, here denoted by $\rA\rB$.  A test with input (output) system
$\rA \rB$ ($\rC \rD$), represents one use of a physical device with two input (output) ports, labeled by $\rA$ and $\rB$ ($\rC$ and $\rD$), respectively.
\begin{definition}[Composite system]\label{def:compsyst} If $\rA, \rB$ are systems, the corresponding composite system is $\rA \rB$. Composition of systems
  enjoys the properties \emph{i)} $\rA = \rI\rA = \rA\rI$, \emph{ii)}
  $\rA \rB \simeq \rB \rA$, and \emph{iii)} $\rA (\rB \rC) = (\rA \rB)
  \rC:= \rA\rB \rC$.

  Diagrammatically, an event from $\rA \rB$ to $\rC \rD$ is
  represented as a box with multiple wires:
\begin{equation}
\begin{aligned}
  \Qcircuit @C=1em @R=.7em @! R {& \qw \poloFantasmaCn \rA & \multigate{1} {\tC_i } & \qw \poloFantasmaCn \rC &\qw\\
& \qw \poloFantasmaCn \rB & \ghost{\tC_i} & \qw \poloFantasmaCn \rD &\qw}
\end{aligned} :=  
\begin{aligned}
\Qcircuit @C=1em @R=.7em @! R {& \qw \poloFantasmaCn {\rA\rB} & \gate {\tC_i } & \qw \poloFantasmaCn {\rC\rD} &\qw} 
\end{aligned} 
\end{equation}\end{definition}

The property \emph{i)} in Def.  \ref{def:compsyst} expresses the fact
that system $\rA$ together with ``nothing'' is still system $\rA$,
while properties \emph{ii)} and \emph{iii)} express the fact that the
specification of a composite system depends only on the list of
component systems, and not on how the elements of the list are ordered
(up to operational equivalence, implemented by a deterministic test
that permutes the component systems), nor on how they are grouped.

In general, we will represent the $N$-partite composite system $\rA_1
\dots \rA_N $ with $N$ wires, as follows: \begin{equation}
\begin{aligned} \Qcircuit @C=1em @R=.7em @! R {&\qw & \qw \poloFantasmaCn {\rA_1} & \qw &\qw \\
    &\qw & \qw \poloFantasmaCn {\rA_2} & \qw &\qw\\
    &  & \vdots &  \\
    &&&\\
    &\qw & \qw \poloFantasmaCn {\rA_N} & \qw &\qw}
\end{aligned} ~:=~
\begin{aligned}  \Qcircuit @C=1em @R=.7em @! R {&\qw &\qw & \qw \poloFantasmaCn {\rA_1 \rA_2\dots \rA_N} & \qw &\qw&\qw} 
\end{aligned} 
\end{equation} 
In the case of trivial systems, we will typically omit the wire. 
In  the sequential composition of two boxes with multiple wires we will always match the output wires of the first box with the input wires of the second.

Physical devices can be run in parallel on different systems, thus
performing a test on the composite system, as in the following
\begin{definition}[Parallel composition of tests] If $\{\tC_i\}_{i \in \rX}$ is a test from $\rA$ to $\rB$
  and $\{\tD_j\}_{j \in \rY}$ is a test from $\rC$ to $\rD$, then their parallel composition is the
  test from $\rA \rC$ to $\rB \rD$, with outcomes $(i,j) \in \rX \times \rY$, and events
  $\{\tC_i\otimes\tD_j\}_{(i,j) \in \rX \times \rY}$.  Diagrammatically the events $\tC_i\otimes\tD_j$ are
  represented as follows
\begin{equation}
\begin{aligned}
  \Qcircuit @C=1em @R=.7em @! R {& \qw \poloFantasmaCn \rA & \gate {\tC_i} & \qw \poloFantasmaCn \rB &\qw\\
& \qw \poloFantasmaCn \rC & \gate {\tD_i} & \qw \poloFantasmaCn \rD &\qw}
\end{aligned}~:=~
\begin{aligned} \Qcircuit @C=1em @R=.7em @! R {& \qw \poloFantasmaCn \rA & \multigate{1} {\tC_i\otimes \tD_j} & \qw \poloFantasmaCn \rB &\qw\\ 
& \qw \poloFantasmaCn \rC & \ghost {\tC_i\otimes \tD_j} & \qw \poloFantasmaCn \rD &\qw} \end{aligned} 
\end{equation} 

If $\tC_i, \tD_j, \tE_k, \tF_l$ are events from $\rA$ to $\rB$, $\rB$ to $\rC$, $\rD$ to $\rE$, and $\rE$ to $\rF$, respectively, their parallel composition enjoys the property

\begin{equation}\label{prop:commutano}
\begin{split}
  \begin{aligned} 
\Qcircuit @C=1em @R=.7em @! R {& \qw \poloFantasmaCn {\rA} &\gate {\tD_j \circ \tC_i}  & \qw \poloFantasmaCn{\rC} &\qw \\
& \qw \poloFantasmaCn {\rD} &\gate {\tF_l \circ\tE_k}  & \qw \poloFantasmaCn{\rF} &\qw} & 
\end{aligned}
= 
\begin{aligned} \Qcircuit @C=1em @R=.7em @! R {& \qw \poloFantasmaCn {\rA} & \gate {\tC_i} & \qw \poloFantasmaCn {\rB} & \gate {\tD_j} & \qw \poloFantasmaCn{\rC} & \qw \\
& \qw \poloFantasmaCn {\rD} & \gate {\tE_k} & \qw \poloFantasmaCn{\rE} &  \gate {\tF_l} & \qw \poloFantasmaCn {\rF} &\qw}
\end{aligned}
\end{split}  
\end{equation}
\end{definition}

Note that property (\ref{prop:commutano}) implies that tests on different systems commute, that is, for every couple of events $\tC_i, \tD_j$
\begin{equation}
\begin{split}
\begin{aligned}  \Qcircuit @C=1em @R=.7em @! R {& \qw \poloFantasmaCn {\rA} &\gate { \tC_i}  & \qw \poloFantasmaCn{\rB} &\qw \\
& \qw \poloFantasmaCn {\rC} &\gate {\tD_j}  & \qw \poloFantasmaCn{\rD} &\qw}
\end{aligned} & =\begin{aligned}  \Qcircuit @C=1em @R=.7em @! R {& \qw \poloFantasmaCn {\rA} & \gate {\tC_i} & \qw \poloFantasmaCn {\rB} & \gate {\tI} & \qw \poloFantasmaCn{\rB} & \qw \\
& \qw \poloFantasmaCn {\rC} & \gate {\tI} & \qw \poloFantasmaCn{\rC} &  \gate {\tD_j} & \qw \poloFantasmaCn {\rD} &\qw}
\end{aligned} \\
 & = \begin{aligned} \Qcircuit @C=1em @R=.7em @! R {& \qw \poloFantasmaCn {\rA} & \gate {\tI} & \qw \poloFantasmaCn {\rA} & \gate {\tC_i} & \qw \poloFantasmaCn{\rB} & \qw \\
& \qw \poloFantasmaCn {\rC} & \gate {\tD_j} & \qw \poloFantasmaCn{\rD} &  \gate {\tI} & \qw \poloFantasmaCn {\rD} &\qw}
\end{aligned}
\end{split}  
\end{equation}
From now on, in diagrams like the above we will typically omit the box
with identity test, leaving just a wire for the corresponding system.
Also in formulae we will often omit the identity, e.g. for
$\tC\in\Trnset (\rA, \rB)$ and $\rho \in \Stset (\rA\rC)$ we will
often write $\tC \K{\rho}_{\rA\rB}$ in place of $(\tC \otimes \tI_{\rC}) \K
\rho_{\rA\rC}$.

Note that the difference between parallel and sequential composition of two
tests is already encoded in their input and output spaces: if the
input of a test is the output of the other the composition is
sequential, if all spaces are distinct the composition is parallel.
For this reason, when the kind of composition is evident we will omit the symbols $\circ$ and $\otimes$. For example, if $\rho$ is a preparation-event for $\rA$ and $\tC$ is an event from $\rA$ to $\rB$  we will write $\tC \K \rho_\rA$ in place of $\tC \circ \K{\rho}_\rA$, whereas if $\rho$ and $ \sigma$ are preparation-events for $\rA$ and $\rB$, respectively, we will write $\K \rho_\rA \K{\sigma}_\rB$ in place of $\K \rho_\rA \otimes \K \sigma_\rB$.

\subsection{Operational theories}
We are now in position to make more precise the notion ``operational theory": 
\begin{definition}[Operational theory] An operational theory is
  specified by a collection of systems, closed under composition, and by a collection of tests, closed under parallel and sequential composition.
\end{definition}

In an operational theory one can draw circuits that \emph{i)} represent the connections of
physical devices in an experiment, like e.g. the circuit 
\begin{equation}
  \Qcircuit @C=1em @R=.7em @! R {& \prepareC {\{\rho_i\}} &\qw \poloFantasmaCn {\rA} &\gate {\{\tC_j\}}  & \qw \poloFantasmaCn{\rB} & \measureD{\{a_k\}}} 
\end{equation}
and \emph{ii)} can also represent which specific set of events took
place in the experiment, like e.g. the circuit
\begin{equation}\label{scalar}
  \Qcircuit @C=1em @R=.7em @! R {& \prepareC {\rho_i} &\qw \poloFantasmaCn {\rA} &\gate {\tC_j}  & \qw \poloFantasmaCn{\rB} & \measureD{a_k}} 
\end{equation}
In particular, the latter circuit represents the preparation-event
$\rho_i$ followed by the event $\tC_j$ from system $\rA $ to system
$\rB$, which is in turn followed by the observation-event $a_k$ on
system $\rB$.  The whole sequence can be seen as single event $
p_{kji} := \B{a_k}_{\rB} \tC_j \K{\rho_i}_{\rA}$ from the trivial
system to itself.

\subsection{Relation with category theory}

In the previous Subsections we presented in an informal way the basic
notions pertaining to the use of physical devices in sequences and in
parallel. More formally, these notions can be summarized with the
language of category theory \cite{maclane}, which provides the suitable mathematical
framework capturing the fundamental structure presented so far.  In
this language, an operational theory is a category, where systems and events are respectively  \emph{objects}
and \emph{arrows}.  Every arrow has an input and
an output object, and arrows can be sequentially composed.  A test is
then a collection of arrows labeled by outcomes.

The fact that in an operational theory we have a parallel composition
of systems, and that such a composition is symmetric (i.e.  $\rA \rB
\simeq \rB \rA$) is expressed in technical words by saying that we
have a \emph{ strict symmetric monoidal category} \cite{maclane}.
In the next Subsection we will specify more requirements on this
category, imposing that the scalars (arrows from the trivial system to
itself) are probabilities.

\subsection{Probabilistic structure: states, effects, and transformations}\label{subsect:probint}
An operational theory is a language, whose words are diagrams representing circuits.  
With this language one can give instructions
to build up experiments or, alternatively, one can graphically represent which particular
outcomes took place in an experiment. However, in a physical theory one wants more: one wants to give
probabilistic predictions about the occurrence of  possible
outcomes.  To have this, there must be a rule assigning a probability
to every event from the trivial system to itself \cite{nota:rule}.  More directly, we
can say that in a probabilistic theory the events from the trivial
system to itself are probabilities, as in the following


\begin{definition}[Operational-probabilistic theory] An operational
  theory is \emph{probabilistic} if for every test $\{ p_i\}_{i \in
    \rX}$ from the trivial system $\rI$ to itself one has $p_i \in
  [0,1]$ and $\sum_{i \in \rX} p_i =1$, and the composition of two
  events from the trivial system to itself is given by the product of
  probabilities: $p_i\otimes q_j= p_i \circ q_j= p_i q_j$.
\end{definition}  


For short, we will often refer to operational-probabilistic theories simply
as probabilistic theories.

In a probabilistic theory, a preparation-event $\rho_i$ for system $\rA$ defines a function $\hat \rho_i$ sending observation-events of $\rA$ to probabilities:
\begin{equation}
\hat \rho_i: \Cntset (\rA) \to [0,1], \quad \B{a_j}\mapsto \SC{a_j}{\rho_i}.
\end{equation}
Likewise, an observation-event $a_j$ defines a function $\hat a_j$
from preparation-events to probabilities
\begin{equation}
\hat a_j:   \Stset (\rA)  \to [0,1], \quad \K{\rho_i}\mapsto \SC{a_j}{\rho_i}.  
\end{equation}    

From a probabilistic point of view, two observation-events
(preparation-events) corresponding to the same function are
indistinguishable. This leads to the notions of states and effects
(see \cite{ludwig,holevobook}):
\begin{definition}[States and effects]
  Equivalence classes of indistinguishable preparation-events are
  called \emph{states}. Equivalence classes of indistinguishable
  observation-events are called \emph{effects}.
\end{definition}   

From now on we will identify preparation-events with states and
observation-events with effects, without keeping the distinction
between an event $\rho_i$ ($a_j$) and the corresponding function $\hat
\rho_i$ ($\hat a_j$).  Accordingly, a preparation(observation)-test
will be a collection of states (effects), and the sets $\Stset (\rA),
\Cntset (\rA)$ will be the set of states and and the set of effects of
system $\rA$, respectively.  

\medskip 
{\bf Remark (states and effects in quantum theory).}  In quantum theory systems are associated with Hilbert spaces. The deterministic states of a system $\rA$ are represented by density matrices on the corresponding Hilbert space: a deterministic state $\rho$ is a matrix satisfying $\rho \ge 0$ and $\Tr [\rho]=1$.  A non-deterministic preparation-test $\{\rho_i\}_{i\in \rX}$, sometimes called a quantum information source, is a collection of positive operators with the property $\sum_{i \in \rX}  \Tr[\rho_i]=1$.  Accordingly, the set  $\Stset(\rA)$ of all states of system $\rA$ is the collection of all unnormalized density matrices $\rho$ with $\Tr[\rho]\le 1$.  An effect  is represented by positive operator $P$ with $P\le I_\rA$ ($I_\rA$ being the identity operator), and the probability resulting from the pairing between a state $\rho$ and and effect $P$ is given by the Born rule: $\SC P \rho_\rA = \Tr[P \rho]$.

\medskip 

Notice that according to the definition of states and effects as
equivalence classes, states are \emph{separating} for effects and effects are
\emph{separating} for states, that is,
\begin{equation}
\begin{split}
\K{\rho_0}_\rA = \K{\rho_1}_\rA &  \Longleftrightarrow  \SC a {\rho_0}_\rA = \SC a {\rho_1}_\rA \quad \forall a \in \Cntset (\rA)\\
\B {a_0}_\rA  = \B {a_1}_\rA & \Longleftrightarrow \SC  {a_0} \rho_\rA = \SC  {a_1} \rho_\rA \quad \forall \rho \in \Stset (\rA).
\end{split} 
\end{equation}
Since states (effects) are functions from effects (states) to
probabilities, one can take linear combinations of them. This defines
two real vector spaces $\Stset_\Reals (\rA)$ and $\Cntset_\Reals
(\rA)$, one dual of the other (we recall that the dual of a real vector space $V$ is the real vector space $V^*$ of all linear functions from $V$ to $\Reals$).   In this paper we will always restrict
our attention to the case of set of states that span finite dimensional vector spaces.  In this
case, by construction one has
\begin{equation}
\dim (\Stset_\Reals (\rA)) = \dim (\Cntset_\Reals (\rA)).
\end{equation}  
Notice that a spanning set for $\Stset _\Reals(\rA)$ is a separating set for $\Cntset_\Reals(\rA)$, while a spanning set for $\Cntset_\Reals (\rA)$ is a separating set for $\Stset_\Reals (\rA)$. 

Moreover, linear combinations with positive coefficients define two
convex cones $\Stset_{+} (\rA)$ and $\Cntset_{+} (\rA)$ (we recall that a set $S$ is a cone if for every $x\in S$ and for every $\lambda \ge 0$ one has $\lambda x \in S$, whereas the set is convex if for every $x,y \in S$ and for every $p\in [0,1]$ one has $p x + (1-p) y \in S$). Since the
pairing between states and effects yields positive numbers, one has
the inclusions 
\begin{equation}\label{coneinclusion} 
\begin{split}\Cntset_+ (\rA) &\subseteq \Stset_+ (\rA)^*  \\ \Stset_+ (\rA) &\subseteq \Cntset_+ (\rA)^*, 
\end{split} 
\end{equation}
where $\Stset_+ (\rA)^*$ and $\Cntset_+ (\rA)^*$ are the dual cones of
$\Stset_+ (\rA)$ and $\Cntset_+ (\rA)$, respectively.  We recall that the dual of a cone $S$ in some vector space $V$ is the cone $S^*$ defined by $S^*: = \{\lambda \in V^*, \lambda (x) \ge 0 ~\forall x\in S\}$.

\medskip
   
We conclude this Subsection by noting that every event $\tC_k$ from $\rA$ to
$\rB$ induces a linear map $\hat {\tC_k}$ from $\Stset_\Reals (\rA)$ to
$\Stset_\Reals (\rB)$, uniquely defined by \cite{nota:linearextension}   
\begin{equation}
\hat{\tC_k}: \K{\rho} \in \Stset (\rA)
\mapsto \tC_k \K{ \rho}_\rA \in \Stset (\rB).
\end{equation}

Likewise, for every system $\rC$ the event $\tC_i \otimes \tI_\rC$
induces a linear map from $\Stset_\Reals (\rA\rC)$ to $\Stset_\Reals
(\rB\rC)$. From a statistical point of view, if two events $\tC_i$ and
$\tC_i'$ induce the same maps for every possible system $\rC$, then
they are indistinguishable.
\begin{definition}[Transformations]
Equivalence classes of indistinguishable events from $\rA$ to $\rB$ are called \emph{transformations} from $\rA$ to $\rB$. 
\end{definition}
Again, we will assume that the equivalence classes have been already
done since the start, and, consequently, we will identify events with
transformations, without introducing new notation.  Accordingly, a
test will be a collection of transformations. 
\medskip 

{\bf Remark (transformations and tests in quantum theory)}.   In quantum theory, a transformation is usually called \emph{quantum operation}. Technically speaking, a quantum operation from $\rA$ to $\rB$ is a linear, completely positive, trace non-increasing map sending density matrices of system $\rA$ to (unnormalized) density matrices of system $\rB$.   A test $\{\tC_i\}_{i\in \rX}$ from $\rA$ to $\rB$ is typically referred to as a \emph{quantum instrument} \cite{davlew}, and is a collection of  quantum operations  with the property that $\sum_{i\in \rX}\tC_i$ is trace-preserving, namely $\sum_{i\in \rX}  \Tr [\tC_i (\rho)] =\Tr[\rho]$ for every  state $\rho$.
\medskip

{\bf Remark (different transformations).}  Note that two
transformations $\tC,\tD \in \Trnset (\rA , \rB)$ can be different
even if $\tC\K{ \rho}_\rA = \tD \K{ \rho}_\rA$ for every $\rho \in \Stset (\rA)$:
indeed to make $\tC$ different from $\tD$  it is enough that there exists an ancillary system $\rC$ and a joint state $\K{\rho}_{\rA \rC}$ such that $(\tC \otimes \tI_\rC)\K{\rho}_{\rA \rC} \not = (\tD \otimes \tI_\rC)\K{ \rho}_{\rA \rC}$. We will come back on this point when discussing local discriminability in Sect. \ref{sec:localdiscr}.

The following definitions will be used in the following
\begin{definition}[Channel]\label{def:channel}
A deterministic transformation $\tC \in \Trnset (\rA, \rB)$ is called \emph{channel}.
\end{definition}

\begin{definition}[Reversible channel]
A channel $\tU \in \Trnset (\rA, \rB)$ is called \emph{reversible} if there is another channel $\tW \in \Trnset (\rB, \rA)$ such that
\begin{equation}
\begin{split}
  \Qcircuit @C=1em @R=.7em @! R {& \qw \poloFantasmaCn \rA & \gate  \tU & \qw \poloFantasmaCn {\rB} & \gate \tW &\qw \poloFantasmaCn \rA &\qw} &=~   \Qcircuit @C=1em @R=.7em @! R {& \qw \poloFantasmaCn \rA &\gate \tI  & \qw \poloFantasmaCn\rA &\qw}  \\
  &\\
  \Qcircuit @C=1em @R=.7em @! R {&\qw \poloFantasmaCn {\rB}
    &\gate{\tW} & \qw\poloFantasmaCn\rA & \gate {\tU} & \qw
    \poloFantasmaCn {\rB} &\qw} &=~ \Qcircuit @C=1em @R=.7em @! R {&
    \qw \poloFantasmaCn {\rB} &\gate{\tI} & \qw\poloFantasmaCn{\rB}
    &\qw}
\end{split} 
\end{equation}
\end{definition}

If there exists a reversible channel $\tU$ from $\rA$ to $\rB$, then the systems $\rA$ and $\rB$ are operationally equivalent, in the sense of Def. \ref{def:opeq}. Note that the reversible channels from $\rA$ to itself form a group. We will denote this group  by $\grp G_\rA$.

We can now consider states that are invariant under the group of reversible transformations $\grp G_\rA$:
\begin{definition}[Invariant states] A state $\rho\in \Stset (\rA)$ is \emph{invariant} under the action of the group $\grp G_\rA$ if
\begin{equation}
\begin{aligned}
 \Qcircuit @C=1em @R=.7em @! R {\prepareC \rho & \qw \poloFantasmaCn \rA &\gate \tU & \qw \poloFantasmaCn \rA &\qw} 
\end{aligned}
~=~ 
\begin{aligned}
 \Qcircuit @C=1em @R=.7em @! R { \prepareC \rho & \qw \poloFantasmaCn \rA &\qw }
\end{aligned}
\qquad \forall \tU \in \grp G_\rA.
\end{equation}
\end{definition}
Similarly, we can consider channels with invariant output, that we call \emph{twirling channels}.
\begin{definition}[Twirling channels/Twirling tests]\label{def:twirling}
A channel $\tT \in \Trnset (\rA)$  is a \emph{twirling-channel} if 
\begin{equation}
\begin{aligned} \Qcircuit @C=1em @R=.7em @! R {& \qw \poloFantasmaCn \rA &\gate \tT & \qw \poloFantasmaCn \rA &\gate \tU & \qw \poloFantasmaCn \rA &\qw} 
\end{aligned}
~=~ 
\begin{aligned}
 \Qcircuit @C=1em @R=.7em @! R {& \qw \poloFantasmaCn \rA &\gate \tT &  \qw \poloFantasmaCn \rA &\qw }
\end{aligned}
\qquad \forall \tU \in \grp G_\rA. 
\end{equation}
If a test $\{\tC_i\}_{i \in \rX}$ is such that $\sum_{i \in \rX}
\tC_i$ is a twirling channel, we call it a \emph{twirling test}.
\end{definition}

We will see that in a theory with purification there is a unique invariant state and a unique
twirling channel for every system. 

\subsection{Relation with the convex sets framework}

The standard assumption in the literature is that, since the
experimenter is free to randomize the choice of devices with arbitrary
probabilities, all sets of states, effects, and transformations are
convex. We will call the theories satisfying this assumption ``convex".   
The assumption of convexity will be clarified in Subsect.
\ref{subsect:closeconv} in the context of \emph{causal theories}. 
Nevertheless,  for many of our results the assumption
of convexity is not essential, and we will discuss the validity of our
results in non-convex theories, like the toy-theories considered by
Spekkens in Ref.  \cite{toyspekkens}.  Bearing this in mind, whenever
possible we will present our results in a convexity-independent
language. 
We will add the specification ``convex'' to the theory for
those particular results in which convexity is essential.

In addition to the convexity of all sets of states, effects, and
transformations, the usual convex sets framework (see e.g. Refs.
\cite{mackey,ludwig,holevobook}, and, more recently, Refs.
\cite{hardy5,info-processingBarrett}) includes an assumption of
mathematical simplicity. The assumption is that every binary probability rule describes the statistics of a possible two-outcome experiment.  
Precisely, with the expression ``probability rule" we mean a collection of positive linear functionals $\{a_i\}_{i\in \rX} \subset \Stset_+^* (\rA)$ such that $\sum_{i\in \rX} \SC{a_i}{\rho}_\rA =1$ for every deterministic state $\rho \in \Stset(\rA)$.  We will refer to this assumption as ``no-restriction hypothesis", as it states that there is no restriction on the  set of (binary) probability rules that can be implemented in actual experiments. 
\begin{definition}[No-restriction hypothesis]\label{def:no-restriction}
A probabilistic theory satisfies the \emph{no-restriction hypothesis} if every binary probability rule $\{a_0,a_1\} \subset \Stset_+^* (\rA)$  is an observation-test.
\end{definition}  

In this paper we will not make this assumption.  However, we will
discuss a few implications of it in subsections \ref{subsect:no-info} and
\ref{subsect:completeness}.

\subsection{Coarse-graining and  refinement}
Here we give some definitions that will be often used in this paper. 

\begin{definition}[Coarse-graining] A test $\{\tC_i\}_{i \in \rX}$
  is a coarse-graining of the test $\{\tD_j\}_{j \in \rY}$ is there is
  a partition of $\rY$ into disjoint sets $\rY_i$ such that $\tC_i =
  \sum_{j \in \rY_i} \tD_j$ for every $i \in \rX$.  
\end{definition}
Since we can always decide to join two (or more) outcomes in a single
outcome, the set of all tests must be closed under coarse-graining.  

The inverse of coarse-graining is refinement: 
\begin{definition}[Refinement of a test] If $\{\tC_i\}_{i \in \rX}$ is a
  \emph{coarse-graining} of $\{\tD_j\}_{j \in \rY}$,  we say that
  $\{\tD_j\}_{j \in \rY}$ is a \emph{refinement} of $\{\tC_i\}_{i \in
    \rX}$.
\end{definition}  
\begin{definition}[Refinement of an event]
  A \emph{refinement} of the event $\tC $ is given by
  a test $\{\tD_j\}_{j \in \rY}$ and a subset $\rY_0\subseteq \rY$
  such that $\tC = \sum_{j \in \rY_0} \tD_i$.
\end{definition}
\begin{definition}
 We say that an event $\tD \in\Trnset(\rA, \rB)$ refines
  $\tC\in\Trnset(\rA, \rB)$, and write $\tD \prec\tC$, if there exist a refinement of $\tC$  such that  $\tD  \in \{\tD_j\}_{j \in \rY_0}$.
\end{definition}
\begin{definition}[Refinement set]\label{def:refset}   The refinement set $D_\tC$ of an event $\tC\in \Trnset (\rA, \rB)$ is the set of all events $\tD$ that refine $\tC$, namely $D_{\tC} := \{ \tD \in \Trnset (\rA, \rB)~|~ \tD \prec \tC\}$.
\end{definition}
\begin{definition}[Atomic vs refinable events]\label{def:atomic}
  An event $\tC$ is called \emph{atomic} if it admits only trivial refinements,---equivalently, if
  $\tD \prec \tC$ implies $\tD = \lambda \tC$ for some $\lambda \in [0,1]$. An event is \emph{refinable} if it is not atomic. 
\end{definition}

In the case of preparation-events the notion of refinement gives rise to the definitions of pure and mixed states:
\begin{definition}[Pure vs mixed states] An atomic preparation-event $\rho\in \Stset (\rA)$ is called
  \emph{pure state}. A refinable preparation-event is called \emph{mixed state}.
\end{definition}
Clearly, in a convex theory a state $\rho$ is pure if and only if it
is an extreme point of the convex set $\Stset (\rA)$. Moreover, in a
convex theory the refinement set $D_\rho$  is a convex subset
of the state space.
 For example, in quantum theory the refinement set of a density matrix $\rho$ is the set of all (unnormalized)  density matrices $\sigma$ such that $\sigma \le \rho$, and is clearly convex.   Note that the condition $\sigma\le \rho$  implies that the support of $\sigma$ is contained in the support of $\rho$. In fact, any density matrix $\sigma$ with ${\rm Supp} (\sigma) \subseteq {\rm Supp} (\rho)$, is proportional to a matrix in $D_\rho$.   In particular, if the support of $\rho$ is the whole Hilbert space (that is, if $\rho$ is a full-rank matrix), then any density matrix is proportional to a matrix in $D_\rho$. In this case $D_\rho$ is a spanning set for the set of all hermitian operators.  The analogue of a full rank density matrix in the general context is given by the notion of \emph{internal state}:

\begin{definition}[Internal state]\label{def:intst}
  A state $\omega\in\Stset(\rA)$ is {\em internal} if its refinements span the whole state space, i.e. if
  $\Span(D_\omega)=\Stset_\Reals (\rA)$.
\end{definition}

In the probabilistic theories considered in this paper every
preparation-test $\{\rho_i\}_{i \in \rX}$ for system $\rA$ admits an
ultimate refinement $\{\varphi_j\}_{j \in \rY}$, such that each state
$\varphi_j$ is pure.  Using the states-transformations isomorphism we
will also prove in Sect. \ref{sec:iso} that in a theory with purification
this property is enough to imply that every test $\{\tC_i\}_{i\in
  \rX}$ from $\rA$ to $\rB$ admits an ultimate refinement
$\{\tD_j\}_{j \in \rY}$, such that each event $\tD_j$ is atomic.

\subsection{Discrimination and distance}  

By making tests one can try to discriminate between different devices.
For example, imagine that we have a black box preparing one of the
two deterministic states $\rho_0, \rho_1 \in \Stset (\rA)$, and that
we want to find out which one.  To discriminate between the two states  we can perform a binary observation-test $\{a_0, a_1\}$. 
 The probabilities of outcomes are then given by 
\begin{equation}\label{eq:condiprob}
  p(j|i) :=  \SC {a_j} {\rho_i}_\rA \qquad i,j = 0,1.
\end{equation}
Assuming prior probabilities $\pi_0, \pi_1$ for the states $\rho_0,\rho_1$, respectively, we can try to maximize the (average) probability of correct discrimination, defined as $p_{succ} := \pi_0~ p(0|0) + \pi_1~ p(1|1)$.   Substituting the expression for the probabilities given in Eq. (\ref{eq:condiprob}) and using the fact probabilities sum up  to unit, we obtain 
\begin{equation}
\begin{split}
  p_{succ} & = \pi_0 +   \SC {a_1}  {\pi_1\rho_1-\pi_0\rho_0}_\rA \\
&= \pi_1 +  \SC {a_0} {\pi_0\rho_0 - \pi_1 \rho_1}_\rA,     
\end{split}
\end{equation}
and, optimizing over all binary tests,
\begin{equation}
\begin{split}
  p^{(opt)}_{succ} & = \pi_0 +   \sup_{a_1\in\Cntset(\rA)} \SC {a_1}  {\pi_1\rho_1-\pi_0\rho_0}_\rA \\
&= \pi_1 +  \sup_{a_0\in\Cntset(\rA)}\SC {a_0} {\pi_0\rho_0 - \pi_1 \rho_1}_\rA.    
\end{split}
\end{equation}
Summing the two expressions above we  finally get
\begin{equation}\label{psuccopt}
p^{(opt)}_{succ} = \frac {  1 +  |\!| \pi_1 \rho_1 - \pi_0 \rho_0 |\!|_\rA  } 2
\end{equation} 
where $|\!|\!  \cdot |\!|_\rA $ is the \emph{operational norm} defined by 
\begin{equation}\label{opnormstates}
|\!|   \delta  |\!|_{\rA} =     \sup_{a_1 \in \Cntset (\rA)}    \SC {a_1} \delta_{\rA}  - \inf_{a_0 \in \Cntset (\rA)}  \SC {a_0} \delta_\rA  \qquad \delta \in \Stset_\Reals (\rA).    
\end{equation}
Note that the norm $ |\!| \pi_1 \rho_1 - \pi_0 \rho_0 |\!|_\rA $
ranges between $0$ (when the two states and the prior
probabilities are equal) and 1 (when the two states
are perfectly discriminable).  For real numbers $x \in \Stset_{\Reals} (\rI) \equiv \Reals$ one has $|\!|  x |\!|_{\rI}  = |x|$.  
\medskip 

{\bf Remark (operational norm in quantum theory).} In quantum theory the operational norm is the usual \emph{trace-norm}  $|\!| \cdot |\!|_1$: Indeed, if we denote by $\delta_+$ and $\delta_-$ the positive and negative part of the hermitian operator $\delta = \pi_1 \rho_1 - \pi_0 \rho_0$, respectively, we obtain $|\!|  \delta |\!|_\rA  = \Tr[\delta_+] - \Tr[\delta_-] =|\!|  \delta |\!|_1$.
\medskip

In addition to the defining properties of a norm, the operational norm has a simple monotonicity property: 
\begin{lemma}[Monotonicity of the operational norm]\label{lem:monotonicitystates} If $\tC\in\Trnset(\rA,\rB)$  is a channel from $\rA$ to $\rB$, then for every $\delta \in \Stset_\Reals (\rA)$ one has
\begin{equation}
|\!|  \tC \delta |\!|_\rB  \le |\!|  \delta  |\!|_\rA.
\end{equation} 
If $\tC$ is reversible one has the equality.
\end{lemma}
\Proof By definition, $|\!|  \tC \delta |\!|_\rB  =\sup_{b_1\in\Cntset (\rB)}  \B {b_1}_\rB \tC \K\delta_\rA- \inf_{b_0\in\Cntset (\rB)}  \B {b_0}_\rB \tC \K\delta_\rA$. Since $\B {b_1} _\rB \tC$ and $\B{b_0}_\rB \tC$ are effects on system $\rA$, one has $|\!|  \tC \delta |\!|_\rB  \le\sup_{a_1\in\Cntset (\rA)}  \B {a_1}_\rB \K\delta_\rA- \inf_{a_0\in\Cntset (\rA)}  \B {a_0}_\rA  \K\delta_\rA = |\!|  \delta |\!|_\rA$. Clearly, if $\tC$ is reversible one has the converse bound $|\!|\delta |\!|_\rA =  |\!|\tC^{-1} \tC\delta|\!|_\rA \le |\!|  \tC \delta |\!|_\rB$, thus proving the equality $|\!| \delta |\!|_\rA = |\!| \tC \delta |\!|_\rB$.\qed
\medskip

For a generic state  $\rho \in \Stset (\rA)$, Eq.
(\ref{opnormstates}) reduces to
\begin{equation}\label{eq:rednorm}
|\!| \rho |\!|_\rA = {\sup_{e \in \Cntset (\rA)}}'~  \SC e \rho,
\end{equation}  
where $\sup'$ denotes the supremum restricted to the set of
deterministic effects.  
We can now give the notion of \emph{normalized states}:
\begin{definition}[Normalized states]
A state $\rho\in\Stset (\rA)$ is \emph{normalized} if $|\!| \rho |\!|_\rA =1$.  We will denote the set of normalized states by $\Stset_1 (\rA)$.
\end{definition}
Clearly, if $\rho$ is deterministic, then Eq. (\ref{eq:rednorm}) implies that  it is normalized (since $\rho$ corresponds to a single-outcome preparation-test and $e$ to a single-outcome observation-test, the probability of the only possible outcome, given by $\SC e \rho_\rA$, must be unit).  
In Sect. \ref{sec:causaltheo} we will
consider causal theories, where the deterministic effect $e \in
\Cntset (\rA)$ is unique, and, therefore one has $|\!| \rho |\!|_\rA = \SC
e \rho_{\rA}$. In this context one also has the converse: if a state is normalized, then it is deterministic.

\begin{definition}{\bf (Distinguishable states, discriminating
    tests)}\label{def:distinguishable}
  The states $\{\rho_i\}_{i \in \rX} $ are \emph{perfectly
    distinguishable} if there is a test $\{a_i\}_{i \in \rX}$ such
  that
\begin{equation}
\SC {a_j} {\rho_i} = |\!|  \rho_i   |\!|_\rA ~ \delta_{ij}.  
\end{equation}
The test $\{a_i\}_{i \in \rX}$ is called \emph{discriminating test}.
\end{definition}

{\bf Remark (Distinguishable states and discriminating test in quantum
  theory).}  In quantum theory a set of distinguishable states
$\{\rho_i\}_{i=1}^n$ is a set of density matrices with orthogonal
support. An example of discriminating test for this set is the
collection of orthogonal projectors $\{P_i\}_{i =1}^n$, where $P_i$ is
the projector on the support of $\rho_i$ for all $i <n$, while $P_n =
I - \sum_{i=1}^{n-1} P_i$. Clearly, the maximum number of
distinguishable states available for a certain system is the dimension
$d$ of the corresponding Hilbert space. In this case, the
distinguishable states are rank-one projectors on  an
orthonormal basis, and the corresponding discriminating test is the projective measurement on the same basis.

\medskip

If we want a theory that can describe the exchange of classical
messages, we need at least two states $\rho_0$ and $\rho_1$ that are
deterministic and perfectly distinguishable. In this case, a sender
can encode a classical bit $b=0,1$ in these two states and a receiver
can decode perfectly  the message by using the
binary discriminating test $\{a_0, a_1\}$. Indeed, one has $p(j|i ) =
\delta_{ij}$. Clearly, using this encoding for any bit in a string
allows perfect deterministic decoding of the whole string.
\medskip

We conclude this Subsection with a simple Lemma that will be useful in
the discussion of the general no-cloning theorem for probabilistic theories
(see Theorem \ref{lemma:clon-disc}):
\begin{lemma}\label{lemma:worstcase}
In any convex theory, if two deterministic states $\rho_0, \rho_1\in \Stset (\rA)$ are distinct (\emph{i.e.} $\rho_0 \not = \rho_1$), then there exists a binary test $\{a_0, a_1\}$ such that 
\begin{equation}
p(1|0) = p(0|1) < \frac 1 2 .
\end{equation}
 \end{lemma} 

 \Proof Since the states are distinct there exists at least an effect
 $ a$ such that $\SC a {\rho_0} > \SC a {\rho_1}$. Moreover, since the
 theory is convex we can choose without loss of generality $\SC a
 {\rho_1}\ge 1/2$ (if $a$ does not meet this condition, we can replace
 it with the convex combination $a' = 1/2 (a + e)$).  Now define the
 binary test $\{a_0, a_1\}$ by the convex combination
\begin{equation} 
\left\{ 
\begin{array}{ll}  
  a_0 &= q a + (1-q) 0\\
  a_1 & = e-a_0
\end{array} \right. \qquad  q =
\frac 1 {\SC a {\rho_0} +\SC a {\rho_1}} 
\end{equation} 
where $0$ is the null effect, defined by $\SC 0 \rho_\rA = 0, \forall \rho\in\Stset (\rA)$.  For this test one has  $p (1|0) = p(0|1) = \SC a
{\rho_1} / [\SC a {\rho_0} +\SC a {\rho_1}] <1/2$. \qed
\medskip

The above Lemma states that if two states are different, then the
worst-case error probability, defined as $p_{wc} := \max\{ p(1|0),
p(0|1)\}$, can be reduced to a value that is strictly smaller than
$1/2$. In other words, if two states are different, then in the
worst-case scenario we can always distinguish between them better than
with a random guess.

\subsection{Closure}\label{subsect:closure}

The closure of $\Stset (\rA)$ with respect to the operational norm
contains all the elements of $\Stset_{\Reals}(\rA)$ that can be approximated arbitrarily well by
physical states: a vector $\rho\in \Stset_{\Reals}(\rA)$ is in the closure if there is a sequence of states $\{\rho_n\}$ such that $\lim_{n\to \infty} |\!| \rho - \rho_n  |\!|_\rA =0$.  Since $\Stset_\Reals (\rA)$ is finite dimensional,
it is natural to assume that all such vectors correspond to physical states.  We will make this assumption in the paper.  In
particular, assuming that the set $\Stset (\rI)$ of states of the
trivial system is closed with respect to the operational norm means
assuming that the probabilities appearing in the theory form a closed
subset of the interval $[0,1]$. 
In fact, we have the following:
\begin{lemma}\label{lemma:densein01}  If an operational-probabilistic theory is not  deterministic, then $\Stset (\rI)$ is dense in the interval $[0,1]$.
\end{lemma} 
\Proof If the theory is not deterministic there is a binary
test giving outcomes $0,1$ with probabilities $q_0, q_1 \not = 0$,
respectively.  Now, this test provides a biased coin, which can be
tossed many times, thus allowing for the
approximation of any coin with bias $p\in [0,1]$ \cite{ranextr}.  \qed
\medskip

Therefore, if we assume that the set of states $\Stset(\rI)$ is closed, then the previous Lemma implies the following:
\begin{corollary}
If $\Stset (\rI)$ is closed, then it is the whole interval $[0,1]$. 
\end{corollary}
In Subsect. \ref{subsect:closeconv} we will discuss the relation
between closure and convexity in the context of causal theories.


\section{Causal theories}\label{sec:causaltheo}  

In this Section we restrict our attention to causal theories, in which
the probability of outcomes of an experiment at a given time does not depend on the
choice of experiments performed at later times.

\subsection{Definition and main properties}

Although in the circuits discussed until now we had sequences of
tests, such sequences were not necessarily \emph{causal sequences}.
The input-output arrow determined by the connections of
physical devices was not necessarily the causal arrow defined a signalling structure.   In fact, one can formulate
operational-probabilistic theories even in the absence of a pre-defined causal
arrow, and this is a crucial point to formulate a quantum theory of
gravity (see e.g. Hardy in Ref. \cite{causaloid}).  A concrete example
of non-causal theory is the theory studied in Refs.
\cite{supermaps,comblong}, where the states are quantum operations,
and the transformations are ``supermaps'' transforming quantum
operations into quantum operations. In this case, transforming a
``state'' means inserting the corresponding quantum operation in a
larger circuit, and the sequence of two such transformations is not a
causal sequence.  However, the analysis of non-causal theories is not
the scope of the present work. We now give the condition that allows
us to interpret sequential composition as a causal cascade:
\begin{definition}[Causal theories] A theory is \emph{causal} if for
  every preparation-test $\{\rho_i\}_{i \in \rX}$ and every
  observation-test $\{a_j\}_{j \in \rY}$ on system $\rA$ the marginal
  probability $p_i := \sum_{j\in\rY} \SC {a_j} {\rho_i}_{\rA}$ is
  independent of the choice of the observation-test $\{a_j\}_{j \in
    \rY}$.  Precisely, if $\{a_j\}_{j \in \rY}$ and $\{b_k\}_{k \in
    \rZ}$ are two different observation-tests, then one has
  \begin{equation}\label{no-sig}
   \sum_{j \in \rY}  \SC {a_j} {\rho_i}_\rA = \sum_{k \in \rZ} \SC {b_k} {\rho_i}_{\rA}.
  \end{equation}
\end{definition}

Loosely speaking, we may say that the condition of Eq. (\ref{no-sig})
expresses the principle of ``no-signaling from the future''.  

Causal theories have a simple characterization:
\begin{lemma}[Characterization of causal theories]\label{lem:charcaus} A theory is causal if and only if
  for every system $\rA$ there is a unique deterministic effect
  $\B{e}_\rA$.
\end{lemma}
\Proof Suppose that $e$ and $e'$ are two deterministic effects for
system $\rA$. Since deterministic effects belong to single-outcome
tests, Eq. (\ref{no-sig}) gives $\SC e {\rho_i}_\rA = \SC {e'}
{\rho_i}_\rA$ for every state $\rho_i$. Therefore, $e = e'$.
Conversely, suppose that the deterministic effect is unique and take
an observation-test $\{a_j\}_{j \in \rY}$ on system $\rA$.  Then by
coarse-graining one obtains a single-outcome test, with
deterministic effect $\B {e'}_\rA = \sum_{j \in \rY} \B {a_j}_{\rA}$,
and, by uniqueness of the deterministic effect,  $\B e_\rA = \B
{e'}_\rA = \sum_{j\in \rY} \B{a_j}_\rA$.  Therefore, for every state
$\rho_i$ we have $\sum_{j \in \rY} \SC {a_j} {\rho_i}_\rA = \SC e
{\rho_i}_\rA$, independently of the choice of the observation-test
$\{a_j\}_{j \in \rY}$. This proves Eq. (\ref{no-sig}). \qed
 
\medskip
 
{\bf Remark (quantum theory as an example of causal theory)} 
Ordinary quantum theory is an example of causal theory.  
Indeed, there is a unique deterministic effect, corresponding to the (trace with the) identity operator on the system's Hilbert space. In other words, the only operator $P$ satisfying the equation $\Tr_\rA[P \rho] =1$ for every density matrix is $P=I_\rA$, the identity on $\rA$.       
\medskip

An immediate consequence of causality is that the deterministic
effect of a composite system $\rA \rB$ is the product of the deterministic effects of $\rA$ and $\rB$, as expressed by the following
\begin{corollary}{\bf (Factorization of the deterministic effect on product systems)} Let $\rA$ and $\rB$ be two arbitrary systems. In a causal theory one has 
\begin{equation}
\B e_{\rA \rB} = \B e_\rA \B e_\rB.
\end{equation}
\end{corollary}
\Proof Since the parallel composition of two single-outcome tests is a single-outcome test, the effect $\B e_\rA \B e_\rB$ is deterministic, according to Def. \ref{def:deterministic}.  Since the deterministic effect $\B e_{\rA\rB}$ is unique, one must have $\B e_\rA \B e_\rB= \B e_{\rA\rB}$. \qed

\medskip

Note that in a causal theory there is a unique way of defining marginal states:
\begin{definition}[Marginal state]  The marginal state of  $\K\sigma_{\rA \rB}$ on system $\rA$ is the state $\K \rho_\rA := \B e_\rB \K\sigma_{\rA \rB}$. 
\end{definition}

In a causal theory the channels (deterministic transformations corresponding to single-outcome tests) are characterized as follows:
\begin{lemma}[Characterization of channels]\label{lem:charchan}
In a causal theory a transformation $\tC \in \Trnset(\rA, \rB)$ is a channel (Def. \ref{def:channel} ) if and only if $\B e_\rB  \tC = \B e_\rA$.  Diagrammatically,
\begin{equation}\label{channorm}
  \Qcircuit @C=1em @R=.7em @! R {& \qw \poloFantasmaCn \rA & \gate {\tC} & \qw \poloFantasmaCn \rB & \measureD e } ~=~  \Qcircuit @C=1em @R=.7em @! R { & \qw \poloFantasmaCn \rA & \measureD e }
\end{equation}
In particular, a state $\rho\in \Stset (\rB)$ is deterministic if and only if $\SC e \rho_\rB =1$.
\end{lemma}
\Proof If $\tC$ is a channel, then $\B e_\rB \tC$ is a deterministic
effect. By uniqueness of the deterministic effect,  Eq. (\ref{channorm}) holds. Conversely, suppose that
$\{\tC_i\}_{i \in \rX}$ is a test from $\rA$ to $\rB$ and $\tC \equiv
\tC_{i_0}$ is a transformation such that Eq. (\ref{channorm}) holds.
By coarse-graining, we can define the channel $\tC' :=\sum_{i \in \rX}
\tC_i$. Since $\tC'$ is a channel, we must have $\B e_\rA =\B e_\rB
\tC' = \B e_\rA + \B e_\rB (\sum_{i \not = i_0} \tC_i)$, whence $\B
e_\rB (\sum_{i \not = i_0} \tC_i) =0$. But this implies $\sum_{i \not = i_0 }\tC_i =0$, and, therefore, $\tC = \tC'$.  Hence, $\tC$ is a
channel. Finally, a deterministic state is nothing but a channel with trivial input system $\rA=\rI$. Since the deterministic effect of the trivial system $\rI$ is the number $1$, the normalization of Eq. (\ref{channorm}) becomes $\SC e \rho_\rB =1$. 
\qed
\medskip
 
Lemma \ref{lem:charchan} also leads to the following
\begin{corollary}[Normalization of tests]\label{obstestnorm} A test $\{\tC_i\}_{i\in \rX}$ from $\rA$ to $\rB$ satisfies the normalization condition 
\begin{equation}\label{eq:testnorm}
\sum_{i \in \rX} \B e_\rB  \tC_i = \B e_\rA.
\end{equation}
In particular, an observation-test $\{a_i\}_{i \in \rX}$ on system $\rA$ must satisfy the normalization condition 
\begin{equation}\label{eq:obstestnorm}
\sum_{i\in \rX}  \B{a_i}_\rA = \B e_{\rA}.
\end{equation}
\end{corollary}

In quantum theory, the normalization condition of Eq. (\ref{channorm}) means that any quantum channel must be trace-preserving (identity preserving in the Heisenberg picture). Indeed, the deterministic effect is the identity operator, and Eq. (\ref{channorm}) implies that, for every quantum state $\rho$, one has $\Tr_\rB[ \tC (\rho)] = \Tr_\rA[\rho]$.    The normalization condition for observation-tests given in Eq. \ref{eq:obstestnorm} is instead the normalization of quantum measurements: a quantum measurement is a POVM, that is a collection of positive operators $\{A_i\}_{i\in \rX}$ satisfying the condition $\sum_{i\in \rX}  A_i = I_\rA$, where $I_\rA$ is the identity operator on the system's Hilbert space.   

\medskip

Moreover, in a causal theory we have a simple characterization of the normalized states:
\begin{corollary}[Characterization of normalized states]
Let $\rho$ be a state of system $\rA$. In a causal theory the following are equivalent 
\begin{enumerate} 
\item $\rho$ is normalized
\item    $\SC e \rho_\rA = 1$
\item $\rho$ is deterministic. 
\end{enumerate}
\end{corollary} 
\Proof Since there is a unique deterministic effect,  the expression of the norm given in Eq. (\ref{eq:rednorm}) yields $|\!| \rho |\!|_\rA = \SC e \rho_\rA$.  This proves the  equivalence $1\Leftrightarrow 2$.  The equivalence 2 $\Leftrightarrow$ 3 was already proved in Lemma \ref{lem:charchan}. \qed
 \medskip
  
For every  state $\K\rho_\rA$  we can consider the normalized state  
\begin{equation}
\K{\bar \rho}_{\rA} : = \frac{\K{\rho}_\rA}{\SC e \rho_{\rA}}.
\end{equation} 
Operationally, this means that we can always make  \emph{rescaled
  preparations}: we can perform a preparation-test $\{\rho_i\}_{i \in
  \rX}$, and, if the test gives outcome $i_0$ we can claim that we
prepared the normalized state $\bar \rho_{i_0}$.  In other words, in a causal theory any preparation-event can be promoted to a single-outcome preparation-test. Following this observation, in a causal
theory there is no reason to forbid that every normalized state can be
actually produced in some single-outcome test.  This implies that every
state is proportional to a deterministic one.  In the following we will always assume this fact as a property of causal theories.

Note that also the converse is true:
\begin{lemma}{\bf (Causality is necessary for rescaled preparations)}
  A theory where every state is proportional to a deterministic one is
  causal.
\end{lemma}
\Proof Let $\K{\rho}_\rA$ be an arbitrary state and $e$ and $e'$ be
two deterministic effects. By hypothesis, we have $\K \rho_\rA = k
\K{\bar\rho}_\rA$, where $\bar \rho$ is deterministic. This implies $\SC e
\rho_\rA = k = \SC {e'} \rho$, and, since $\rho$ is arbitrary $e =
e'$. By lemma \ref{lem:charcaus}, this implies that the theory is
causal. \qed

\medskip

Remarkably, the causal principle of ``no-signalling from the future''
implies the impossibility of signalling in space without exchange of physical
systems:

\begin{theorem}{\bf (No-signalling without exchange of physical
    systems)} In a causal theory it is impossible to have signalling
  without exchanging systems.
\end{theorem} 
\Proof Suppose that two distant parties Alice and Bob share a
bipartite state $\K{\Psi}_{\rA\rB}$, and that  Alice (Bob) performs
a local test $\{\tA_i\}_{i \in \rX}$ ($\{\tB_j\}_{j \in \rY}$) on the
system at her (his) disposal.  Let us define the joint probability
$p_{ij} := \B e_{\rA \rB} (\tA_i \otimes \tB_j ) \K\Psi_{\rA \rB} $
and its marginal $p^{(\rA)}_i := \sum_j p_{ij}$ ($p^{(\rB)}_j :=
\sum_i p_{ij}$) on Alice's (Bob's) side.  It is immediate to verify
that the marginal $p_i^{(\rA)}$ on Alice's side does not depend on the
test $\{\tB_j\}$ on Bob's side: indeed, one has
\begin{equation}
\begin{split}
p_i^{(\rA)} &= \sum_{j} \B e_\rA \B e_\rB (\tA_i \otimes \tB_j ) \K{\Psi}_{\rA \rB}  \\
&=\B e_\rA  \left(\tA_i \otimes  \left[\sum_{j}  \B e_\rB \tB_j \right]\right) \K{\Psi}_{\rA \rB}\\
&= \B e_\rA \tA_i  \K\rho_\rA ,  
\end{split}
\end{equation}
having used the normalization condition $\sum_j \B e_\rB \tB_j = \B
e_\rB$ (Corollary \ref{obstestnorm}), and having defined the marginal state $\K\rho_\rA := \B e_\rB \K \Psi_{\rA \rB}$.
The same reasoning holds for the marginal on Bob's side. \qed

\subsection{Conditioning}

In a causal sequence the choice of a device can depend on the outcomes
of previous devices. This gives rise to the notion of
\emph{conditioned test}, which generalizes the notion of sequential
composition:
\begin{definition}[Conditioned test]  
  If $\{\tC_i\}_{i \in \rX}$ is a test from $\rA$ to $\rB$ and, for
  every  $i$, $\left\{\tD^{(i)}_{j_i}\right\}_{j_i \in \rY_i}$ is
  a test from $\rB$ to $\rC$, then the \emph{conditioned test} is a test
  from $\rA$ to $\rC$, with outcomes $(i,j_i) \in \rZ :=\bigcup_{i\in\rX}
  \{i\} \times \rY_i$, and events $\{\tD^{(i)}_{j_i}\circ\tC_i\}_{(i,j_i)
    \in \rZ}$.  Diagrammatically, the events $\tD^{(i)}_{j_i} \circ \tC_i$
  are represented as follows
\begin{equation}
  \Qcircuit @C=1em @R=.7em @! R {& \qw \poloFantasmaCn \rA & \gate {\tC_i} & \qw \poloFantasmaCn \rB &\gate{\tD^{(i)}_{j_i}} & \qw\poloFantasmaCn\rC &\qw} ~:=~ \Qcircuit @C=1em @R=.7em @! R {& \qw \poloFantasmaCn \rA & \gate {\tD^{(i)}_{j_i}\circ \tC_i} & \qw \poloFantasmaCn \rC  &\qw}
\end{equation}
\end{definition} 

The above definition of conditioning makes sense in a causal theory,
where the uniqueness of the deterministic effect ensures that the test
$\{ \tD_{j_i}^{(i)} \circ \tC_i\}_{i \in \rX, j_i \in \rY_i}$
  satisfies the normalization condition required by Corollary \ref{obstestnorm}: 
\begin{equation}\sum_{i \in \rX}
  \sum_{j_i \in \rY_i}  \B e_\rC \tD_{j_i}^{(i)} \circ \tC_i  = \sum_{i \in \rX} \B e_\rB \tC_i = \B
  e_\rA.
\end{equation}

Conditioning expresses the possibility of choosing what to do at a
certain step using the classical information generated in the previous
steps. In a causal operational theory there is no reason to forbid 
an experimenter to perform conditioned tests. Accordingly, in the following we will assume that in a causal theory any conditioned test is allowed.
In fact, the possibility to perform conditioned tests is essentially equivalent to causality. Indeed, one has also the converse statement:
\begin{lemma}{\bf (Causality is necessary for conditioned tests)}
A theory where every conditioned test is possible is causal.  
\end{lemma}
\Proof To prove that the theory is causal we show that for every system $\rA$ the deterministic effect $\B e_\rA$ is unique. Suppose that $\B e_\rA$ and $\B {e'}_\rA$ are two deterministic effects, and let $\rho\in \Stset(\rA)$ be an arbitrary state. By definition, there is a preparation-test $\{\rho_i\}_{i \in \rX}$ that contains $\rho$, that is, $\rho = \rho_{i_0}$ for some outcome $i_0 \in \rX$. Moreover, using coarse-graining we obtain the two-outcome preparation-test $\{\rho_0,\rho_1\}$, where $\rho_0 = \rho$ and $\rho_1 := 
\sum_{i \not = i_0} \rho_i$.   Now, consider the conditioned test $\{ \SC e {\rho_{0}}_\rA, \SC {e'} {\rho_1}_\rA  \}$, defined  by the following procedure: first perform the preparation-test $\{\rho_{0},\rho_{1}\}$, and then, if the outcome is $0$ apply the effect $\B e_\rA$, otherwise apply  $\B {e'}_\rA$.    Since $\{ \SC e {\rho_{0}}_\rA, \SC {e'} {\rho_{1}}_\rA  \}$ is a test from the trivial system to itself one must have  
\begin{equation} 
\SC e {\rho_{0}}_\rA + \SC {e'} {\rho_1}_\rA  =1
\end{equation}
On the other hand, since the effect $e'$ is deterministic, one must have $\SC {e'} {\rho_{0}}_\rA + \SC {e'} {\rho_{1}}_\rA  =1$. By comparison, this implies $\SC e {\rho_{0}}_\rA = \SC {e'} {\rho_{0}}_\rA $, and, since $\rho_{0}$ was a generic state,  $e = e'$. \qed

\medskip

{\bf Remark (conditioning with different outputs and ``direct sum" systems).}  
In principle, one could also consider a conditioning where the output
system of each test $\left\{ \tD^{(i)}_{j_i} \right\}$ is a system
$\rC_i$ that depends on the outcome $i$.  In this case the output
 of the conditioned test would be a ``direct sum'' system ``$\rC
:=\bigoplus_{i \in \rX} \rC_i$''.  In quantum theory, this situation can be described introducing a superselection rule, according to which the possible states of the ``direct sum" system are the block-diagonal density matrices  of the form $\rho = \bigoplus_{i \in \rX}  \rho_i$, where each $\rho_i$ is a density matrix on the Hilbert space associated to system $\rC_i$.    
This kind of extension would also require treating
the outcome spaces $\rX$ as a \emph{classical systems} that can be the
input or the output of some classical information-processing device.
However, we will not consider here this generalization as it is not
needed for the main purpose of the paper.  

\medskip

A particular case of conditioning is \emph{randomization}: 
\begin{definition}[Randomization] If $\{p_i\}_{i \in \rX}$ is a preparation-test for the trivial system and, for every outcome $i$, $\{\tC_{j_i}^{(i)}\}_{j_i \in \rY_i}$ is a test from $\rA$ to $\rB$,   the \emph{randomized test} $\{ p_i  \tC^{(i)}_{j_i}\}_{i \in \rX, j_i \in \rY_i}$ is the test from $\rA$ to $\rB$ with events defined by
\begin{equation}
p_i ~\begin{aligned} \Qcircuit @C=1em @R=.7em @! R {& \qw \poloFantasmaCn \rA & \gate { \tC^{(i)}_{j_i}} & \qw \poloFantasmaCn \rB  &\qw}
\end{aligned}  ~:=~ 
\begin{aligned}\Qcircuit @C=1em @R=.7em @! R {&\qw  \poloFantasmaCn \rA& \qw &  \qw &  \gate{\tC^{(i)}_{j_i}} & \qw \poloFantasmaCn\rB &\qw\\
    & \qw \poloFantasmaCn \rI & \gate {p_i} & \qw & \qw  & \qw \poloFantasmaCn \rI& \qw} 
\end{aligned}
 \end{equation}
(on the left-hand side we used the fact that that the composition with trivial systems is trivial, and, therefore, one has $\rA \rI = \rA ,\rB \rI = \rB)$.
\end{definition}

If a causal theory is not deterministic (i.e. if the possible values of probabilities are not only 0 and 1)  then randomization and coarse-graining always allows one to construct an internal state (see Def. \ref{def:intst}): it is enough to take a spanning set of
states $\{\rho_i\}_{i \in\rX}$, to randomize them with some
non-zero probabilities $\{p_i\}_{i \in \rX}$, and then to coarse-grain, thus getting the internal state
$\omega = \sum_{i \in \rX} p_i \rho_i$.


\medskip 

Finally, conditioning allows one to prove that a causal theory contains
all possible \emph{measure-and-prepare} channels, defined as follows

\begin{definition}[Measure-and-prepare channels] A channel $\tC \in \Trnset (\rA, \rB)$ is \emph{measure-and-prepare} if there exists an observation-test $\{a_i\}_{i \in \rX}$ on $\rA$, and a collection of normalized states $\{ \beta_i\}_{i \in \rX}  \subset \Stset_1(\rB)$ such that  
\begin{equation}
\tC = \sum_{i \in \rX} \K{\beta_i}_\rB  \B {a_i}_\rA.
\end{equation} 
\end{definition}

\subsection{Distance between transformations}
Here we introduce a norm for transformations that has a direct operational interpretation:  it quantifies the maximum probability of success in the discrimination of two channels in a causal theory.  Suppose that we are given two channels $\tC_0, \tC_1 \in \Trnset (\rA, \rB)$
with prior probabilities $\pi_0, \pi_1$, respectively. In a causal
theory, the most general way to discriminate is to prepare a bipartite
input state $\rho \in\Stset_1 (\rA\rC)$, to apply the unknown channel, and to
perform a binary test that distinguishes between the two possible
output states $\tC_0 \K \rho_{\rA \rC}$ and $\tC_1 \K \rho_{\rA\rC}$.
Optimizing over all binary tests and using Eq. (\ref{psuccopt}) we
obtain the success probability $p_{succ} = 1/2 (1 + |\!| (\pi_1\tC_1
-\pi_0\tC_0) \rho |\!|_{\rB\rC})$ Moreover, optimizing the input state and
the extension we find the maximum probability of success
\begin{equation}\label{eq:psuccchan}
p^{opt}_{succ} = \frac 1 2 (1 + |\!| \pi_1 \tC_1  - \pi _0 \tC_0   |\!|_{\rA,\rB} )
\end{equation}
where the operational norm for transformations is defined by
\begin{equation}\label{opnormtrans}
  |\!| \Delta |\!|_{\rA, \rB}  = \sup_{\rC} \sup_{\rho \in \Stset_1 (\rA\rC)} |\!|  \Delta  \rho  |\!|_{\rB\rC} \qquad \Delta \in \Trnset_\Reals (\rA, \rB).
\end{equation} 



In quantum theory our expression for the operational norm reduces to the diamond norm  in Schr\"odinger picture \cite{diamond}, or equivalently, to the completely bounded (CB)  norm in Heisenberg picture  \cite{paulsen}. 

\medskip  
 
In the case of trivial input system $\rA = \rI$, Eq. (\ref{opnormtrans}) gives back the norm of states introduced in Eq. (\ref{opnormstates}).    In the case of trivial output system $\rB =\rI$, it provides an operational norm for effects, given by
\begin{equation}\label{opnormeff}
|\!|  \delta |\!|_{\rA,\rI} =  \sup_{\rC} \sup_{\rho\in\Stset_1 (\rA\rC)}  |\!|    \delta \rho   |\!|_\rC \qquad \delta \in \Cntset_\Reals (\rA).  
\end{equation} 
In fact, the extension with the ancillary system $\rC$ is not needed in this case: 
\begin{lemma}
The operational norm of an element of the vector space $\delta\in\Cntset_{\Reals} (\rA)$ spanned by the effects for system $\rA$  is given by the expression
\begin{equation}\label{opnormeffects1}
|\!| \delta |\!|_{\rA,\rI} = \sup_{\rho\in\Stset_1 (\rA)} |\SC \delta \rho_\rA|.
\end{equation}
\end{lemma}
\Proof  Taking $\rC = \rI$ in Eq. (\ref{opnormeff}) yields the lower bound $|\!|  \delta |\!|_{\rA,\rI} \ge \sup_{\rho\in\Stset_1(\rA)} |\!|\SC \delta \rho_\rA |\!|_\rI = \sup_{\rho\in\Stset_1 (\rA)}| \SC \delta \rho_\rA|$, where we used the fact that the norm of a real number $x \in \Reals \equiv \Stset_\Reals (\rI)$ is given by its modulus: $|\!|x|\! |_{\rI}  = |x|$.  To prove the equality of Eq. (\ref{opnormeffects1}) we now prove converse bound. By the definition of the operational norm for states in Eq. (\ref{opnormstates}), for every $\sigma\in\Stset_1(\rA\rC)$ we have 
\begin{equation}
\begin{split} 
|\!|  \delta \sigma|\!|_\rC &= \sup_{c_1\in\Cntset (\rA)}   \B \delta_\rA \B {c_1}_\rC \K \sigma_{\rA\rC} -\inf_{c_0\in\Cntset (\rA) } \B \delta_\rA \B {c_0}_\rC \K\sigma_{\rA\rC}\\
&= \sup_{\{c_0,c_1\}}  \B \delta_\rA \B{c_1-c_0}_\rC \K\sigma_{\rA\rC},
\end{split}
\end{equation}
where the optimization in the last equation is over all possible binary tests $\{c_0, c_1\}$ for system $\rC$.  Now, applying the observation-test $\{c_0, c_1\}$ to the bipartite state $\K \sigma_{\rA\rC}$ we obtain a preparation-test $\{\rho_0,\rho_1\}$ for system $\rA$, defined by  $\K {\rho_i}_\rA = \B {c_i}_\rC  \K\sigma_{\rA\rC}, i = 0,1$.  Defining the probabilities $p_i = \SC e {\rho_i}_\rA$ and the normalized states $\bar \rho_i = \rho_i/ \SC e {\rho_i}_\rA$ we then have
\begin{equation}
\begin{split} 
 \B \delta_\rA \B{c_1-c_0}_\rC \K\sigma_{\rA\rC}  & = p_1 \SC \delta {\bar \rho_1}_\rA - p_0 \SC \delta{\bar \rho_0}_\rA\\
 &\le \max\{  \SC \delta {\bar \rho_1}_\rA, -\SC \delta {\bar \rho_0}_\rA \}\\
 &\le \sup_{\rho\in\Stset_1 (\rA)}   |\SC \delta \rho_\rA|.
\end{split}
\end{equation}
\qed

In quantum theory the norm $|\!| D|\!|_{\rA,\rI}$ of an hermitian operator on the Hilbert space of system $\rA$  coincides with the operator norm $|\!| D|\!|_\infty = \sup_{\rho\ge 0, \Tr[\rho]=1} |\Tr[D \rho]| =\max_i\{|d_i|\}$, where $\{d_i\}$ are the eigenvalues of $D$. 
\medskip

We conclude by mentioning a monotonicity property of the operational norm of transformations: 
\begin{lemma}\label{lem:monotonicitytransformations}{\bf (Monotonicity of the operational norm for transformations)}
If $\tC\in \Trnset(\rA,\rB)$  and $\tE \in \Trnset (\rC,\rD)$ are two channels, then for every $\Delta \in \Trnset_{\Reals} (\rB,\rC)$ one has
\begin{equation}
|\!|  \tE \Delta \tC |\!|_{\rA,\rD} \le  |\!|  \Delta  |\!|_{\rB,\rC}. 
\end{equation}
If $\tC$ and $\tE$ are reversible one has the equality.
\end{lemma} 

\Proof Let $\rR$ be an ancillary system, and $\rho\in\Stset_1 (\rA\rR)$ be a normalized state of $\rA\rR$. Then, since $\K\sigma_{\rB\rR} =\tC \K\rho_{\rA\rR}$ is a normalized state of $\rB\rR$,  we have $|\!|  \tE \Delta \tC |\!|_{\rA,\rD} = \sup_{\rR} \sup_{\rho\in\Stset_1 (\rA\rR)}  |\!|  \tE \Delta \tC \rho |\!|_{\rD\rR} \le \sup_\rR \sup_{\sigma\in\Stset_1 (\rB\rR)}  |\!|  \tE \Delta \sigma |\!|_{\rD\rR}$. Now, using Lemma \ref{lem:monotonicitystates} we obtain $  |\!|  \tE \Delta \sigma |\!|_{\rD\rR}\le |\!|  \Delta \sigma |\!|_{\rC\rR}$. Hence, $|\!|  \tE \Delta \tC |\!|_{\rA,\rD} \le  \sup_\rR \sup_{\sigma \in \Stset_1 (\rB\rR)}  |\!| \Delta \sigma|\!|_{\rC\rR} = |\!| \Delta |\!|_{\rB,\rC}$.  
Clearly, if $\tC$ and $\tE$ are reversible, one has  the converse bound $|\!|  \Delta |\!|_{\rB, \rC} = |\!|\tE^{-1} (\tE  \Delta \tC )\tC^{-1} |\!|_{\rB,\rC} \le |\!| \tE  \Delta \tC |\!|_{\rA,\rD} $, thus proving the equality. \qed

\subsection{Closure and convexity in causal theories}\label{subsect:closeconv}

In Subsect. \ref{subsect:closure} we saw that if a theory is not
 deterministic, then one can construct a circuit that
simulates (with arbitrary precision) a coin with arbitrary bias $p \in [0,1]$.
  
In causal theories the possibility of conditioning  gives directly the following:
\begin{lemma}{\bf (Approximation of convex combinations)}\label{lemma:closconv}
  If a causal theory is not deterministic,  then
  any convex combination of states, effects, and transformations can be approximated with arbitrary precision.
\end{lemma}

\Proof Let $p\in [0,1]$ be an arbitrary probability and $p_n \in
\Stset (\rI)$ be such that $|p-p_n| < 1/n$ (such a probability exists because $\Stset(\rI)$ is dense in the interval $[0,1]$, as stated by Lemma \ref{lemma:densein01}). Consider two arbitrary tests
$\{\tC_i\}_{i \in \rX}$ and $\{\tD_j\}_{j \in \rY}$ from $\rA$ to $\rB$.   By randomization, we get the test $
\{p_n \tC_i\}_{i \in \rX} \cup \{(1-p_n)\tD_j\}_{j \in \rY}$.  Then,
by coarse-graining we can obtain the convex combination $p_n\tC_i +
(1-p_n) \tD_j$. The distance with the desired convex
combination $p\tC_i + (1-p) \tD_j$ is bounded by $ (|\!| \tC_i |\!|_{\rA,\rB} +
|\!| \tD_j |\!|_{\rA,\rB} )/n < 2/n$.  \qed

As a simple consequence we have the following
\begin{corollary}[Closure implies convexity]\label{cor:closconv}
  If a causal theory is not deterministic and the set of states of
  the trivial system is closed, then all sets of states, effects, and
  transformations are convex.
\end{corollary}

In this paper for simplicity we will always work with closed sets of
states.  Our attention will be devoted to non-deterministic causal theories, and, therefore, by the previous  Corollary \ref{cor:closconv} closure implies convexity.    Note that, however, most results hold independently of the
assumption of convexity, since in the context of non-deterministic causal theories any
desired combination can be approximated with arbitrary
precision.

\subsection{No-restriction hypothesis in causal theories}

In a causal theory the no-restriction hypothesis of Def. \ref{def:no-restriction} implies that for every system $\rA$ the cone generated by the effects coincides with the dual of the cone generated by the states: 
\begin{lemma}\label{lem:no-restrictioncausal}
In a causal theory the no-restriction hypothesis of Def. \ref{def:no-restriction} implies  the condition $\Cntset_+ (\rA) = \Stset_+^* (\rA)$ for every system $\rA$.
\end{lemma} 
\Proof  
Suppose that $a$ is an element of $\Stset_+^* (\rA)$ and let $|\!| a |\!|_{\rA,\rI}$ be the operational norm of $a$, as defined in Eq. (\ref{opnormeffects1}).  If $|\!| a |\!|_{\rA,\rI} =0$ , then $a$ is the null effect, which is trivially an element of $\Cntset_+ (\rA)$.  If $|\!| a |\!|_{\rA,\rI} \not = 0$,  then define the normalized effect $a_0 = a /|\!|  a |\!|_{\rA,\rI}$.  Upon defining $a_1 = e -a_0$, we  now have $\SC {a_1} \rho \ge 0$ for all $\rho \in \Stset_+(\rA)$, i.e. $a_1 \in \Stset_+^* (\rA)$.  Moreover, $\SC {a_0} \rho_\rA + \SC{a_1}\rho_{\rA} =\SC e \rho_\rA=1$ for every normalized state $\rho\in\Stset_1 (\rA)$. Hence, $\{a_0,a_1\}$ is a probability rule. By the no-restriction hypothesis, we then have that $\{a_0, a_1\}$ is an observation-test, and, therefore, $a_0$ and $a_1$ are effects.  This proves that every $a\in\Stset_+^*(\rA) $ is proportional to an effect $a_0$, that is, $\Stset^*_+ (\rA) \subseteq \Cntset_+ (\rA)$. On the other hand, all effects are positive functionals on states, and, therefore $\Stset^*_+ (\rA) \supseteq \Cntset_+ (\rA)$. 
\qed 
\medskip 

The above condition will be useful when discussing the implications of the no-restriction hypothesis in subsections \ref{subsect:no-info} and
\ref{subsect:completeness}.

\section{Local discriminability}\label{sec:localdiscr} 

Here we discuss the property of local discriminability, which
expresses the possibility of distinguishing multipartite states using only
local devices.

\subsection{Definition and main properties}

A common assumption in the literature on probabilistic theories is what we will call here \emph{local discriminability} (see e.g.  Refs. \cite{hardy5,nobroad,info-processingBarrett,swapping,tele,infocaus,maurolast}). 
\begin{definition}[Local discriminability]
A theory enjoys local discriminability if whenever two states $\rho, \sigma \in \Stset (\rA \rB)$ are distinct, there are two local effects $a \in \Cntset (\rA)$ and $b \in \Cntset (\rB)$ such that
\begin{equation}
\begin{aligned} \Qcircuit @C=1em @R=.7em @! R {\multiprepareC{1}{\rho}& \qw \poloFantasmaCn \rA &\measureD a \\
\pureghost\rho & \qw \poloFantasmaCn \rB &\measureD b} \end{aligned}
~\not =~
\begin{aligned}
 \Qcircuit @C=1em @R=.7em @! R {\multiprepareC{1}{\sigma}& \qw \poloFantasmaCn \rA &\measureD a \\
\pureghost\sigma & \qw \poloFantasmaCn \rB &\measureD b} 
\end{aligned}
\end{equation}
\end{definition}

Note that local discriminability on bipartite states implies local
discriminability on multipartite states, as can be seen by simple
iteration.  

The meaning of the local discriminability condition is that if two
bipartite states are different, then there is a chance of distinguishing
between them by using only local devices.  Of course, the resulting
discrimination may not be optimal, but at least it is strictly
better than the random guess.  Indeed, in the next Lemma we show that
in a convex theory with local discriminability two parties Alice and Bob,
holding systems $\rA$ and $\rB$, respectively, can always find a
discrimination protocol that uses only local operations and classical
communication (LOCC) and outperforms the random guess.

\begin{lemma}[LOCC discrimination]
  In a convex theory with local discriminability, if two states
  $\rho_0, \rho_1\in \Stset_1 (\rA\rB)$ are distinct, then there
  exists a LOCC discrimination protocol, described by a binary test
  $\{A_0,A_1\}$, such that the probability $p_{wc} := \max \{p(0|1),
  p(1|0)\}, \quad p(i|j) = \SC {A_i} {\rho_j}_{\rA\rB}$ is strictly
  smaller than $1/2$.
 \end{lemma} 
 \Proof If $\rho \not = \sigma$, then by local discriminability there
 are always two effects $a, b$ such that $\SC {a \otimes b}
 \rho_{\rA\rB} > \SC {a \otimes b} \sigma_{\rA\rB}$.  The binary test
 $\{A, e_{\rA\rB}-A\}$ defined by $A := a \otimes b$ can be obtained
 by performing the local tests $\{a, e_\rA-a\}$ and $\{b, e_\rB-b\}$
 and taking a coarse-graining. If the theory is convex, exploiting the
 construction of Lemma \ref{lemma:worstcase} (which only requires
 randomization and coarse-graining) we obtain a binary test $\{A_0 ,
 A_1\}$ satisfying $p(0|1) = p(1|0) < 1/2$ and, therefore $p_{wc} <
 1/2$.
 \qed

\medskip

Local discriminability is an enormous advantage in experiments. For example it allows one to perform tomography of multipartite states with only local measurements. Indeed,  every bipartite effect $\B E_{\rA\rB}$ can be written as linear combination of product effects, and, therefore every probability $\SC E \rho_{\rA\rB}$  can be computed as a linear combination of the probabilities $\SC {a_i \otimes b_j} {\rho}_{\rA\rB}$ arising from a finite set of product effects:    
\begin{lemma}[Local tomography]
  Let $\{\rho_i\}$ and $\{\tilde \rho_j\}$ be two bases for the vector
  spaces $\Stset_\Reals (\rA)$ and $\Stset_\Reals (\rB)$,
  respectively, and let $\{a_i\}$ and $\{b_j\}$ be two bases for the
  vector spaces $\Cntset_\Reals (\rA)$ and $\Cntset_\Reals (\rB)$,
  respectively. A theory enjoys local discriminability if and only if every state $\sigma \in \Stset (\rA \rB)$ (every
  effect $E \in \Cntset (\rA\rB)$) can be written as
\begin{equation}\label{linearcomb}
\begin{split}
&\K{\sigma}_{\rA \rB} = \sum_{i,j} ~A_{ij} \K{\rho_i}_{\rA} \K{\tilde \rho_j}_{\rB} \\
&\left( \B{E}_{\rA \rB} = \sum_{i,j} ~B_{ij} \B{a_i}_{\rA} \B{b_j}_{\rB}\right)
\end{split}
\end{equation}
for some suitable real matrix $A_{ij}$ ($B_{ij}$).
\end{lemma}  

\Proof Suppose that local discriminability holds. By definition, the
product effects $a \otimes b$ are a separating set for
$\Stset_{\Reals} (\rA \rB)$, and, therefore, they are a spanning set
for $\Cntset_\Reals (\rA\rB)$. Since states and effects span vector
spaces of equal dimension, this also implies that the product states
are a spanning set for $\Stset_\Reals (\rA \rB)$. Conversely, if Eq.
(\ref{linearcomb}) holds, then the product effects are a spanning set
for the vector space $\Cntset_\Reals (\rA\rB)$. Clearly, if $ \SC {a
  \otimes b} \rho_{\rA\rB} = \SC {a \otimes b} \sigma_{\rA\rB} $ for
all product effects, then one has $\rho = \sigma$, and this proves local discriminability.  \qed
\medskip

This also implies:

\begin{theorem}[Product of internal states is internal]\label{theo:internalprod} In a causal theory with local discriminability if the
  states $\omega_{\rA}$ and $ \omega_{\rB}$ are internal in
  $\Stset (\rA) $ and $\Stset(\rB)$, respectively, then the product
  $\omega_\rA\otimes\omega_\rB$ is internal in $\Stset(\rA\rB)$.
\end{theorem}
\Proof By definition, one has $\Span (D_{\omega_\rA \otimes \omega_\rB})
\supset \Span (D_{\omega_\rA}) \otimes \Span (D_{\omega_\rB}) = \Stset_\Reals (\rA)
\otimes \Stset _\Reals(\rB)$. Since local discriminability holds, this
is also equal to $\Stset_\Reals (\rA\rB)$.  \qed

Moreover, local discriminability allows one to distinguish two
different transformations $\tC, \tD \in \Trnset (\rA, \rB)$ without
considering their extension with an arbitrary ancilla system $\rC$:
\begin{lemma}
If two transformations $\tC, \tD \in \Trnset (\rA, \rB)$ are different and local discriminability holds, then there exist a state $\rho\in \Stset (\rA)$ such that
\begin{equation}
 \begin{aligned}
\Qcircuit @C=1em @R=.7em @! R { \prepareC \rho & \qw\poloFantasmaCn \rA & \gate \tC &\qw \poloFantasmaCn \rB &\qw}
\end{aligned} ~\not =~
\begin{aligned}  \Qcircuit @C=1em @R=.7em @! R { \prepareC \rho & \qw\poloFantasmaCn \rA & \gate \tD &\qw \poloFantasmaCn \rB &\qw}
\end{aligned}
\end{equation}
\end{lemma}  
\Proof By definition, if $\tC$ and $\tD$ are different there exist a system $\rC$ and a joint state $\sigma \in \Stset (\rA \rC)$ such that $\tC  \K{\sigma}_{\rA \rC} \not = \tD \K{\sigma}_{\rA \rC}$. Now, since local discriminability holds, there are two effects $b,c$ on systems $\rB, \rC$, respectively such that
\begin{equation}
 \begin{aligned}
\Qcircuit @C=1em @R=.7em @! R {\multiprepareC{1}{\sigma}& \qw \poloFantasmaCn \rA & \gate \tC & \qw \poloFantasmaCn \rB & \measureD b \\
\pureghost\sigma & \qw \poloFantasmaCn \rC & \qw & \qw  &\measureD c}
\end{aligned} 
~\not =~
\begin{aligned} \Qcircuit @C=1em @R=.7em @! R {\multiprepareC{1}{\sigma}& \qw \poloFantasmaCn \rA & \gate \tD & \qw \poloFantasmaCn \rB & \measureD b\\
\pureghost \sigma & \qw \poloFantasmaCn \rC & \qw &\qw & \measureD c} 
\end{aligned}
\end{equation} 
Defining $\K{\rho} := \B c_\rC \K{\sigma}_{\rA \rC}$ we then obtain $\B b_{\rB} \tC \K \rho_\rA \not =\B b_{\rB} \tD \K \rho_\rA $. This implies  $ \tC \K \rho_\rA \not = \tD \K \rho_\rA $. \qed

\subsection{Causal theories with local discriminability}

The results of this paper can be formulated in the simplest way for
causal theories that enjoy local discriminability. In this case one
has the following useful properties:

\begin{lemma}\label{lem:spanningmarginals} Let $\K{\sigma}_{\rA \rB} $
  be a state of $\rA\rB$ and $\K \rho_\rA := \B e_\rB \K \sigma_{\rA
    \rB}$, $\K {\tilde \rho}_\rB := \B e_\rA \K\sigma_{\rA \rB}$ be
  its marginals on systems $\rA$, $\rB$, respectively. In a causal
  theory with local discriminability one has
\begin{equation}
\sigma \in \Span (D_{\rho \otimes \tilde \rho}),
\end{equation}
where $D_{\rho \otimes \tilde \rho}$ is the refinement set of $\rho \otimes \tilde \rho$, as defined in Def. \ref{def:refset}.
\end{lemma} 
\Proof Take a basis $\{\rho_i\}_{i=1}^{n}$ ($\{\tilde
\rho_j\}_{j=1}^{\tilde n}$) of states for the (span of) the refinement set of
$\rho$ ($\tilde \rho$), and extend it to a basis
$\{\rho_i\}_{i=1}^{D_\rA}$ ($\{\tilde\rho_j\}_{j=1}^{D_\rB}$) of
$\Stset_\Reals (\rA)$ (of $\Stset_\Reals (\rB)$). By local
discriminability, we can write $\sigma$ as a linear combination as in
Eq. (\ref{linearcomb}) for some coefficients $A_{ij}$. Now, for every
effect $\B a_{\rA}$ the state $\K{\tilde \rho_a}_\rB := \B a_{\rA} \K
\sigma_{\rA \rB}$ is clearly in $D_{\tilde \rho}$. Therefore, we must
have $A_{ij} =0$ for all $j > \tilde n$. Likewise, applying an
arbitrary effect $\B b_\rB$ on system $\rB$ we find that we must have
$A_{ij} =0$ for all $i > n$. This implies
\begin{equation}\label{eq:a}
  \K\sigma_{\rA \rB} = \sum_{i=1}^n \sum_{j =1}^{\tilde n}  A_{ij} \K\rho_i \K{\tilde \rho}_j,
\end{equation}
that is, $\sigma \in \Span (D_{\rho \otimes \tilde \rho})$. \qed 

Since in a non-deterministic causal theory the set of states $\Stset (\rA)$ is convex (Corollary \ref{cor:closconv} along with the assumption that $\Stset (\rI)$ is closed), we also have the following:  

\begin{theorem}\label{theo:marginaldecomp}
  Let $\K{\sigma}_{\rA \rB} $ be a state of $\rA\rB$ and $\K \rho_\rA
  := \B e_\rB \K \sigma_{\rA \rB}$, $\K {\tilde \rho}_\rB := \B e_\rA
  \K\sigma_{\rA \rB}$ be its marginals on systems $\rA$, $\rB$,
  respectively. In a non-determinisitc causal theory with local discriminability
  there exists a non-zero probability $k>0$ such that
\begin{equation}\label{eqcolk1}
  k \sigma \in D_{\rho \otimes \tilde \rho}.
\end{equation}
\end{theorem}

The proof of the Theorem is immediate using Lemma
\ref{lem:spanningmarginals} along with the following
\begin{lemma}\label{lem:inthedecompset} In a non-deterministic causal theory, for every couple of states
  $\sigma ,\rho \in \Stset_1(\rA)$ one has
\begin{equation}\label{eqcolk2}
  \sigma \in \Span (D_{\rho}) \Longrightarrow k\sigma \in D_{\rho},
\end{equation}
\end{lemma} 
for some non-zero probability $k>0$.  \Proof Take a basis
$\{\rho_i\}_{i=1}^n$ of states in $D_\rho$. By hypothesis, we can
write $\sigma = \sum_i~ s_i \rho_i$ with suitable real coefficients
$s_{i}$. Moreover, since we are in finite dimensions, there is surely
a maximum coefficient $s_{\max} = \max_{i} s_{i}$.  On the other hand,
since $\rho_i$ belongs to $D_\rho$, there is surely a state $\chi_i$
such that $\rho = \rho_i + \chi_i$. This implies
\begin{equation}\label{eq:b}
\rho = \frac 1 {n}  \sum_{i} (  \rho_i + \chi_i ).
\end{equation} 
Let us define $\tau: = \rho - k \sigma$, with $k = \frac 1 {2 n
  s_{\max}}$, and normalize it as $\bar \tau := \tau / \SC e
\tau_{\rA}$. Using Eq. (\ref{eq:b}) it is easy to verify that $\bar
\tau$ is a state, since it is a convex combination of states (recall that in a non-deterministic causal theory the set of states is convex).
Moreover, we have $\rho = k\sigma +(1- k) \bar \tau$, which implies
the thesis. \qed

\medskip

{\bf Remark.}  In the previous Lemma \ref{lem:inthedecompset} we used the fact that in a non-deterministic causal theory a set of states is convex (Corollary \ref{cor:closconv} along with the assumption that $\Stset (\rI)$ is closed).  In fact, we can weaken this assumption in the proofs of Theorem
\ref{theo:internalprod} and Lemma \ref{lem:inthedecompset}. Indeed, in
any non-deterministic causal theory we can approximate the convex combinations needed for
the proof of Lemma \ref{lem:inthedecompset} with arbitrary precision
(Lemma \ref{lemma:closconv}), thus proving Eqs.  (\ref{eqcolk1}) and
(\ref{eqcolk2}) with a non-zero probability $k >0$ that arises from a
test allowed by the theory.

\medskip

Theorems \ref{theo:internalprod} and \ref{theo:marginaldecomp} state
two very natural properties. Even when discussing the extension of our
results beyond the framework of local
discriminability we will assume these properties to hold.

Finally, causal theories with local discriminability enjoy a nice characterization of states that are invariant under the group of reversible transformations:

  \begin{theorem}\label{theo:uniqueLI} 
  In a causal theory with local discriminability if systems $\rA$ and
  $\rB$ have unique invariant states $\K\chi_{\rA} \in \Stset_1 (\rA)$
  and $\K\chi_{\rB} \in \Stset_1 (\rB)$, respectively, then $\K \chi_\rA
  \K\chi_\rB \in \Stset_1 (\rA\rB)$ is the unique locally invariant
  state of system $\rA \rB$.
\end{theorem}
\Proof Suppose that $\K \sigma_{\rA \rB}$ is a locally invariant state, namely
\begin{equation}
\begin{aligned}
 \Qcircuit @C=1em @R=.7em @! R {\multiprepareC{1}{\sigma}& \qw \poloFantasmaCn \rA &\gate \tU & \qw \poloFantasmaCn \rA &\qw \\
\pureghost\sigma & \qw \poloFantasmaCn \rB &\gate \tV & \qw \poloFantasmaCn \rB &\qw} \end{aligned}
~=~
\begin{aligned} \Qcircuit @C=1em @R=.7em @! R {\multiprepareC{1}{\sigma}& \qw \poloFantasmaCn \rA &\qw \\
\pureghost\sigma & \qw \poloFantasmaCn \rB &\qw} 
\end{aligned}
\end{equation}
for all $\tU \in \grp G_\rA$ and $\tV \in \grp G_\rB$. If we apply two arbitrary effects $\B a_\rA$ and $\B b_\rB$ we then get
\begin{equation}\label{basta}
\begin{aligned}  \Qcircuit @C=1em @R=.7em @! R {\multiprepareC{1}{\sigma} &\qw \poloFantasmaCn \rA &\measureD a \\
    \pureghost\sigma & \qw \poloFantasmaCn \rB &\measureD b} 
\end{aligned}~=~ 
\begin{aligned}\Qcircuit @C=1em @R=.7em @! R { \prepareC {\tilde \rho_a} & \qw \poloFantasmaCn \rB & \measureD b }\end{aligned}\\
   ~=~ 
\begin{aligned}\Qcircuit @C=1em @R=.7em @! R { \prepareC {\rho_b}
    & \qw \poloFantasmaCn \rA & \measureD a}
\end{aligned}
\end{equation}
having defined $\K{\tilde \rho_a}_{\rB}:= \B a_{\rA} \K\sigma_{\rA\rB}$ and
$\K{\rho_b}_{\rA}:= \B b_\rB \K\sigma_{\rA\rB}$.  Now, $\tilde \rho_a$ and
$\rho_b$ are invariant (unnormalized) states.  Since $\chi_\rA$ is the
unique state of $\rB$ that is invariant and normalized, one must have 
\begin{equation}
\begin{split}
  \K\chi_\rA &= \frac{\K{\rho_b}_\rA} { \SC e {\rho_b}_\rA} = \frac {
    \K{\rho_b}_\rA} { \SC{ e \otimes b }
    \sigma_{\rA \rB}} := \frac {\K{\rho_b}_\rA} {\SC b {\tilde \rho}_\rB}\\
  \K\chi_\rB &= \frac{\K{\tilde \rho_a}_\rB} { \SC e {\tilde
      \rho_a}_\rB }=\frac{ \K{\tilde \rho_a}_{\rB}} {\SC{ a \otimes e}
    \sigma_{\rA \rB}} := \frac {\K{\tilde\rho_a}_\rB} {\SC a {\rho}_\rA},
\end{split}
\end{equation}
$\K\rho_\rA$, $\K{\tilde \rho}_\rB$ being the marginal states on
systems $\rA$, $\rB$, respectively. Inserting the above relations in Eq. (\ref{basta}), we then obtain

\begin{equation}
\begin{split}
\begin{aligned}  \Qcircuit @C=1em @R=.7em @! R {\multiprepareC{1}{\sigma}&  \qw \poloFantasmaCn \rA &\measureD a \\
    \pureghost\sigma & \qw \poloFantasmaCn \rB &\measureD b } \end{aligned}
  & =
\begin{aligned} \Qcircuit @C=1em @R=.7em @! R {  \prepareC {\chi}
    & \qw \poloFantasmaCn \rA & \measureD a\\
\prepareC{\tilde \rho} & \qw \poloFantasmaCn \rB & \measureD b } \end{aligned} \\
 & =
\begin{aligned}\Qcircuit @C=1em @R=.7em @! R { \prepareC {\rho} & \qw \poloFantasmaCn \rA & \measureD a \\
\prepareC {\chi} & \qw \poloFantasmaCn \rB & \measureD b }
\end{aligned}
\end{split}
\end{equation}
for every $a,b$. By local discriminability, this implies
$\K\sigma_{\rA \rB} =\K \chi_\rA \K{\tilde \rho}_\rB = \K\rho_\rA
\K\chi_\rB$, and, therefore, $\K\sigma_{\rA \rB} =\K \chi_\rA \K{\chi}_\rB$. \qed


\section{Beyond local discriminability and convexity}  

Although the results of this paper take their simplest form for causal
theories with local discriminability, most of them are valid in causal
theories under weaker requirements.  For example, they hold for
quantum theory on real Hilbert spaces, which is a well known
example of theory without local discriminability. Moreover, although
convexity is very well motivated in the context of causal theories,
most results of this paper hold even in non-convex theories.  In this
Section we briefly discuss these generalizations.

\subsection{Relaxing local discriminability}  A weaker requirement than local
discriminability is local discriminability on pure states: 
\begin{definition}{\bf (Local discriminability on pure states)} A theory enjoys \emph{local discriminability on pure states} if whenever two states $\Psi, \sigma \in \Stset (\rA \rB)$ are different, and one of the two states  (say $\Psi$) is pure, there are two effects $a \in \Cntset (\rA)$ and $ b\in\Cntset (\rB)$ such that  
\begin{equation}
 \Qcircuit @C=1em @R=.7em @! R {\multiprepareC{1}{\Psi}& \qw \poloFantasmaCn \rA &\measureD a \\
\pureghost\Psi & \qw \poloFantasmaCn \rB &\measureD b} \ugualonelike{\not =}
 \Qcircuit @C=1em @R=.7em @! R {\multiprepareC{1}{\sigma}& \qw \poloFantasmaCn \rA &\measureD a \\
\pureghost\sigma & \qw \poloFantasmaCn \rB &\measureD b} 
\end{equation}
\end{definition}

An example of theory with this property is quantum theory on real Hilbert spaces:
\begin{lemma}
Quantum theory on real Hilbert spaces enjoys local discriminability on pure states.
\end{lemma}
\Proof Let $\rho = \sum_i p_i |\Phi_i \>\<\Phi_i|$ be a density matrix
on the real Hilbert space $\spc H_\rA \otimes \spc H_\rB$ with $\spc
H_\rA =\mathbb R^{m} $ and $\spc H_\rB = \mathbb R^n $ and $|\Psi\>\in
\mathbb R^{m} \otimes \mathbb R^n$ be a unit vector.  Suppose that $
\Tr [ (\rho - |\Psi \>\< \Psi|) (a \otimes b)] =0$ for every couple of
real matrices $a$ and $b$.  Taking $a =|v\>\<v|$ for some $v \in
\mathbb R^m$ we then obtain $\< v|_\rA |\Phi_i \>_{\rA\rB} = k_{i,v}
\<v|_\rA |\Psi\>_{\rA\rB}$ for some constant $k_{i,v}$. Likewise, taking $b=
|w\>\<w|$ for some $w \in \mathbb R^n$ we obtain $\< w|_\rB |\Phi_i
\>_{\rA\rB} = l_{i,w} \<w|_\rB |\Psi\>_{\rA\rB}$ for some constant
$l_{i,w}$.  Putting the two things together we have 
\begin{equation}
\begin{split}
  \<v|_\rA \<w|_{\rB} |\Phi_i\>_{\rA\rB}& = k_{i,v}
  \<v|_\rA\<w|_\rB |\Psi\>_{\rA\rB}\\
  & = l_{i,w} \<v|_\rA\<w|_\rB |\Psi\>_{\rA\rB}
\end{split}
\end{equation}
hence $k_{i,v} \equiv l_{i,w} := c_i$. Finally, $\< v|_\rA \<w|_\rB|\Phi_i
\>_{\rA\rB} = c_{i} \<v|_\rA\<w|_\rB |\Psi\>_{\rA\rB}$ for every $v,w$ implies
$|\Phi_i\> = c_i |\Psi\>$, and, therefore $\sigma = |\Psi\>\<\Psi|$.
\qed
\medskip

When generalizing our results to theories without local
discriminability we will always assume local discriminability on pure states along with  the theses of Theorems
\ref{theo:internalprod}, \ref{theo:marginaldecomp}, and \ref{theo:uniqueLI}.  Again, all these requirements are met by quantum theory on real Hilbert spaces.

\medskip
An elementary property of causal theories with local discriminability on pure states is that the product of two pure states is pure, as stated in the following Lemma.
\begin{lemma}[Product of pure states is pure]\label{lem:prodpur}
In a causal theory with local discriminability on pure states, if the states $\K \varphi_\rA \in \Stset_1(\rA)$ and $\K \psi_\rB \in \Stset_1(\rB)$ are pure, then their product $\K \varphi_\rA \K \psi_\rB\in \Stset_1 (\rA\rB)$ is pure.
\end{lemma}
\Proof   Suppose that the product can be written as a convex combination $\K \varphi_\rA \K \psi_\rB = \sum_{i\in \rX} p_i  \K{\Psi_i}_{\rA\rB}$, with $\K {\Psi_i}_{\rA\rB} \in \Stset_1 (\rA\rB)$. We now show that $\K {\Psi_i}_{\rA\rB} = \K \varphi_\rA \K \psi_\rB$ for every $i\in \rX$.  Let $\B b_\rB$ be an arbitrary effect for system $\rB$.  We then have 
\begin{equation}
\begin{aligned} \Qcircuit @C=1em @R=.7em @! R {
\prepareC{\varphi}&\qw \poloFantasmaCn{\rA}&\qw\\  
\prepareC{\psi}&\qw \poloFantasmaCn{\rB}&\measureD b  }\end{aligned}~ =~
  \sum_{i \in \rX} ~p_i  \begin{aligned}\Qcircuit @C=1em @R=.7em @! R {
\multiprepareC{1}{\Psi_i}&\qw&\qw\poloFantasmaCn{\rA}&\qw\\
\pureghost{\Psi_i}&\qw&\qw\poloFantasmaCn{\rB}&\measureD{b}} \end{aligned} 
\end{equation}
Since $\K\varphi_\rA$ is pure, this implies
\begin{equation}
\begin{aligned}  \Qcircuit @C=1em @R=.7em @! R {\multiprepareC{1}{\Psi_i}&  \qw \poloFantasmaCn \rA &\qw \\
    \pureghost{\Psi_i} & \qw \poloFantasmaCn \rB &\measureD b} \end{aligned}
  = \lambda_{bi}
\begin{aligned}\Qcircuit @C=1em @R=.7em @! R { \prepareC {\varphi}
    & \qw \poloFantasmaCn \rA &\qw } \end{aligned}
\end{equation}
for some coefficient $\lambda_{bi} \ge 0$. Clearly, for $\B b_\rB =\B e_\rB$ one has $\lambda_{ei} =1$.
Similarly, if $\B a_\rA$ is an arbitrary effect for system $\rA$,  we obtain
\begin{equation}
\begin{aligned}  \Qcircuit @C=1em @R=.7em @! R {\multiprepareC{1}{\Psi_i}&  \qw \poloFantasmaCn \rA &\measureD a \\
    \pureghost{\Psi_i} & \qw \poloFantasmaCn \rB &\qw} \end{aligned}
  = \mu_{ai}
\begin{aligned}\Qcircuit @C=1em @R=.7em @! R { \prepareC {\psi}
    & \qw \poloFantasmaCn \rA &\qw } \end{aligned}
\end{equation}
for some coefficient $\mu_{ai} \ge 0$ satisfying $\mu_{ei} =1$.
Combining the above facts, we obtain
\begin{equation}
\begin{split}
\lambda_{bi}  &= \lambda_{bi}  \begin{aligned}\Qcircuit @C=1em @R=.7em @! R { \prepareC {\varphi}
    & \qw \poloFantasmaCn \rA &\measureD e } \end{aligned} = \begin{aligned}  \Qcircuit @C=1em @R=.7em @! R {\multiprepareC{1}{\Psi_i}&  \qw \poloFantasmaCn \rA &\measureD e \\
    \pureghost{\Psi_i} & \qw \poloFantasmaCn \rB &\measureD b} \end{aligned}\\
   &= \mu_{ei}
\begin{aligned}\Qcircuit @C=1em @R=.7em @! R { \prepareC {\psi}
    & \qw \poloFantasmaCn \rB &\measureD b } \end{aligned} =\begin{aligned}\Qcircuit @C=1em @R=.7em @! R { \prepareC {\psi}
    & \qw \poloFantasmaCn \rB &\measureD b } \end{aligned}.
\end{split}  
\end{equation}
Finally, this implies
\begin{equation}
\begin{aligned}  \Qcircuit @C=1em @R=.7em @! R {\multiprepareC{1}{\Psi_i}&  \qw \poloFantasmaCn \rA &\measureD a \\
    \pureghost{\Psi_i} & \qw \poloFantasmaCn \rB &\measureD b} \end{aligned}
  = \lambda_{bi}
\begin{aligned}\Qcircuit @C=1em @R=.7em @! R { \prepareC {\varphi}
    & \qw \poloFantasmaCn \rA &\measureD a} \end{aligned} = \begin{aligned} \Qcircuit @C=1em @R=.7em @! R {
\prepareC{\varphi}&\qw \poloFantasmaCn{\rA}&\measureD a\\  
\prepareC{\psi}&\qw \poloFantasmaCn{\rB}&\measureD b  }\end{aligned}
\end{equation}
 and, by local discriminability on pure states $\K{\Psi_i}_{\rA\rB} = \K \varphi_\rA \K \psi_\rB$.
 \qed
Clearly, iterating the above reasoning one can also show that the product of $N$ pure states $\K {\varphi_1}_{\rA_1}\K {\varphi_2}_{\rA_2} \dots\K {\varphi_N}_{\rA_N} $ is pure.

\subsection{Relaxing convexity}
If one wants to relax convexity, it is clear from the proof of Lemma
\ref{lemma:closconv} and Corollary \ref{cor:closconv} that one must
have at least one of the following features: \emph{i)} the theory is
 deterministic, i.e. all events have either zero or unit
probability, \emph{ii)} some randomizations or some coarse-grainings
are forbidden, and \emph{iii)} the set of probabilities $\Stset (\rI)$
of the theory is not closed.  For the purposes of this paper, 
deterministic theories are not quite interesting, and theories with 
non-closed sets of transformations are just technically cumbersome, although
most of the conclusions of this paper remain unchanged.   Therefore, in
relaxing convexity we will only consider the case in which some
conditioned tests or some coarse-grained tests are forbidden.  Of course,
if one wants to drop a basic operational requirement like the
possibility of conditioning, one has to take care that some minimal
properties hold.  For example, the existence of internal states, the
fact that every test has an ultimate refinement, and the validity of the theses of
Theorems \ref{theo:internalprod} and \ref{theo:marginaldecomp} have to
be explicitly postulated.  One would also need to assume that is not forbidden
\emph{i)} to attach a distinguishable state $\K{\varphi_i}_\rB$ to
every state in a preparation-test $\{\K{\rho_i}_\rA \}_{i \in \rX}$,
thus getting the new test $\{\K{\rho_i}_\rA \K{ \varphi_i}_\rB\}_{i
  \in \rX}$, and \emph{ii)} to perform a discriminating test
$\{a_i\}_{i \in \rX}$ for the perfectly discriminable states
$\{\rho_i\}_{i \in \rX}$, and to re-prepare state $\rho_i$ when the
outcome is $i$, thus getting the ``measure-and-prepare'' test $\{
\K{\rho_i}_\rA \B {a_i}_{\rA}\}_{i \in \rX}$.

Finally, we will show that the existence of twirling tests is necessary
for deterministic teleportation. If one wants to
consider non-convex theories with deterministic teleportation one has
also to require the existence of a twirling-test and the thesis of Theorem \ref{theo:uniqueLI}.

\section{Summary of the framework}
This short Section concludes the presentation of the general framework
used in this paper. The standing assumptions of the paper are summarized by the following table:

\begin{center}\framebox{\begin{minipage}{.95\columnwidth} In this
      paper, if not otherwise stated, we will consider operational-probabilistic theories satisfying the following requirements:
\begin{enumerate}
\item {\bf the theory is causal (every state is proportional to a normalized one)}
\item {\bf local discriminability holds}
\item {\bf the set of all tests is closed under
    coarse-graining and conditioning}
\item {\bf for every system, the set of states is finite-dimensional
    and closed in the operational norm}
\item {\bf there exist perfectly discriminable states}
\item {\bf the theory is not deterministic}
\end{enumerate}
\end{minipage}}\end{center}
\medskip

Note that the existence of perfectly discriminable states, needed to
describe perfect classical communication, is guaranteed in the usual
convex framework, which contains the no-restriction hypothesis of Def.
\ref{def:no-restriction}. We recall that we don't make this
assumption here.

In most proofs the  background requirement 2.  can be always weakened to: 
\begin{center}
\framebox{\begin{minipage}{.95\columnwidth}
\begin{itemize}
\item[2'.] {\bf local discriminability of pure states and the
    theses of Theorems \ref{theo:internalprod},
    \ref{theo:uniqueLI} and \ref{theo:marginaldecomp} hold}
\end{itemize}
\end{minipage}}
\end{center}
If a particular results requires local discriminability or convexity this will be mentioned explicitly in its statement. 

\section{Theories with purification}

Here we introduce the purification postulate ``every mixed state has a purification, unique up to reversible transformations on the purifying system'', and we explore its consequences within the general framework outlined in the previous Sections.

\subsection{The purification postulate}
\begin{definition}[Purification]
  A pure state $\Psi\in\Stset_1(\rA\rB)$ is a \emph{purification} of
  $\rho \in \Stset_1(\rA)$ if   $\K{\rho}_\rA=\B{e}_{\rB}\K{\Psi}_{\rm A B}$. Diagrammatically,
\begin{align}\label{purificami}
\begin{aligned}  \Qcircuit @C=1em @R=.7em @! R {
    \prepareC{\rho}&\qw\poloFantasmaCn{\rA}&\qw}\end{aligned}&~ =~ 
\begin{aligned}\Qcircuit @C=1em @R=.7em @! R {
    \multiprepareC{1}{\Psi}&\qw\poloFantasmaCn{\rA}&\qw&\\
    \pureghost{\Psi}&\qw\poloFantasmaCn{\rB}&\measureD{e}}
\end{aligned}
\end{align}
\end{definition} 

\begin{definition}[Purifying system]
  If system $\rA \rB$ contains a purification of $\rho \in \Stset_1 (\rA)$, we call
  system $\rB$ a \emph{purifying system} for $\rho$.
\end{definition}

\begin{definition}[Complementary state]\label{def:complementary} Let $\Psi\in \Stset_1 (\rA\rB)$ be a purification of  $\rho \in \Stset_1 (\rA)$.  The \emph{complementary state} of $\rho$ is the state $\tilde \rho \in \Stset_1 (\rB)$ defined by 
\begin{align}\label{dualstate}
\begin{aligned}  \Qcircuit @C=1em @R=.7em @! R {
    \prepareC{\tilde \rho}&\qw\poloFantasmaCn{\rB}&\qw}\end{aligned}& ~=~ 
\begin{aligned}\Qcircuit @C=1em @R=.7em @! R {
    \multiprepareC{1}{\Psi}&\qw\poloFantasmaCn{\rA}&\measureD{e}\\
    \pureghost{\Psi}&\qw\poloFantasmaCn{\rB}&\qw}
\end{aligned}
\end{align}
\end{definition}

An elementary property of purification is the following

\begin{lemma}\label{lem:purestatepur}
  If  $\psi\in \Stset_1 (\rA )$ is pure and $\Psi \in \Stset_1 (\rA \rB)$ is a purification of $\psi$, then $\Psi$ must be of the form $\Psi =\psi\otimes\tilde \psi$, with $\tilde \psi\in\Stset_1(\rB)$ pure. 
\end{lemma}
\Proof Take an observation-test $\{b_i\}_{i \in \rX}$ on $\rB$. Since $\sum_i b_i = e_\rB$ we have
\begin{equation}\label{purodecomposto}
\begin{aligned} \Qcircuit @C=1em @R=.7em @! R {
\prepareC{\psi}&\qw \poloFantasmaCn{\rA}&\qw }\end{aligned}~ =~
  \sum_{i \in \rX} ~ \begin{aligned}\Qcircuit @C=1em @R=.7em @! R {
\multiprepareC{1}{\Psi}&\qw&\qw\poloFantasmaCn{\rA}&\qw\\
\pureghost{\Psi}&\qw&\qw\poloFantasmaCn{\rB}&\measureD{b_i}} \end{aligned} ~:=~ \sum_{i \in \rX} ~ \begin{aligned}
\Qcircuit @C=1em @R=.7em @! R {
\prepareC{\rho_i}&\qw \poloFantasmaCn{\rA}&\qw }
\end{aligned}
\end{equation}
namely, the states $\{\rho_i\}_{i \in \rX}$ defined by $\rho_i:= \B {b_i}_\rB \K{\Psi}_{\rA\rB}$ form a refinement of
$\psi$.  Since $\psi$ is pure, we necessarily have $\rho_i = p_i \psi$
for some probabilities $\{p_i\}$.  Precisely, we have $p_i = \SC e {\rho_i}_\rA = \SC {e_\rA \otimes b_i} \Psi_{\rA\rB} = \SC{b_i} {\tilde \psi}_{\rB}$, where $\tilde \psi$ is the complementary state
of $\psi$. Therefore, we have
\begin{equation}
 \begin{aligned}\Qcircuit @C=1em @R=.7em @! R {
\multiprepareC{1}{\Psi}&\qw&\qw\poloFantasmaCn{\rA}&\qw\\
\pureghost{\Psi}&\qw&\qw\poloFantasmaCn{\rB}&\measureD{b_i}}\end{aligned}  ~=~
\begin{aligned} \Qcircuit @C=1em @R=.7em @! R {
\prepareC{\psi}&\qw \poloFantasmaCn{\rA}&\qw \\
\prepareC{\tilde \psi} & \qw \poloFantasmaCn{\rB}& \measureD{b_i}}
\end{aligned} \quad \forall i \in \rX.
\end{equation}  
The above equation implies that $\Psi $ cannot be distinguished from $ \psi \otimes \tilde \psi$ by any local test. Since $\Psi$ is pure, this implies  $\Psi= \psi \otimes \tilde \psi$. Clearly,  $\tilde \psi$ has to be pure, otherwise we would have a non-trivial refinement of the pure state $\Psi$.  \qed

It is important to stress that purification is not a physical process:
There is no physical transformation that is able to turn any arbitrary
mixed state $\rho$ into some purification $\Psi$ of it. In quantum
mechanics, this has been noted by Kleinman \emph{et al.}  in Ref. \cite{no-pury}.  Along the same lines, it is easy to prove the
following general Theorem:
 
\begin{theorem}[No-purification of collinear states] Let $\rho_i, i =
  1,2,3$ be three distinct collinear states of system $\rA$---i.e.
  $\rho_1 \not =\rho_3$ and $\rho_2 = p \rho_1 + (1-p) \rho_3$ for
  some $0<p<1$. Suppose that $\K{\Psi_i}_{\rA\rB}, i =1,2,3$ is a
  purification of $\K{\rho_i}_\rA$.  Then for every finite number of
  copies $N$ there is no physical transformation $\tC \in \Trnset
  (\rA^{\otimes N}, \rA\rB)$ such that $\tC \K{\rho_i}^{\otimes N}_\rA
  = \K{\Psi_i}_{\rA\rB} $ for every $i = 1,2,3$.
\end{theorem}
\Proof The proof is by contradiction. Suppose that such a transformation $\tC$ exists for some finite $N$. Then, expanding the product $\rho_2^{\otimes N} = [p\rho_1 + (1-p) \rho_3]^{\otimes N}$, and applying the transformation $\tC$, we obtain 
\begin{equation}
\begin{split}
\K{\Psi_2}_{\rA\rB} &=\tC \K{\rho_2}^{\otimes N}_\rA\\
& = p^N   \tC \K{\rho_1}_\rA^{\otimes N} + (1-p)^N  \tC \K{\rho_3}_\rA^{\otimes N} + \K{\rho_{rest}}_{\rA\rB}\\
& = p^N \K{\Psi_1}_{\rA \rB} + (1-p)^N \K{\Psi_3}_{\rA \rB} + \K{\rho_{rest}}_{\rA\rB},
\end{split}
\end{equation}
where $\rho_{rest}$ is a suitable non-normalized state. This is clearly absurd, since we obtained a non-trivial convex decomposition of the pure state $\Psi_2$. \qed

\medskip 

If $\Psi $ is a purification of $\rho$ and $\tU_\rB$ is a reversible transformation on the purifying
system, then also $\K{\Psi'}_{\rA \rB} = \tU_B \K{\Psi}_{\rA\rB}$ is a new purification of $\rho$. Indeed,
$\tU_B \K{\Psi}_{\rA\rB}$ must be pure, otherwise by inverting $\tU_B$ on $\tU_B \K{\Psi}_{\rA\rB}$
by linearity one would find that $\K{\Psi}_{\rA\rB}$ is mixed. 
In the following Postulate we impose that all purifications are of this form:

\begin{postulate}[Purification]\label{pos:pur}
  Every state has a purification, unique up to reversible transformations on the purifying system: if $\Psi, \Psi'\in \Stset_1(\rA \rB)$ are two purifications of the same state, then they are connected by a reversible transformation $\tU \in \Trnset(\rB)$,
  namely
\begin{align}\label{unipur}
\begin{aligned}  \Qcircuit @C=1em @R=.7em @! R {
    \multiprepareC{1}{\Psi'}&\qw\poloFantasmaCn{\rA}&\qw\\
    \pureghost{\Psi'}&\qw\poloFantasmaCn{\rB}&\measureD{e} }\end{aligned}& ~=~ \begin{aligned}
  \Qcircuit @C=1em @R=.7em @! R {
    \multiprepareC{1}{\Psi}&\qw\poloFantasmaCn{\rA}&\qw\\
    \pureghost{\Psi}&\qw\poloFantasmaCn{\rB}&\measureD{e}}
\end{aligned}
\nonumber\\
~\Longrightarrow~
\begin{aligned}  \Qcircuit @C=1em @R=.7em @! R {
    \multiprepareC{1}{\Psi'}&\qw\poloFantasmaCn{\rA}&\qw\\
    \pureghost{\Psi'}&\qw\poloFantasmaCn{\rB}&\qw }
\end{aligned}& ~=~\begin{aligned} \Qcircuit
  @C=1em @R=.7em @! R {
    \multiprepareC{1}{\Psi}&\qw&\qw&\qw\poloFantasmaCn{\rA}&\qw\\
    \pureghost{\Psi}&\qw\poloFantasmaCn{\rB}&\gate{\tU}&\qw\poloFantasmaCn{\rB}&\qw
  } \end{aligned}
\end{align}
\end{postulate}

{\bf Remark (Uniqueness of the complementary state)} Note that
uniqueness of the purification assumed in the purification postulate is equivalent to the
uniqueness (up to reversible transformations) of the complementary
state defined in Def. \ref{def:complementary}.

\medskip

We now show some simple consequences of the purification postulate.  First, it implies that all pure states of a system are connected by reversible transformations: 
\begin{lemma}{\bf (Transitivity of the group of reversible transformations on the set of pure states)}\label{lem:transitivity}
  For any couple of pure states $\psi, \psi'\in\Stset_1(\rA)$ there is
  a reversible transformation $\tU\in\Trnset(\rA)$ such that
  $\psi'=\tU\psi$.
\end{lemma}
\Proof Every system is a purifying system for the trivial system. Then
just apply Eq. (\ref{unipur}) with $\rA\equiv \rI $. \qed

\medskip

An obvious consequence of the purification postulate is that in a theory with purification there are \emph{entangled states}, according to the usual definition:
\begin{definition}[Separable states/entangled states]\label{def:ent}
A bipartite state $\sigma\in\Stset_1(\rA\rB)$ is \emph{separable} if it can be written as a convex combination of product states, that is, as $\K\sigma_{\rA\rB} = \sum_i p_i \K {\phi_i}_\rA \K{\psi_i}_\rB$ with $p_i\ge 0, \sum_i p_i =1$.  A bipartite state is \emph{entangled} if it is not separable. 
\end{definition}
As already anticipated, one has the following (trivial) Corollary: 
\begin{corollary}[Existence of entangled states]\label{cor:entstates}
If ${\Psi_\rho}\in \Stset_1 (\rA\rB)$ is a purification of $\rho\in\Stset_1 (\rA)$ and $\rho$ is mixed, then $ {\Psi_\rho}$ is entangled.
\end{corollary}
\Proof By contradiction, suppose that $\Psi_\rho$ is separable. Because it is pure, it must be of the form $\K{\Psi_\rho}_{\rA\rB} = \K\varphi_\rA \K \psi_\rB$ with $\K\varphi_\rA$ and $\K\psi_\rB$ pure.   Then the marginal $\K \rho_\rA = \B e_\rB \K{\Psi_\rho}_{\rA\rB} = \K\varphi_\rA$ is pure, in contradiction with the hypothesis.   \qed 
\medskip

{\bf Remark (Purification and classical theories).}  Clearly,  Corollary \ref{cor:entstates} shows that the purification postulate rules out  classical probability theory.  In fact, there is only one possibility for a causal theory to satisfy the purification postulate without having entangled states: the theory must not contain mixed states.  This necessarily implies that the theory is deterministic, that is, that the probabilities of outcomes in any test are either $0$ or $1$ (if the theory were not deterministic one could construct mixed states by randomization). In particular, this also implies that in such a theory the pure states of an arbitrary system $\rA$ are perfectly distinguishable.   In conclusion, the only causal theories that satisfy the purification postulate and have no entanglement are classical deterministic theories. 
\medskip

Another elementary consequence of the purification postulate is that  ``purity implies
independence from the rest of the world":

\begin{corollary}[Purity implies independence]\label{cor:purity-independence}
If $\psi \in \Stset_1 (\rA)$ is pure and $\rho \in \Stset_1(\rA \rB)$ is an extension of $\psi$, namely $\K{\psi}_{\rA} = \B{e}_{\rB}  \K{\rho}_{\rA \rB}$, then $\rho = \psi \otimes \sigma$, for some state $\sigma \in \Stset_1(\rB)$. 
\end{corollary}
\Proof Let $\Psi \in \Stset_1 (\rA\rB \rC)$ be a purification of $\rho$. Since $\Psi$ is also a purification of $\psi$, by the Lemma \ref{lem:prodpur} we have $\K{\Psi}_{\rA\rB\rC} = \K{\psi}_{\rA} \K \eta_{\rB \rC}$, for some pure state $\eta \in \Stset_1(\rB \rC)$. But since $\Psi$ is a purification of $\rho$ we have $\K{\rho} = \B{e}_{\rC}  \K{\Psi}_{\rA\rB\rC}  = \K{\psi}_{\rA} \K{\sigma}_{\rB}$, with $\K{\sigma}_{\rB} := \B{e}_{\rC} \K{\eta}_{\rB \rC}$. \qed


We conclude this subsection  with an important Lemma that extends the uniqueness of purification to the case of purifications with different purifying systems:

\begin{lemma}\label{lem:purichan}{\bf (Uniqueness of the purification up to channels on the purifying systems)}
  Let $\Psi\in\Stset_1(\rA\rB)$ and $\Psi'\in\Stset_1(\rA\rC)$ be two
  purifications of $\rho \in \Stset_1 (\rA)$. Then  there exists a channel $\tC \in \Trnset (\rB , \rC)$ such that\begin{equation}\label{uniquenessBC}
\begin{aligned}\Qcircuit @C=1em @R=.7em @! R {
\multiprepareC{1}{\Psi'}&\qw\poloFantasmaCn{\rA}&\qw\\
\pureghost{\Psi'}&\qw\poloFantasmaCn{\rC}&\qw} \end{aligned}
 ~=~
\begin{aligned}
\Qcircuit @C=1em @R=.7em @! R {
\multiprepareC{1}{\Psi}&\qw&\qw&\qw\poloFantasmaCn{\rA}&\qw\\
\pureghost{\Psi}&\qw\poloFantasmaCn{\rB}&\gate{\tC} &\qw \poloFantasmaCn \rC &\qw}
\end{aligned}
\end{equation}
Moreover, channel $\tC$ has the form
\begin{equation}
\begin{aligned}
  \Qcircuit @C=1em @R=.7em @! R {&\poloFantasmaCn{\rB}\qw &\gate{\tC}&\qw\poloFantasmaCn{\rC}&\qw}
\end{aligned}\;=\;
\begin{aligned}  \Qcircuit @C=1em @R=.7em @! R {\prepareC{\varphi_0}&\qw&\qw\poloFantasmaCn{\rC}&\qw &
    \multigate{1}{\tU}&\qw&\qw\poloFantasmaCn{\rB}&\measureD{e} \\
     &\qw &\qw \poloFantasmaCn{\rB}&\qw&\ghost{\tU} &
    \qw&\qw\poloFantasmaCn{\rC}&\qw }
\end{aligned}
\end{equation}
 for some pure state
$\varphi_0 \in \Stset_1 (\rC)$ and some reversible channel $\tU \in
\grp G_{\rB \rC}$.
\end{lemma}

\Proof Let $\K{\eta}_\rB$ and $\K{\varphi_0}_\rC$ be an arbitrary pure state of $\rB$ and $\rC$, respectively. Then $\K{\Psi'}_{\rA \rC} \K{ \eta}_{\rB}$ and  $\K{\Psi}_{\rA \rB}\K{\varphi_0}_{\rC}$ are two purifications of $\rho$ with the same purifying system $\rB \rC$. Due to Eq. (\ref{unipur}), we have 
\begin{equation}
\begin{aligned}\Qcircuit @C=1em @R=.7em @! R {
\multiprepareC{1}{\Psi'}&\qw\poloFantasmaCn{\rA}&\qw\\
\pureghost{\Psi'}&\qw\poloFantasmaCn{\rC}&\qw \\
\prepareC{\eta}&\qw\poloFantasmaCn{\rB}&\qw}\end{aligned}~=~
\begin{aligned}
\Qcircuit @C=1em @R=.7em @! R {
\multiprepareC{1}{\Psi}&\qw&\qw&\qw\poloFantasmaCn{\rA}&\qw\\
\pureghost{\Psi}&\qw\poloFantasmaCn{\rB}&\multigate{1}{\tU}&\qw\poloFantasmaCn{\rC}&\qw \\
\prepareC{\varphi_0}&\qw\poloFantasmaCn{\rC}&\ghost{\tU}&\qw\poloFantasmaCn{\rB}&\qw}
\end{aligned}
\end{equation}
Applying the deterministic effect $e$ on system $\rB$ we obtain Eq.
(\ref{uniquenessBC}), with $\tC := \B e_\rB \tU \K{\varphi_0}$.  \qed

\subsection{Purification of preparation-tests}

We now show that the purification of normalized states implies the purification of preparation-tests.
 
\begin{theorem}[Purification of preparation-tests]\label{theo:purificami-ens}
  Let $\{\rho_i\}_{i \in \rX}$ be a preparation-test for system $\rA$,
  and let $\Psi \in \Stset_1 (\rA \rB)$ be a purification of the
  coarse-grained state $\rho:=\sum_{i\in \rX} \rho_i$. Then there exists an
  observation-test $\{b_i\}_{i\in \rX}$ on system $\rB$ such that
\begin{align}\label{purificami-ens}
 \begin{aligned} \Qcircuit @C=1em @R=.7em @! R {
    \prepareC{\rho_i}&\qw\poloFantasmaCn{\rA}&\qw}& \end{aligned}~=~ 
\begin{aligned}\Qcircuit @C=1em @R=.7em @! R {
    \multiprepareC{1}{\Psi}&\qw\poloFantasmaCn{\rA}&\qw&\\
    \pureghost{\Psi}&\qw\poloFantasmaCn{\rB}&\measureD{b_i}} 
\end{aligned}
\end{align}
for any outcome $i \in \rX$.  By suitably choosing the purifying system $\rB$, the observation-test
$\{b_i\}_{i\in \rX}$ can be taken to be discriminating (Definition \ref{def:distinguishable}). 
\end{theorem}

\Proof Take a set of $|\rX|$ perfectly distinguishable states
$\{\varphi_i\}_{i \in \rX}\subset \Stset_1 (
\rC)$ for some system $\rC$. By definition of
perfect distinguishability, there exists a discriminating test
$\{c_i\}_{i \in X }$ such that
\begin{align}\label{discri}
  \Qcircuit @C=1em @R=.7em @! R {
    \prepareC{\varphi_i}&\qw\poloFantasmaCn{\rC} &\measureD{c_j}}&= \delta_{ij} 
\end{align}
for all $i,j \in \rX$.  Now consider the state
\begin{equation}
\sigma := \sum_{i\in \rX}  \left(\rho_i \otimes \varphi_i\right) \in \Stset_1(\rA \rC),  
\end{equation}
which is clearly an extension of $\rho$, namely $\K{\rho}_A = \B{e}_{\rC} \K{\sigma}_{\rA \rC}$.
Let $\Psi_\sigma \in \Stset_1(\rA \rC \rD)$  be a purification of $\sigma$. By definition, $\Psi$ is also a purification of $\rho$. Using Eq.(\ref{discri}) we obtain for every outcome $i\in \rX$
\begin{align}
\begin{aligned}  \Qcircuit @C=1em @R=.7em @! R {
    \prepareC{\rho_i}&\qw\poloFantasmaCn{\rA}&\qw}\end{aligned}& ~=~
\begin{aligned}  
\Qcircuit @C=1em @R=.7em @! R {
    \multiprepareC{1}{\sigma}&\qw\poloFantasmaCn{\rA}&\qw&\\
    \pureghost{\sigma}&\qw\poloFantasmaCn{\rC}&\measureD{c_i}}\end{aligned}
~=~ \begin{aligned}\Qcircuit @C=1em @R=.7em @! R {
    \multiprepareC{2}{\Psi_\sigma}&\qw\poloFantasmaCn{\rA}&\qw&\\
    \pureghost{\Psi_\sigma}&\qw\poloFantasmaCn{\rC}&\measureD{c_i} \\
\pureghost{\Psi_\sigma}&\qw\poloFantasmaCn{\rD}&\measureD{e}}\end{aligned} \\
&~=~\begin{aligned}  \Qcircuit @C=1em @R=.7em @! R {
    \multiprepareC{1}{\Psi_\sigma}&\qw\poloFantasmaCn{\rA}&\qw&\\
    \pureghost{\Psi_\sigma}&\qw\poloFantasmaCn{\rC\rD}&\measureD{b_i}}
\end{aligned} 
\end{align}
having defined the discriminating test $\B{b_i}_{\rC \rD} :=
\B{c_i}_{\rC} \B{e}_{\rD}$.  This proves that there exists a
purification of $\rho$ with purifying system $\rB := \rC \rD$, and a
discriminating test $\{b_i\}_{i \in \rX}$ on $\rB$ such that the thesis holds.

Finally, if $\Psi \in \Stset_1(\rA \rB')$ is any other purification of $\rho$,  using Lemma \ref{lem:purichan} we have
\begin{align}
\begin{aligned}\Qcircuit @C=1em @R=.7em @! R {\prepareC{\rho_i}&\qw\poloFantasmaCn{\rA}&\qw}\end{aligned} &~= ~ \begin{aligned} 
\Qcircuit @C=1em @R=.7em @! R {
\multiprepareC{1}{\Psi_\sigma}&\qw\poloFantasmaCn{\rA}&\qw\\
\pureghost{\Psi_\sigma}&\qw\poloFantasmaCn{\rB}&\measureD{b_i}}\end{aligned} ~=~
\begin{aligned}\Qcircuit @C=1em @R=.7em @! R {
\multiprepareC{1}{\Psi}&\qw&\qw&\qw\poloFantasmaCn{\rA}&\qw\\
\pureghost{\Psi}&\qw\poloFantasmaCn{\rB'}&\gate{\tC}&\qw\poloFantasmaCn{\rB}\qw&\measureD{b_i}}
\end{aligned}\\
& ~=~\begin{aligned} \Qcircuit @C=1em @R=.7em @! R {
\multiprepareC{1}{\Psi}&\qw\poloFantasmaCn{\rA}&\qw\\
\pureghost{\Psi}&\qw\poloFantasmaCn{\rB'}&\measureD{b'_i}}
\end{aligned}
\end{align}
where $\{b'_i\}_{i \in \rX}$ is the observation-test on $\rB'$ defined
by $\B{b_i'}:= \B{b_i} \tC$. 
\qed

\medskip

The property stated by Theorem \ref{theo:purificami-ens} is sometimes called \emph{steering} in quantum theory, with a terminology that dates back to Schr\"odinger   
\cite{shroed} (see also Ref. \cite{steer}, for a very recent discussion in the general probabilistic framework): one says that a bipartite state $\K\sigma_{\rA\rB}$ steers its marginal $\K \rho_\rA = \B e_\rB \K\sigma_{\rA\rB}$ on system $\rA$, if every convex decomposition  $\K \rho_\rA = \sum_{i\in \rX} p_i \K{\rho_i}_{\rA}$ is induced by a suitable observation-test on system $\rB$.   
 Using the notion of steering, we may state the following: 
\begin{corollary}{\bf (Pure bipartite states are steering for their marginals)}
In a theory with purification any pure state $\K\Psi\in\Stset_1(\rA\rB)$ steers its marginal states $\K{\rho}_\rA =\B e_\rB \K\Psi_{\rA\rB}$ and $\K{\tilde \rho}_\rA =\B e_\rA \K\Psi_{\rA\rB}$.   
\end{corollary}

\medskip
We now present a few other corollaries of the purification of preparation-tests stated by Theorem \ref{theo:purificami-ens}.
\begin{corollary}\label{cor:remote-prep}
  Let $\Psi \in \Stset_1 (\rA \rB)$ be a purification of $\rho $. Then, a state $\sigma$ is in the
  refinement set $D_\rho$ if and only if there is an effect $b_{\sigma} \in \Cntset(\rB)$ such that
\begin{equation}
\begin{array}{cc}
\begin{aligned}\Qcircuit @C=1em @R=.7em @! R {\prepareC{\sigma} & \qw \poloFantasmaCn{\rA} & \qw}  \end{aligned}& ~=~\begin{aligned} \Qcircuit @C=1em @R=.7em @! R {\multiprepareC{1}{\Psi}&\qw\poloFantasmaCn{\rA}& \qw \\
\pureghost{\Psi} &\qw \poloFantasmaCn{\rB} &\measureD{b_{\sigma}}}
\end{aligned}
\end{array}
\end{equation}    
\end{corollary}
\Proof The ``if'' part is trivial. Conversely, if $\sigma$ is in
$D_\rho$, by definition there exists a preparation-test $\{\rho_i\}_{i \in
  \rX}$ and an outcome $i_0$ such that $\rho_{i_0} =  \sigma$. Using Theorem \ref{theo:purificami-ens} and
taking the effect $b_{\sigma}:= b_{i_0}$ one proves the thesis.\qed

\begin{corollary}[Bound on dimensions]\label{cor:dimpurifier}
Let $\Psi \in \Stset_1 (\rA \rB)$ be a purification of $\rho \in \Stset_1 (\rA)$. Then, one has the bound
\begin{equation}\label{sonoallostremo}
\dim   \Stset_\Reals  (\rB) \ge \dim \Span (D_\rho).
\end{equation}
In particular, if $\rho$ is an internal state,  one has 
\begin{equation}\label{eq:dimpurifierinternal}
\dim   \Stset_\Reals  (\rB)) \ge  \dim \Stset_\Reals  (\rA).
\end{equation}
\end{corollary}

\Proof Consider the map $\hat \omega : \Cntset_\Reals  (\rB) \to
\Stset_\Reals  (\rA)$ defined by $b \mapsto \K{\hat \omega_b}_{\rA}
:= \B{b} \K{\Psi}_{\rA \rB}$. By the previous corollary, the range of $\hat \omega$ contains $D_\rho$. Since $\hat \omega$ is linear, this implies $\dim \Cntset_\Reals  (\rB) \ge \dim \Span (D_\rho)$.  On the other hand, since states and effects span dual vector spaces, one has $\dim \Cntset_\Reals  (\rB) \equiv \dim \Stset_\Reals  (\rB)$, thus proving Eq. (\ref{sonoallostremo}). \qed

\medskip

Theorem \ref{theo:purificami-ens} implies the existence of pure
bipartite states exhibiting perfect correlations in the statistics of
independent observations:
\begin{corollary}{\bf (Pure states with perfect correlations)}
  Let $\rho = \sum_{i\in \rX}~ p_i \varphi_i$ be a mixture of
  perfectly distinguishable states $\{\varphi_i\} \subset \Stset_1 (\rA)$,
  and let $\Psi\in \Stset_1 (\rA \rB)$ be a purification of $\rho$. Then
  there exist two observation-tests $\{a_i\}_{i \in \rX}$ and
  $\{b_j\}_{j \in \rX}$ on systems $\rA$ and $\rB$, respectively, such
  that
\begin{equation}
\begin{array}{cc}
\begin{aligned}\Qcircuit @C=1em @R=.7em @! R {\multiprepareC{1}{\Psi}&\qw\poloFantasmaCn{\rA}& \measureD{a_i} \\
\pureghost{\Psi} &\qw \poloFantasmaCn{\rB} &\measureD{b_j}}\end{aligned} &~=~ p_i \delta_{ij}
\end{array}
\end{equation}
\end{corollary}
\Proof Consider the preparation-test $\{\rho_i\}_{i\in \rX}$ with
$\rho_i = p_i \varphi_i$. Since its coarse-grained state is $\rho$, by
Theorem \ref{theo:purificami-ens} there exists an observation-test
$\{b_i\}$ such that $\K{\rho_i}_{\rA} =\B{b_i}_{\rB}
\K{\Psi}_{\rA\rB}$. On the other hand, the states $\{\varphi_i\}$ are
perfectly distinguishable with a test $\{a_i\}_{i \in \rX}$. Hence, we have $\B{a_i}_{\rA} \B{b_j}_{\rB} \K{\Psi}_{\rA\rB} = \B{a_i}_{\rA}  \K{\rho_j} = p_i \delta_{ij}$. \qed

\medskip

This directly implies the following property
\begin{corollary}
Let $\rho = \sum_{i \in \rX} ~p_i \varphi_i\in\Stset_1(\rA)$ be a mixture of perfectly distinguishable states,  $\Psi \in \Stset_1 (\rA\rB)$ be a purification of $\rho$, and $\tilde \rho = \B{e}_{\rA} \K{\Psi}_{\rA\rB}$ be the complementary state of $\rho$. Then, one has
\begin{equation}
\tilde \rho = \sum_{i \in \rX}  ~p_i \tilde\varphi_i,
\end{equation} 
where $\{\tilde \varphi\}_{i \in \rX}$ are perfectly distinguishable
states of $\rB$.
\end{corollary}

We conclude this subsection with a crucial consequence of the purification of preparation-test stated by Theorem
\ref{theo:purificami-ens}, namely that if two transformations coincide
on a purification of $\rho$, they also coincide \emph{upon input of $\rho$}, according to the following definition: 
\begin{definition}[Equality upon input of $\rho$] 
Two transformations $\tA, \tA' \in \Trnset (\rA, \rB)$ are \emph{equal upon input of $\rho$}, denoted by $\tA =_\rho \tA'$, if one has
\begin{equation}
\tA \K{\sigma}_\rA = \tA' \K\sigma_\rA \qquad \forall \K \sigma_\rA \in D_\rho.
\end{equation}
\end{definition} 
In quantum theory two quantum operations $\tA,\tA'$ are equal upon input of $\rho$ if and only of one has $\tA(\sigma)=\tA'(\sigma)$ for every density matrix $\sigma$ whose support is contained in the support of $\rho$.
\medskip

We then have the following:
\begin{theorem}\label{theo:uponinput}{\bf (Equality upon input of $\rho$ vs equality on purifications)}
  Let $\Psi \in \Stset_1(\rA \rC)$ be a purification of $\rho \in \Stset_1
  (\rA)$, and let $\tA, \tA' \in \Trnset (\rA,\rB)$ be two
  transformations. Then one has
\begin{equation}\label{equalpuri} 
  \tA \K{\Psi}_{\rA\rC} = \tA' \K{\Psi}_{\rA \rC}   \qquad \Longrightarrow \qquad \tA =_{\rho}  \tA'~.
\end{equation} 
If local discriminability holds, one has the equivalence
\begin{equation}\label{equalpuriLO} 
  \tA \K{\Psi}_{\rA\rC} = \tA'  \K{\Psi}_{\rA \rC}  \qquad \Longleftrightarrow \qquad \tA =_{\rho}  \tA'~.
\end{equation} 
If one of the two transformations is proportional to a reversible transformation the equivalence of Eq. (\ref{equalpuriLO}) holds under the weaker assumption of local discriminability on pure states.
\end{theorem}
\Proof By definition, a state $\sigma$ is in the refinement set
$D_\rho$ iff there exists a preparation-test $\{\rho_i\}_{i \in \rX}$
and an outcome $i_0$ such that $\rho_{i_0} = \sigma$.  Using Corollary \ref{cor:remote-prep},
we have that $\sigma$ is in $D_\rho$ iff there exist an effect $c$ on
$\rC$ such that
\begin{equation}
\begin{array}{cc}
 \quad \begin{aligned} \Qcircuit @C=1em @R=.7em @! R {\prepareC{\sigma} & \qw \poloFantasmaCn{\rA} & \qw}\end{aligned}  & ~=~ \begin{aligned} \Qcircuit @C=1em @R=.7em @! R {\multiprepareC{1}{\Psi}&\qw\poloFantasmaCn{\rA}& \qw \\
\pureghost{\Psi} &\qw \poloFantasmaCn{\rC} &\measureD{c}}
\end{aligned}
\end{array}
\end{equation} 
Therefore,  we have that $\tA =_\rho \tA'$ if and only if 
\begin{equation}
\begin{array}{cc}
  \begin{aligned}\Qcircuit @C=1em @R=.7em @! R {\multiprepareC{1}{\Psi}&\qw\poloFantasmaCn{\rA}&\gate{\tA}& \qw\poloFantasmaCn{\rB} &\qw \\
    \pureghost{\Psi} &\qw \poloFantasmaCn{\rC} &\measureD{c}} 
\end{aligned} & ~= ~\begin{aligned} \Qcircuit @C=1em @R=.7em @! R {\multiprepareC{1}{\Psi}&\qw\poloFantasmaCn{\rA}&\gate{\tA'}& \qw\poloFantasmaCn{\rB}&\qw \\
    \pureghost{\Psi} &\qw \poloFantasmaCn{\rC} &\measureD{c}}
\end{aligned}
\end{array}
\end{equation}   
that is, if and only if the states $\tA \K{\Psi}_{\rA\rC}$ and
$\tA'\K{\Psi}_{\rA \rC}$ cannot be distinguished by local tests, that is, if and only if 
\begin{equation}
\begin{array}{cc}
  \begin{aligned}\Qcircuit @C=1em @R=.7em @! R {\multiprepareC{1}{\Psi}&\qw\poloFantasmaCn{\rA}&\gate{\tA}& \qw\poloFantasmaCn{\rB} &\measureD b \\
      \pureghost{\Psi} &\qw \poloFantasmaCn{\rC} &\qw &\qw &\measureD{c}}
\end{aligned} & ~= ~\begin{aligned} \Qcircuit @C=1em @R=.7em @! R {\multiprepareC{1}{\Psi}&\qw\poloFantasmaCn{\rA}&\gate{\tA'}& \qw\poloFantasmaCn{\rB}&\measureD b \\
    \pureghost{\Psi} &\qw \poloFantasmaCn{\rC} &\qw &\qw &\measureD{c}}
\end{aligned}
\end{array}
\end{equation}
for every product effect $\B b_\rB \B c_{\rC}$. Clearly, if $\tA
\K{\Psi}_{\rA \rC} = \tA' \K{\Psi}_{\rA\rC}$, this condition is
verified: this proves Eq. (\ref{equalpuri}). When local
discriminability holds, equality on local tests implies equality on
global tests, hence Eq. (\ref{equalpuriLO}). Finally, if
$\tA'=\lambda \tU$ with $\tU$ reversible, then the state $\tA' \K{\Psi}_{\rA\rC} =
\lambda \tU \K{\Psi}_{\rA \rC}$ is pure, and, by local discriminability of
pure states, equality on local tests implies equality. \qed

\subsection{Dynamically faithful pure states}\label{ssect:dynfaith}

We show now an important feature of theories with purification: the possibility of imprinting physical transformations into states in an injective way (that is, if two transformations differ, then the corresponding states are differ).  This feature reduces the tomography of a physical process to the tomography of the corresponding state.   Technically speaking, we call \emph{dynamically faithful} any state that allows for the tomography of physical processes.

\begin{definition}[Dynamically faithful state]\label{def:dynfaith} We say that a state
  $\sigma \in \Stset(\rA\rC)$ is \emph{dynamically faithful} for
  system $\rA$ if for any couple of transformations $\tA, \tA' \in
  \Trnset (\rA, \rB)$ on has
\begin{equation}
\tA \K{\sigma}_{\rA \rC} = \tA' \K{\sigma}_{\rA \rC} \qquad \Longrightarrow \qquad \tA = \tA'.   
\end{equation}
\end{definition}
The existence of dynamically faithful mixed states is a quite generic fact: for example, in any theory with local discriminability if one takes a basis $\{\rho_i\}\subset \Stset_1 (\rA)$ for $\Stset_\Reals (\rA)$ and a set $\{\varphi_i\}\subset \Stset_1(\rC)$ of perfectly distinguishable states of some system $\rC$, than any mixture $\sigma=\sum_i p_i  \rho_i \otimes \varphi_i$ is dynamically faithful.  
The remarkable fact in a theory with purification is that there exist dynamically faithful states, which, in addition, are pure. 
\begin{theorem}\label{theo:dynfaith}{\bf (Existence of dynamically faithful pure states)} Let $\omega \in \Stset_1(\rA)$ be an internal state, and
  let $\Psi_{\omega}\in\Stset_1(\rA\rC)$ be a purification of $\omega$.
  Then $\Psi_\omega$ is dynamically faithful for system $\rA$.
\end{theorem}

\Proof Suppose that
$\tA \K{\Psi_\omega}_{\rA\rC}=\tA'\K{\Psi_\omega}_{\rA\rC}$.
Then take an arbitrary system $\rD$, an internal state $\sigma \in
\Stset_1 (\rD)$, and a purification of $\sigma$, say $\Psi_{\sigma}
\in \Stset_1 (\rD \rE)$.  Clearly, we have $\tA
\K{\Psi_{\omega}}_{\rA\rC} \otimes\K{\Psi_\sigma}_{\rD \rE}=\tA'
\K{\Psi_\omega}_{\rA\rC}\otimes\K{\Psi_{\sigma}}_{\rD \rE}$. According
to Theorem \ref{theo:uponinput}, this implies that $\tA$ and $\tA'$
coincide upon input of $\omega \otimes \sigma$. Since $\omega$ and
$\sigma$ are internal in $\Stset (\rA)$ and $\Stset (\rD)$,
respectively, by Theorem \ref{theo:internalprod} $\omega \otimes
\sigma$ is internal in $\Stset (\rA \rD)$, that is, the refinement set
$D_{\omega \otimes \sigma}$ is a spanning set for $\Stset_\Reals
(\rA\rD)$.  Now, $\tA$ and $\tA'$ coincide on a spanning set, and,
therefore, they coincide on every state of $\Stset (\rA\rD)$. Since
the ancillary system $\rD$ is arbitrary, this implies $\tA = \tA'$. \qed

The converse of the previous Theorem \ref{theo:dynfaith} also holds: 
\begin{theorem}\label{theo:chardynfaith}{\bf (Characterization of dynamically faithful pure states)} A pure state $\Psi \in \Stset_1 (\rA\rC)$ is dynamically faithful for system $\rA$ if and only if the marginal state $\K \omega_\rA := \B e_\rC \K\Psi_{\rA\rC}$ is internal. 
\end{theorem}
\Proof The ``if" part has been just shown in Theorem \ref{theo:dynfaith}.  To prove the ``only if" part, let $a, a'$ be two distinct effects for system $\rA$. Since
$\Psi$ is dynamically faithful one has $\B a_\rA \K\Psi_{\rA\rC} \not
=\B a'_\rA \K\Psi_{\rA\rC}$. This means that there exists an effect
$\B c_\rC$ such that $\B a_\rA \B c_\rC \K\Psi_{\rA\rC} \not =\B
a'_\rA \B c_\rC \K\Psi_{\rA\rC}$.  Defining the state $\K{\omega_c}_\rA
:= \B c_\rC \K \Psi_{\rA\rC}$, this implies $\SC a {\omega_c}_\rA \not =
\SC {a'} {\omega_c}_\rA$. Since $\omega_c$ is in the refinement set of
$\omega$, such a refinement set is separating for $\Cntset_{\Reals}
(\rA)$. But a separating set for $\Cntset_{\Reals} (\rA)$ must be a
spanning set for the dual vector space $\Stset_\Reals (\rA)$. Hence,
$\omega$ is internal.  \qed

\medskip
Using this characterization it is immediate to show that the product of dynamically faithful pure states  is dynamically faithful:
\begin{corollary}\label{cor:dynfaithprod}{\bf (Product of dynamically faithful states is dynamically faithful)}
Let $\Psi^{(\rA)}\in\Stset_1(\rA\rC)$ and $\Psi^{(\rB)}\in\Stset_1(\rB\rD)$ be dynamically faithful for systems $\rA$ and $\rB$, respectively.   Then $\Psi^{(\rA)}\otimes \Psi^{(\rB)}$ is dynamically faithful for the compound system $\rA\rB$.
\end{corollary}

\Proof Since the product of two internal states is internal (Theorem \ref{theo:internalprod}), the thesis trivially follows from the previous Theorem. \qed  


\medskip 

The existence of dynamically faithful pure states has remarkable
consequences, among which the ``no-information without disturbance''
and the ``no-cloning'' Theorems, that will be analyzed in the
following Subsections.
\subsection{No information without disturbance}\label{subsect:no-info}

\begin{definition}[Non-disturbing tests]
  We say that a test $\{\tA_i\}_{i \in \rX}$ on system $\rA$ is
  \emph{non-disturbing} upon input of $\rho \in \Stset (\rA)$ if
\begin{equation}
\sum_{i \in \rX} \tA_i \K{\sigma}_{\rA} = \K{\sigma}_{\rA} \qquad \forall \sigma \in D_{\rho}, 
\end{equation}
or, equivalently, if $\sum_{i \in \rX}\tA_i =_\rho \tI_\rA$. If $\rho$
is an internal state, we say that the test is \emph{non-disturbing}, because in this case one has
\begin{equation}
  \sum_{i \in \rX} \tA_i \K{\sigma}_{\rA} = \K{\sigma}_{\rA} \qquad \forall \sigma \in \Stset (\rA).
\end{equation}
\end{definition}
\begin{theorem}[No information without disturbance]\label{theo:no-info}
  In a theory with purification, a test $\{\tA_i\} $ on system $\rA$
  is non-disturbing upon input of $\rho$, if and only if each
  transformation $\tA_i$ is proportional to the identity upon input of
  $\rho$, namely $\tA_i =_\rho ~ p_i \tI_\rA$.
\end{theorem}
\Proof Let $\Psi_{\rA \rB}$ be a purification of $\rho$. By Theorem \ref{theo:uponinput},  the no-disturbance condition $\sum_{i \in \rX} \tA_i =_\rho ~ \tI_\rA$ holds if and only if 
\begin{equation}
\begin{array}{cc}
  \sum_{i \in \rX}~ \begin{aligned}\Qcircuit @C=1em @R=.7em @! R {\multiprepareC{1}{\Psi}&\qw\poloFantasmaCn{\rA}&\gate{\tA_i}& \qw\poloFantasmaCn{\rA} &\qw \\
    \pureghost{\Psi} &\qw \poloFantasmaCn{\rB} &\qw&\qw &\qw} \end{aligned} &~=~ \begin{aligned} \Qcircuit @C=1em @R=.7em @! R {\multiprepareC{1}{\Psi}&\qw\poloFantasmaCn{\rA}& \qw \\
    \pureghost{\Psi} &\qw \poloFantasmaCn{\rB} &\qw}
\end{aligned}
\end{array}
\end{equation}   
Since $\Psi$ is pure, this implies $\tA_i \K{\Psi}_{\rA\rB} = p_i  \K{\Psi}_{\rA\rB}= (p_i \tI_{\rA}) \K{\Psi}_{\rA \rB}$.  Now, since the identity is trivially  a reversible transformation, according to Theorem \ref{theo:uponinput} this is equivalent to $\tA_i =_\rho~ p_i \tI_\rA$. \qed

\begin{theorem}{\bf (No joint discrimination of a spanning set of
    states)}\label{th:discr} In a theory with purification the states
  in a spanning set cannot be perfectly discriminated in a single
  observation-test.
\end{theorem}
\Proof By contradiction, suppose  that a collection of states
$\{\rho_i\}_{i\in \rX}$ is a spanning set---namely $\Span
\{\rho_i\}_{i\in\rX} = \Stset_{\Reals} (\rA)$---and there exists an observation-test  $\{a_i\}_{i \in \rX}$ such that $\SC {a_i}
{\rho_j}_\rA = \delta_{ij}$.  Then, since perfectly distinguishable
states are linearly independent, and they must span a finite dimensional vector space, the number of perfectly distinguishable states must be finite.  Now consider the measure-and-prepare test $\{\tA_i\}_{i
  \in \rX}$ defined by $\tA_i = \K{\rho_i}_\rA \B{a_i}_\rA$.  Since
the states of the spanning set are perfectly distinguishable, the test
$\{\tA_i\}$ is non-disturbing. Indeed, expanding an arbitrary state
$\rho$ on the spanning set, one has
\begin{equation}
\sum_i\tA_i\K{\rho}_A=\sum_i\tA_i\left(\sum_jc_j\K{\rho_j}_A\right)=
\sum_jc_j\K{\rho_j}_A=\K{\rho}.
\end{equation}
Since $\tA_i \not = p_i \tI_\rA$, this is in contradiction with the
no-information without disturbance Theorem \ref{theo:no-info}.  \qed

\begin{corollary}{\bf (No joint discrimination of pure states)}\label{c:discrpur} In a theory with
  purification for every system the pure states cannot be perfectly
  discriminated in a single observation-test.
\end{corollary}
\Proof Since pure states are a spanning set, they cannot be perfectly
discriminated in a single test, according to Theorem
(\ref{th:discr}).\qed

Corollary \ref{c:discrpur} provides a simple alternative way to
see that classical probability theory is excluded by the purification
Postulate.

\begin{corollary}\label{cor:maxdiscr}{\bf (Maximum number of perfectly distinguishable states)} For every system $\rA$ the maximum cardinality  of a set of perfectly distinguishable states is strictly smaller than $\dim \Stset_\Reals (\rA)$.
\end{corollary}

\Proof Since perfectly distinguishable states are linearly independent, if one could find $\dim \Stset_\Reals (\rA)$ perfectly distinguishable states, then they would form a spanning set, in contradiction with Theorem \ref{th:discr}. \qed

\medskip

Note that the maximum number of distinguishable states in quantum theory satisfies a much stronger bound:  such a number is given by the dimension $d_\rA$ of the system's Hilbert space, while the dimension of the vector space spanned by the density matrices is $\dim \Stset_\Reals(\rA) = d_\rA^2$. 
\medskip

\begin{corollary}{\bf (Non-unique convex decomposition on pure
    states)} In a theory with purification satisfying the
  no-restriction hypothesis of Def. \ref{def:no-restriction}, for
  every system $\rA$ there is a mixed state $\rho \in \Stset_1 (\rA)$
  with a non-unique convex decomposition on pure states. In other
  words, the convex set $\Stset_1 (\rA)$ cannot be a simplex.
\end{corollary}
\Proof By contradiction, suppose that $\Stset_1 (\rA)$ is a simplex.  Then
the pure states $\{\varphi_i\}$ of $\rA$ are a finite set, and for
each of them there is a functional $a_i \in\Cntset_\Reals
(\rA)$ such that $\SC {a_i}{\varphi_j} = \delta_{ij}$. Clearly, $a_i$
is positive on every state, namely $a_i \in \Stset_+ (\rA)^*$. Hence,
by the consequence of the no-restriction hypothesis stated by Lemma \ref{lem:no-restrictioncausal}, we have $a_i \in \Cntset_+ (\rA)$.
Moreover, one has $\sum_i \B {a_i}_\rA = \B e_\rA$.  In Corollary  \ref{cor:evm} we will show that any such collection $\{a_i\}$ is an observation-test.   But this test  discriminates all pure states, in
contradiction with Corollary \ref{c:discrpur}. This proves that $\Stset_1(\rA)$ cannot be a simplex. \qed

\subsection{No-cloning}
   
\begin{definition}[Cloning channels] 
Let $\rA, \rA'$ be two operationally equivalent systems, and let $\{\rho_i\}_{i \in \rX}$ be a set of states of $\rA$. A channel  $\tC$ from $\rA$ to $\rA \rA'$ is a \emph{cloning channel} for the set $\{\rho_i\}_{i \in \rX}$ if 
\begin{equation}
\tC \K{\rho_i}_\rA = \K{\rho_i}_\rA \K{\rho_i}_{\rA'}.  
\end{equation}
If there is a cloning channel, we say that the states $\{\rho_i\}_{i \in \rX}$ are \emph{perfectly cloneable}. 
\end{definition}

We now show that a spanning set of states (in particular, the set of
pure states) cannot be perfectly cloned. To see this we use the
equivalence between perfect cloning and perfect discrimination, which
was originally proved in Refs. \cite{noclon,nobroad} for causal
theories with local discriminability using the tomographic limit. Here
we use the stronger result of Ref. \cite{clonunit}, which proves the equivalence in  any convex theory where all ``measure-and-prepare'' channels are allowed, without requiring causality and local
discriminability, and without resorting to the tomographic limit.  For
convenience of the reader, the argument of Ref. \cite{clonunit} is
reproduced here using the notation of the present paper:

\begin{theorem}\label{lemma:clon-disc} {\bf (Cloning/discrimination equivalence)}
  In a convex theory where all ``measure-and-prepare'' channels are
  allowed, the deterministic states $\{\rho_i\}_{i \in \rX}\subset
  \Stset (\rA)$ are perfectly cloneable if and only if they are
  perfectly distinguishable.
\end{theorem}

\Proof   Suppose that the
states $\{\rho_i\}_{i \in \rX}$ can be perfectly cloned and consider
the binary discrimination between two states $\rho_i , \rho_j, i \not = j $
with a binary observation-test $\{a_i,a_j\}$.  Define the worst-case
error probability as
\begin{equation}
p_{wc}:= \max \{ p (i|j), p(j|i)\}\qquad  p(k|l):= \SC{a_k}{\rho_l}_\rA,
\end{equation}     
and take its minimum over all binary tests
\begin{equation}
p_{wc}^{(opt)} := \min_{ a_i , a_j}   p_{wc}.
\end{equation}
Now, if a cloning channel
exists, we can apply it twice to the unknown state, thus getting three
identical copies of it.  Performing three times the optimal test, and
then using majority voting we obtain the new error probabilities given
by
\begin{equation}
p'(i|j) =  f( p^{(opt)}(i|j)) \qquad f(x) =  x^2  (3-2 x),
\end{equation}   
where $p^{(opt)} (i|j) := \SC {a_i^{(opt)}} {\rho_j}_\rA$.  Since $f$
is a non-decreasing function for $x\in [0,1]$, we also have $p_{wc}' =
f\left(p^{(opt)}_{wc}\right)$, and, since $p_{wc}^{(opt)}$ is the minimum error
probability, by definition $p_{wc}' \ge p_{wc}^{opt}$. The only
solutions of the inequality $f(x)\ge x$ are $x=0$ and $x \in [1/2,1]$,
and, since $p_{wc}^{(opt)}$ must be in the interval $[0,1/2)$ (see Lemma \ref{lemma:worstcase}), we
obtain $p_{wc}^{(opt)} = 0$. This proves that any pair of states from
the set $\{\rho_i\}_{i \in \rX}$ can be perfectly distinguished.  But
this implies that using $|\rX| -1$ pairwise tests we can perfectly
discriminate all the states $\{\rho_i\}_{i\in \rX}$.  This proves the
implication ``perfect cloning $\Rightarrow$ perfect discrimination''
in any convex theory. If the theory contains all possible
``measure-and-prepare'' channels, the converse is obviously true: If
the states can be perfectly discriminated by an observation-test
$\{a_i\}_{i \in \rX}$, then the measure-and-prepare channel $\tC:=
\sum_{i \in \rX} \K{\rho_i}_\rA \K{\rho_i}_{\rA'} \B{a_i}_{\rA}$ is a
cloning channel. \qed

Since measure-and-prepare channels can be obtained by conditioning the choice of a preparation-test on the outcome of an observation-test, any causal theory satisfies the hypotheses of the previous Theorem, which becomes  

\begin{corollary}\label{cor:cloncaus}{\bf (Cloning/discrimination equivalence in causal
    theories)} In a causal theory the states $\{\rho_i\}_{i \in
    \rX}\subset \Stset_1 (\rA)$ are perfectly cloneable if and only if
  they are perfectly distinguishable.
\end{corollary}

{\bf Remark (Non-causal theories with all measure-and-prepare
  channels).}  Note that there are also non-causal theories that
contain all measure-and prepare channels. An example can be
constructed by starting from a causal theory $\Theta$, and by
regarding the set of transformations $\Trnset (\rA, \rB)$ from $\rA$
to $\rB$ as the set of ``states'' $\Stset' (\rA \to \rB)$ of the
system ``$ \rA \to \rB$'' in a new \emph{second-order} theory
$\Theta'$.  Performing an observation-test on a ``state'' $\tC \in
\Stset' (\rA \to \rB)$ is then interpreted in the underlying causal
theory $\Theta$ as applying the transformation $\tC \in \Trnset (\rA,
\rB)$ to an input state $\sigma \in \Stset_1 (\rA\rC)$, and subsequently
performing an observation-test $\{b_i\}_{i \in \rX}$ on the output
state $(\tC \otimes \tI_\rC) \K \sigma_{\rA \rC}$.  Of course, since
the theory $\Theta$ is causal, one can use conditioning and perform a
channel $\tC_i$ that depends on the outcome $i $.  This provides the
realization of an arbitrary measure-and-prepare channel in the
non-causal theory $\Theta'$.  \medskip

Coming back to causal theories with purification, the results proved so far imply the following no-cloning statement:

\begin{corollary}{\bf (No-cloning of states in a spanning set)}
  In a theory with purification, a cloning channel for a spanning set of states cannot exist. In
  particular, pure states cannot be cloned.
\end{corollary}
\Proof Immediate consequence of Corollary \ref{cor:cloncaus} combined
with Theorem \ref{th:discr} and Corollary \ref {c:discrpur}.\qed

\section{Probabilistic teleportation} \label{sec:probtele} 

\subsection{Entanglement-swapping and teleportation}

As we previously showed, in a theory with purification there must be entangled states (according to the usual definition, see Def.  \ref{def:ent}). We now show the possibility of probabilistic entanglement swapping:

\begin{theorem}\label{theo:entswap}{\bf (Probabilistic entanglement-swapping)}
  Let $\Psi \in \Stset_1 (\rA \rB)$ be a pure state, and let $\rA'$ and
  $\rB'$ be operationally equivalent to $\rA$ and $\rB$, respectively.
  Then there exist an atomic effect $E_\Psi \in \Cntset
  (\rB\rA')$ (see Def. \ref{def:atomic})and a non-zero probability $p_{\Psi}$ such that
\begin{equation}\label{ent-swap}
  \begin{aligned}
\Qcircuit @C=1em @R=.7em @! R { 
&\multiprepareC{1}{\Psi} &\qw \poloFantasmaCn{\rA} &\qw \\
&\pureghost{\Psi} &\qw \poloFantasmaCn{\rB} & \multimeasureD{1}{E_{\Psi}} \\
&\multiprepareC{1}{\Psi} &\qw \poloFantasmaCn{\rA'} &\ghost{E_{\Psi}}  \\
&\pureghost{\Psi} &\qw \poloFantasmaCn{\rB'} &\qw  }
\end{aligned} ~= p_{\Psi}~
  \begin{aligned}
\Qcircuit @C=1em @R=.7em @! R {\multiprepareC{1}{\Psi} &\qw \poloFantasmaCn{\rA} &\qw \\
    \pureghost{\Psi} &\qw \poloFantasmaCn{\rB'} &\qw }
\end{aligned}
\end{equation}       
\end{theorem}
 
\Proof Let us define the marginal states 
\begin{equation}
\begin{split}
\K{\rho}_{\rA} &:=
\B{e}_{\rB} \K{\Psi}_{\rA\rB}\\
\K {\tilde \rho}_{\rB} &:= \B{e}_{\rA}
\K{\Psi}_{\rA\rB}
\end{split}
\end{equation}  
By Theorem \ref{theo:marginaldecomp} we have that there exists a
non-zero probability $p_\Psi$ such that $p_{\Psi} \Psi \in D_{\rho
  \otimes \tilde \rho}$. Since $\K{\Psi}_{\rA\rB} \K{\Psi}_{\rA'\rB'}$
is a purification of $\K{\rho}_{\rA}\K{\tilde \rho}_{\rB'}$, using
corollary \ref{cor:remote-prep} we get the thesis. The effect $E_\Psi$
can be taken to be atomic: indeed, if it were refinable, i.e.
$E_{\Psi} = \sum_i E_i $, since the right hand side of Eq.
(\ref{ent-swap}) is a pure state, each effect $E_i$ would achieve
entanglement swapping.  \qed

\medskip

{\bf Remark (PR boxes are excluded by the purification Postulate).}
The possibility of probabilistic entanglement swapping shows that the
purification Postulate excludes the theory of Popescu-Rohrlich boxes (see Ref.
\cite{info-processingBarrett} for the definition of transformations on boxes and
states of multipartite boxes). Indeed, Refs. \cite{swapping,tele} showed that
probabilistic entanglement swapping is impossible in this theory.

\medskip

\begin{corollary}[Probabilistic teleportation]\label{cor:probtele} 
  Let $\Psi \in \Stset_1 (\rA \rB)$ be a pure state, and let $\rho \in \Stset_1(\rA)$ and $\tilde \rho \in \Stset_1(\rB)$ be its marginals. Let $\rA'$ and
  $\rB'$ be operationally equivalent to $\rA$ and $\rB$, respectively. Then, there exists an atomic effect $E_{\Psi} \in \Cntset(\rB \rA')$ and a non-zero probability $p_\Psi$ such that
\begin{equation}\label{telerho}
\begin{aligned}  
\Qcircuit @C=1em @R=.7em @! R {
\multiprepareC{1}{\Psi} &\qw \poloFantasmaCn{\rA} &\qw \\
\pureghost{\Psi} &\qw \poloFantasmaCn{\rB} & \multimeasureD{1}{E_\Psi} \\
   &  \qw  \poloFantasmaCn{\rA'}  & \ghost{E_\Psi}  }
\end{aligned}
 ~=_\rho ~ p_{\Psi} ~ 
\begin{aligned}
\Qcircuit @C=1em @R=.7em @! R { &\qw  \poloFantasmaCn{\rA'}  &  \gate{ \tI} &\qw \poloFantasmaCn{\rA}&\qw }
\end{aligned}
\end{equation}       
and 
\begin{equation}\label{telerhotilde}
 \begin{aligned} \Qcircuit @C=1em @R=.7em @! R {
     &\qw \poloFantasmaCn{\rB} & \multimeasureD{1}{E_\Psi} \\
    \multiprepareC{1} {\Psi} &\qw  \poloFantasmaCn{\rA'} &\ghost{E_\Psi} \\
 \pureghost{\Psi} &\qw \poloFantasmaCn{\rB'} &\qw }
\end{aligned}
~=_{\tilde \rho} ~ p_{\Psi} ~
\begin{aligned} \Qcircuit @C=1em @R=.7em @! R { & \poloFantasmaCn{\rB} \qw & \gate{ \tI} &\qw \poloFantasmaCn{\rB'}&\qw }
\end{aligned}
\end{equation}
In particular, if $\rho$ is an internal state, one has the probabilistic teleportation scheme       
\begin{equation}\label{probtele}
  \begin{aligned}
\Qcircuit @C=1em @R=.7em @! R { \multiprepareC{1}{\Psi} &\qw \poloFantasmaCn{\rA} &\qw \\
    \pureghost{\Psi} &\qw \poloFantasmaCn{\rB} & \multimeasureD{1}{E_\Psi} \\
      &  \qw  \poloFantasmaCn{\rA'}  &\ghost{E_\Psi}}
\end{aligned}
 ~= ~ p_{\Psi} ~ 
\begin{aligned}
\Qcircuit @C=1em @R=.7em @! R { & \poloFantasmaCn{\rA'} \qw &  \gate{ \tI} &\qw \poloFantasmaCn{\rA}&\qw }
\end{aligned}
\end{equation}       
\end{corollary}

\Proof Just combine Theorems \ref{theo:entswap} and \ref{theo:uponinput}. \qed

The diagram of probabilistic teleportation (\ref{probtele}) is one of the
main axioms in the categorial approach by Abramsky and Coecke
\cite{abracoecke}.   In the present approach, this property is derived from the purification postulate,
rather than being assumed from the start.

For theories with local discriminability the probability of
teleportation is related to the dimension of the state space as
follows:
\begin{lemma}[Maximum teleportation probability]
 If local discriminability holds, then
  the probability of teleportation  $p_{\Psi}$ in Eq. (\ref{probtele}) satisfies
  the bound
\begin{equation}\label{maxprob}
p_\Psi \le \frac 1 {\dim \Stset_\Reals  (\rA)}~.
\end{equation}
\end{lemma}

\Proof Let us choose two bases $\{\rho_i\}$ and $\{\tilde \rho_j\}$
for the vector spaces $\Stset_\Reals  (\rA)$ and $\Stset_{\mathbb
  R} (\tilde \rA)$, respectively, and write $\Psi$ as $\K{\Psi}_{\rA \tilde \rA} =
\sum_{i,j} A_{ij} \K{\rho_i}_\rA \K{\tilde \rho_j}_{\tilde \rA}$. Now take
the dual bases $\{\rho^*_i\}$ and $\{\tilde \rho^*_j\}$ for the dual
vector spaces $\Cntset_\Reals  (\rA)$ and $\Cntset_\Reals 
(\tilde \rA)$, respectively---so that $\SC{\rho_i^*} {\rho_j}_{\rA} =
\delta_{ij}$ and $\SC{\tilde \rho_k^*} {\tilde \rho_l}_{\tilde \rA} =
\delta_{kl}$---, and write $E_\Psi$ as $\B{E_\Psi}_{\tilde \rA \rA'}= \sum_{k,l} B_{kl}
 \B{\tilde \rho^*_k}_{\tilde \rA} \B{\rho^*_l}_{\rA'}$.  The teleportation diagram
(\ref{probtele}) is then equivalent to the matrix equality 
\begin{equation}
A B = p_{\Psi}  I_{\rA},
\end{equation}
where $I_{\rA}$ is the identity matrix of size $\dim(\Stset_\Reals  (\rA))$. 
Finally, since probabilities are bounded by unit, we obtain
\begin{equation}
1 \ge \SC{E_{\Psi} } {\Psi}_{\rA \tilde \rA} = \Tr [A B] = p_{\Psi}  \dim(\Stset_\Reals  (\rA)),
\end{equation}
which is the desired bound. \qed 

\medskip

{\bf Remark (quantum theory achieves the bound)}. Note that in
quantum theory the teleportation probability achieves the maximum
value allowed by the bound of Eq. (\ref{maxprob}): For a
$d$-dimensional Hilbert space, the real vector space spanned by all
density matrices has dimension $d^2$, which is exactly the maximum
probability of conclusive teleportation.

\medskip

A simple consequence of probabilistic teleportation is the possibility of remotely preparing any bipartite state by acting locally on the purifying system only, according to the following definition
\begin{definition}[Preparationally faithful state] A state $\Psi \in \Stset_1 (\rA \rB)$ is \emph{preparationally faithful} for system $\rB$ if for every bipartite state $\sigma \in \Stset_1 (\rA \rC)$ there are a transformation $\tA_{\sigma} \in \Trnset (\rB, \rC)$  and a non-zero probability $p_{\sigma}$ such that 
\begin{equation}
  p_{\sigma} \begin{aligned}\Qcircuit @C=1em @R=.7em @! R { \multiprepareC{1}{\sigma} & \qw \poloFantasmaCn \rA &\qw \\
    \pureghost{\sigma} & \qw \poloFantasmaCn \rC &\qw} 
\end{aligned} ~=~ \begin{aligned}
  \Qcircuit @C=1em @R=.7em @! R { \multiprepareC{1}{\Psi} & \qw \poloFantasmaCn \rA &\qw &\qw &\qw\\
    \pureghost{\Phi} & \qw \poloFantasmaCn \rB &  \gate {\tA_\sigma} &\qw \poloFantasmaCn \rC &\qw}
\end{aligned}
\end{equation}
\end{definition}
\begin{corollary}{\bf (Existence of preparationally faithful pure states)}  Let $\Psi \in
  \Stset_1 (\rA \rB)$ be the purification of an internal state $\omega \in
  \Stset_1 (\rA)$. Then, $\Psi$ is preparationally faithful for system $\rB$.   
\end{corollary}

\Proof Let $E_\Psi$ be the teleportation effect for $\Psi$, as defined in Corollary \ref{cor:probtele}. Define the transformation $\tA_{\sigma}$ as
\begin{equation}
 \begin{aligned} \Qcircuit @C=1em @R=.7em @! R { 
    &\qw \poloFantasmaCn \rB&\gate{\tA_\sigma} &\qw \poloFantasmaCn \rC &\qw } \end{aligned} ~:= ~ 
\begin{aligned}  \Qcircuit @C=1em @R=.7em @! R { & &\qw \poloFantasmaCn \rB & \multimeasureD{1} {E_{\Psi}}  \\
 & \multiprepareC{1} \sigma & \qw \poloFantasmaCn \rA & \ghost{E_{\Psi}} \\
& \pureghost \sigma & \qw  \poloFantasmaCn \rC &\qw}
\end{aligned}
\end{equation}
Applying $\tA_\sigma$ to $\Psi$ and using Eq. (\ref{probtele}) with $\rA ' \equiv \rA$ we then obtain
\begin{equation}
  \begin{split} \begin{aligned} \Qcircuit @C=1em @R=.7em @! R { \multiprepareC{1} \Psi & \qw\poloFantasmaCn \rA &\qw&\qw &\qw \\ 
      \pureghost{\Psi} &\qw \poloFantasmaCn \rB&\gate{\tA_\sigma} &\qw \poloFantasmaCn \rC &\qw }\end{aligned} &~ = ~ 
\begin{aligned}    \Qcircuit @C=1em @R=.7em @! R { \multiprepareC{1} \Psi & \qw \poloFantasmaCn \rA &\qw \\
      \pureghost{\Psi} &\qw \poloFantasmaCn \rB & \multimeasureD{1} {E_{\Psi}}  \\
      \multiprepareC{1} \sigma & \qw \poloFantasmaCn \rA & \ghost{E_{\Psi}} \\
      \pureghost \sigma & \qw  \poloFantasmaCn \rC &\qw } \end{aligned}\\
    &\\
    & ~= ~ p_{\Psi} ~\begin{aligned} 
\Qcircuit @C=1em @R=.7em @! R { \multiprepareC{1}{\sigma} & \qw \poloFantasmaCn \rA &\qw \\
 \pureghost{\sigma} & \qw \poloFantasmaCn \rC &\qw} 
\end{aligned}
\end{split}
\end{equation}
Hence, the thesis holds with $p_\sigma \equiv p_{\Psi}$ independently
of $\sigma$. \qed

\subsection{Storing and probabilistic retrieving of transformations}

Here we consider the task of storing an unknown transformation in the
state of some system. The output state of such a storing protocol  then becomes a ``program'' from which
the transformation can be retrieved at later time.  The task is
achieved probabilistically by a machine that retrieves the
transformation from the program and applies it on a new input state.

\begin{corollary}[Storing and probabilistic retrieving]\label{cor:stor}
Let $\Psi  \in \Stset_1 (\rA \tilde \rA)$ be a pure dynamically faithful state for system $\rA$, according to Def. \ref{def:dynfaith}. The \emph{storing protocol}, consisting in the application of a transformation $\tC \in\Trnset (\rA, \rB)$ to the input state $\Psi$, as in the following diagram
\begin{equation}\label{stor}
\begin{aligned}  \Qcircuit @C=1em @R=.7em @! R {
    \multiprepareC{1}{R_\tC}&\qw\poloFantasmaCn{\rB}&\qw \\
    \pureghost{R_\tC}&\qw\poloFantasmaCn{\tilde \rA}&\qw }
\end{aligned}
  \quad:=\quad
 \begin{aligned}
\Qcircuit @C=1em @R=.7em @! R {
    \multiprepareC{1}{\Psi}&\qw\poloFantasmaCn{\rA}&\gate{\tC} &\qw\poloFantasmaCn{ \rB}&\qw \\
    \pureghost{\Psi}&\qw\poloFantasmaCn{\tilde \rA}&\qw &\qw &\qw}
\end{aligned}
\end{equation}
 defines an injective map from transformations  $\tC \in \Trnset (\rA, \rB)$ to
  bipartite states  $R \in \Stset (\rB \tilde \rA)$ satisfying the property
  \begin{equation}\label{dominance}
\B{e}_{\rB} \K{R}_{\rB \tilde \rA} \in D_{\tilde \omega},
\end{equation}  
where $\tilde \omega$ is the marginal state $\K{\tilde \omega}_{\tilde \rA} =
\B{e}_{\rA} \K{\Psi}_{\rA\tilde \rA}$.  The inverse map is given by
the \emph{probabilistic retrieving protocol}
\begin{equation}\label{retr}
p_\Psi~\begin{aligned} \Qcircuit @C=1em @R=.7em @! R {
 & \qw \poloFantasmaCn{\rA}  &  \gate{ \tC} &\qw \poloFantasmaCn{\rB}&\qw } \end{aligned} 
~=~ 
\begin{aligned}\Qcircuit @C=1em @R=.7em @! R { 
\multiprepareC{1}{R_{\tC}} &\qw \poloFantasmaCn{\rB} &\qw \\
\pureghost{R_{\tC}} &\qw \poloFantasmaCn{\tilde \rA} & \multimeasureD{1}{E_\Psi} \\
 &  \qw  \poloFantasmaCn{\rA}  & \ghost {E_\Psi} }
\end{aligned}
 \end{equation}
where $E_\Psi$ is the teleportation effect for state $\Psi$ and $p_\Psi$ is the corresponding teleportation probability, as defined in Eq. (\ref{probtele}). 
\end{corollary}
\Proof Since the state $\Psi$ is dynamically faithful, the map $\tC
\mapsto R_\tC$ is injective. Now, any transformation $\tC$ is part of
a test $\{\tC_i\}_{i \in \rX}$, namely one has $\tC = \tC_{i_0}$ for some outcome $i_0$. Defining the coarse-grained channel $\tC_{\rX} := \sum_{i\in \rX}
\tC_i$ we have 
\begin{equation}
\begin{split}
\sum_{i\in \rX}  \B e_\rB  \K {R_{\tC_i}}_{\rB \tilde \rA}  &=\sum_{i\in \rX} \B e_\rB  \tC_i \K \Psi_{\rA\tilde \rA}\\
& = \B e_\rB  \tC_\rX  \K \Psi_{\rA\tilde \rA}\\
& = \B e_\rA \K \Psi_{\rA\tilde\rA}\\
&= \K {\tilde \omega}_{\tilde \rA}.
\end{split}
\end{equation}
having use the normalization condition $\B e_\rB \tC_\rX = \B e_\rA$.  
This implies $\B e_\rB \K {R_\tC}_{\rB \tilde \rA}$ is in the refinement set of $\tilde \omega$, thus proving Eq.
(\ref{dominance}).  The identity (\ref{retr}) simply follows by writing
$p_\Psi \tC = \tC \circ (p_\Psi \tI_{\rA})$ and substituting $p_\Psi \tI_\rA$ as in Eq. (\ref{probtele}). \qed

\medskip

In Section \ref{sec:dilation} we will show that the correspondence
$\tC \mapsto R_\tC$ is also surjective on the set of bipartite states
satisfying Eq. (\ref{dominance}). This will provide an isomorphism
between transformations and bipartite states that enjoys all the structural
properties of the Choi-Jamio\l kowski isomorphism of quantum theory
\cite{nota:choijam}.
  
The probabilistic retrieving of Eq. (\ref{retr}) implies a bound on the operational distance between two transformations $\tA_0, \tA_1$ in terms the operational distance between the corresponding states: 
\begin{theorem}\label{theo:bounddist}
Let $\tA_0, \tA_1 \in \Trnset ( \rA, \rB)$ be two transformations and $R_{\tA_0}, R_{\tA_1} \in \Stset (\rB \tilde \rA)$ be the corresponding states as in Eq. (\ref{stor}).  Then one has the bound
\begin{equation}\label{bounddist}
|\!|  \tA_1 - \tA_0 |\!|_{\rA, \rB} \le \frac {|\!|  R_{\tA_1} - R_{\tA_0} |\!|_{\rB \tilde \rA}} {p_\Psi},
\end{equation}
where $p_\Psi$ is the probability of retrieving a transformation from the corresponding state, as defined in Eq. (\ref{retr}).
\end{theorem}
\Proof  Define $\Delta := \tA_1 - \tA_0$ and $R_\Delta := R_{\tA_1} - R_{\tA_0}$. Take  an ancillary system $\rC$ and a state $\rho \in \Stset_1 (\rA\rC)$. Then Eq. (\ref{retr})  implies
\begin{equation}
 \begin{aligned}
 \Qcircuit @C=1em @R=.7em @! R { \multiprepareC{1} { \rho} &\qw \poloFantasmaCn{\rA} & \gate \Delta & \qw \poloFantasmaCn \rB &\qw\\
    \pureghost{\rho} &\qw \poloFantasmaCn {\rC} &\qw& \qw &\qw } 
\end{aligned} 
~= ~ \frac 1 {p_{\Psi}}~ 
\begin{aligned}
  \Qcircuit @C=1em @R=.7em @! R {
    \multiprepareC{1}{R_\Delta} &\qw \poloFantasmaCn{\rB} &\qw \\
    \pureghost{R_\Delta} &\qw \poloFantasmaCn{\tilde \rA} & \multimeasureD{1}{E_{\Psi}} \\
    \multiprepareC{1}{\rho} &  \qw  \poloFantasmaCn{\rA}  & \ghost{E_{\Psi}}\\
    \pureghost{\rho} &\qw \poloFantasmaCn{\rC} & \qw }
\end{aligned}
\end{equation}      
Applying a bipartite effect $\B a_{\rB \rC}$ on both sides we then obtain 
\begin{equation}
\B a_{\rB \rC} \Delta \K \rho_{\rA\rC} = \frac{\SC {b_\rho} {R_{\Delta}}_{\rB \tilde \rA}}{p_\Psi},
\end{equation} 
where $\B {b_\rho}_{\rB \tilde \rA}:=   \left[ \B a_{\rB \rC} \otimes \B{E_{\Psi}}_{\tilde \rA \rA}\right] \K{\rho}_{\rA\rC}$. Since $b_\rho$ is an effect, the above equality implies the bound 
\begin{equation}
\frac{\inf_{b}  \SC {b} {R_{\Delta}}_{\rB \tilde \rA}} {p_\Psi} \le \B a_{\rB \rC} \Delta \K \rho_{\rA\rC} \le  \frac{\sup_{b} \SC {b} {R_{\Delta}}_{\rB \tilde \rA}}{p_\Psi}.
\end{equation} 
By the definition of operational norm in Eq. \ref{opnormtrans}, this implies 
\begin{equation} 
|\!|  \Delta  \rho |\!|_{\rB\rC}  \le  |\!| R_{\Delta}  |\!|_{\rB \tilde \rA}/ p_\Psi.
\end{equation} 
Finally, taking the supremum over the ancillary system $\rC$ we get
the desired bound. \qed
\subsection{Systems of purifications and the link product}\label{subsect:link}

For every system $\rA$  we now fix a dynamically faithful pure state $\K{\Psi^{(\rA)}}_{\rA\tilde \rA}$, where $\tilde \rA$ is some suitable purifying system.  According to the characterization of dynamically faithful pure states given in Theorem \ref{theo:chardynfaith}, the marginal state $\K \omega_\rA := \B e_{\tilde \rA} \K {\Psi^{(\rA)}}_{\rA\tilde\rA}$ must be internal.   The role of the upper index in $\Psi^{(\rA)}$ is precisely to recall that
the marginal is internal for system $\rA$, while it may not be
internal for the purifying system $\tilde \rA$.  Moreover, we denote
by $\B{E^{(\rA)}}_{\tilde \rA \rA }$ and by $p_\rA$ the effect and the
probability appearing in the teleportation scheme (\ref{probtele}),
respectively.

Note that since the product of dynamically faithful pure states is dynamically faithful  (Corollary \ref{cor:dynfaithprod}), for
bipartite systems $\rA\rB$ we can choose $\K{\Psi^{(\rA\rB)}}_{\rA \rB
  \widetilde{\rA \rB}}:= \K{\Psi^{(\rA)}}_{\rA \tilde \rA} \K{\Psi^{(\rB)}}_{\rB
  \tilde \rB}$. Likewise, we can choose $\B{E^{(\rA \rB)}}_{\widetilde {\rA  \rB} \rA \rB} =
\B{E^{(\rA)}}_{\tilde \rA \rA} \B{E^{(\rB)}}_{\tilde \rB \rB}$, and
$p_{\rA \rB} = p_\rA p_\rB$. We call a \emph{system of purifications}
such a choice of bipartite states and effects:

\begin{definition}[System of purifications]\label{def:syspur} A \emph{system of purification} is a choice of a dynamically faithful pure states $\K{\Psi^{(\rA)}}_{\rA \tilde \rA}$ and  teleportation effects $\B {E^{(\rA)}}_{\tilde \rA\rA}$ that satisfies the properties
\begin{equation}
\begin{split}
\K{\Psi^{(\rA\rB)}}_{\rA \rB \widetilde{\rA \rB}} & = \K {\Psi^{(\rA)}}_{\rA \tilde \rA} \K{\Psi^{(\rB)}}_{\rB \tilde \rB}\\
\B{E^{(\rA \rB)}}_{\widetilde {\rA  \rB} \rA \rB} &=
\B{E^{(\rA)}}_{\tilde \rA \rA} \B{E^{(\rB)}}_{\tilde \rB \rB}.
\end{split}
\end{equation}
\end{definition}

Once a system of purifications has been fixed, one can discuss the composition of transformations in terms of composition of states, generalizing the definitions and the results introduced by Refs. \cite{qca,comblong} in the quantum setting. 

\begin{definition}[Link product]  The \emph{link product} of two vectors $ \rho \in \Stset_\Reals (\rB \tilde \rA)$ and $\sigma \in \Stset_\Reals (\rC \tilde \rB)$ is the vector $\rho * \sigma \in \Stset_\Reals (\rC \tilde \rA)$  given by
\begin{equation}
\begin{aligned}
\Qcircuit @C=1em @R=.7em @! R { \multiprepareC{1} {\rho * \sigma} &\qw \poloFantasmaCn{\rC} &\qw\\
    \pureghost{\rho * \sigma} &\qw \poloFantasmaCn {\tilde \rA}&\qw}
\end{aligned} 
 ~:=~ \frac 1 {p_{\rB}}~
\begin{aligned}
 \Qcircuit @C=1em @R=.7em @! R { \multiprepareC{1}{\sigma} &\qw \poloFantasmaCn{\rC} &\qw \\
    \pureghost{\sigma} &\qw \poloFantasmaCn{\tilde \rB} & \multimeasureD{1}{E^{(\rB)}} \\
    \multiprepareC{1}{\rho} &  \qw  \poloFantasmaCn{\rB}  & \ghost{E^{(\rB)}}\\
    \pureghost{\rho} &\qw \poloFantasmaCn{\tilde \rA} & \qw  & } 
\end{aligned}
 \end{equation}
\end{definition}
Note that if $\rho$ and $\sigma$ are proportional to states, then also $\rho * \sigma$ is proportional to a state:  one has $\rho *\sigma \in \Stset_+ (\rC\tilde\rA)$ for any couple $\rho\in\Stset_+(\rB \tilde \rA),  \sigma\in\Stset_+ (\rC\tilde\rB)$.

 
The product and composition of transformations are then given by the following
\begin{corollary}[Composition of states]
Consider the correspondence given by the storing protocol in Eq. (\ref{stor}). For two transformations $\tC \in \Trnset (\rA, \rB)$ and $\tD \in \Trnset (\rC , \rD)$ one has
\begin{equation}
\begin{aligned}  
\Qcircuit @C=1em @R=.7em @! R { \multiprepareC{1} {R_{\tC\otimes \tD}} &\qw \poloFantasmaCn{\rB\rC} &\qw\\
\pureghost{R_{\tD\otimes\tC}} &\qw \poloFantasmaCn {\tilde \rA \tilde \rC}&\qw}
\end{aligned}
 ~=~
\begin{aligned} \Qcircuit @C=1em @R=.7em @! R { \multiprepareC{1}{R_\tD} &\qw \poloFantasmaCn{\rD} &\qw \\
    \pureghost{R_\tD} &\qw \poloFantasmaCn{\tilde \rC} & \qw \\
    \multiprepareC{1}{R_\tC} &  \qw  \poloFantasmaCn{\rB}  &\qw  \\
    \pureghost{R_\tC} &\qw \poloFantasmaCn{\tilde \rA} & \qw }
\end{aligned}
\end{equation}
For two transformations $\tC \in \Trnset (\rA , \rB)$ and $\tD\in \Trnset (\rB, \rC)$ one has 
\begin{equation}\label{linkprod}
\begin{split}
\begin{aligned}
  \Qcircuit @C=1em @R=.7em @! R { \multiprepareC{1} {R_{\tD \circ \tC}} &\qw \poloFantasmaCn{\rC} &\qw\\
    \pureghost{R_{\tD\circ\tC}} &\qw \poloFantasmaCn {\tilde \rA}&\qw}
\end{aligned} & = ~\frac 1 {p_{\rB}}~
\begin{aligned} 
\Qcircuit @C=1em @R=.7em @! R { \multiprepareC{1}{R_\tD} &\qw \poloFantasmaCn{\rC} &\qw \\
    \pureghost{R_\tD} &\qw \poloFantasmaCn{\tilde \rB} & \multimeasureD{1}{E^{(\rB)}} \\
    \multiprepareC{1}{R_\tC} &  \qw  \poloFantasmaCn{\rB}  &  \ghost{E^{(\rB)}}\\
    \pureghost{R_\tC} &\qw \poloFantasmaCn{\tilde \rA} & \qw }
\end{aligned}\\
&\\
& =~
\begin{aligned}
 \Qcircuit @C=1em @R=.7em @! R { \multiprepareC{1} {R_\tC * R_\tD} &\qw \poloFantasmaCn{\rC} &\qw\\
    \pureghost{R_\tC * R_\tD} &\qw \poloFantasmaCn {\tilde \rA}&\qw}
\end{aligned}
\end{split}
 \end{equation}
\end{corollary}
\Proof The first equation follows from the fact that
$\K{\Psi^{(\rA\rC)}}_{\rA \rC \widetilde{\rA \rC}}:= \K{\Psi^{(\rA)}}_{\rA \tilde \rA}
\K{\Psi^{(\rC)}}_{\rC \tilde \rC}$, while the second follows from the probabilistic retrieving of Eq. (\ref{retr}):\begin{equation}
\begin{split}
~\frac 1 {p_{\rB}}~
\begin{aligned} 
\Qcircuit @C=1em @R=.7em @! R { \multiprepareC{1}{R_\tD} &\qw \poloFantasmaCn{\rC} &\qw \\
    \pureghost{R_\tD} &\qw \poloFantasmaCn{\tilde \rB} & \multimeasureD{1}{E^{(\rB)}} \\
    \multiprepareC{1}{R_\tC} &  \qw  \poloFantasmaCn{\rB}  &  \ghost{E^{(\rB)}}\\
    \pureghost{R_\tC} &\qw \poloFantasmaCn{\tilde \rA} & \qw }
\end{aligned}& = 
\begin{aligned} 
\Qcircuit @C=1em @R=.7em @! R { 
    \multiprepareC{1}{R_\tC} &  \qw  \poloFantasmaCn{\rB}  &  \gate\tD &\qw \poloFantasmaCn{\rC} &\qw   \\
    \pureghost{R_\tC} &\qw \poloFantasmaCn{\tilde \rA} & \qw &\qw &\qw}
\end{aligned}\\
& = 
\begin{aligned} 
\Qcircuit @C=1em @R=.7em @! R { 
    \multiprepareC{1}{\Psi^{(\rA)}} &  \qw  \poloFantasmaCn{\rA}  &\gate \tC & \qw \poloFantasmaCn\rB&  \gate\tD &\qw \poloFantasmaCn{\rC} &\qw   \\
    \pureghost{\Psi^{(\rA)}} &\qw \poloFantasmaCn{\tilde \rA} & \qw &\qw &\qw&\qw&\qw}
\end{aligned}\\
& = 
\begin{aligned} 
\Qcircuit @C=1em @R=.7em @! R { 
    \multiprepareC{1}{R_{\tD \circ \tC}} &  \qw  \poloFantasmaCn{\rC}  &\qw   \\
    \pureghost{R_{\tD \circ \tC}} &\qw \poloFantasmaCn{\tilde \rA} & \qw }
\end{aligned}
\end{split}
\end{equation} 
\qed

\section{Dilation of physical processes}\label{sec:dilation}

In this Section we derive dilation Theorems for channels,
observation-test, and general tests. These Theorems extend to all
theories with purification the validity of the theorems by Stinespring
\cite{stine}, Naimark \cite{naimark}, and Ozawa \cite{ozawa},
originally obtained in the setting of operator algebras.

\subsection{Reversible dilation of channels}

In order to derive the reversible dilation of a channel we need the following lemma:

\begin{lemma}\label{lem:fordilation}
Let $R \in \Stset_1 (\rB \tilde \rA)$ be a state such that
\begin{equation}\label{chaoticmarginal}
\begin{aligned}  \Qcircuit @C=1em @R=.7em @! R {\multiprepareC{1}{R} & \qw \poloFantasmaCn{\rB} &\measureD{e}\\
\pureghost{R}& \qw \poloFantasmaCn{\tilde \rA} &\qw}
\end{aligned} ~=~ \begin{aligned} \Qcircuit @C=1em @R=.7em @! R {\multiprepareC{1}{\Psi^{(\rA)}} & \qw \poloFantasmaCn{\rA} &\measureD{e}\\
\pureghost{\Psi^{(\rA)}}& \qw \poloFantasmaCn{\tilde \rA} &\qw}
\end{aligned}
\end{equation}
where $\Psi^{(\rA)}$ is a pure dynamically faithful state for system $\rA$. 
Then there exist a system $\rC$, a pure state $\varphi_0 \in \Stset_1 (\rB \rC)$, and a reversible channel $\tU \in \Trnset (\rA \rB  \rC)$ such that
\begin{equation}\label{eq:detchoi}
 \begin{aligned}
 \Qcircuit @C=1em @R=.7em @! R {\multiprepareC{1}{R} & \qw \poloFantasmaCn{\rB} &\qw  \\
    \pureghost{R}& \qw \poloFantasmaCn{\tilde \rA} &\qw}
\end{aligned}
~=~ 
\begin{aligned} 
  \Qcircuit @C=1em @R=.7em @! R {\prepareC{\varphi_0}&\qw&\qw\poloFantasmaCn{\rB  \rC}&\qw &
    \multigate{1}{\tU}&\qw&\qw\poloFantasmaCn{\rm A \rC}&\measureD{e} \\
    \multiprepareC{1}{\Psi^{(\rA)}}&\qw &\qw \poloFantasmaCn{\rA}&\qw&\ghost{\tU} &
    \qw&\qw\poloFantasmaCn{\rB}&\qw \\
    \pureghost{\Psi^{(\rA)}}&\qw& \qw& \qw& \qw& \qw&\qw\poloFantasmaCn{\tilde \rA}&\qw}
\end{aligned}
\end{equation} 
Moreover, the channel $\tV \in \Trnset (\rA, \rA \rB \rC)$ defined by
$\tV := \tU \K{\varphi_0}_{\rB \rC}$ is unique up to reversible channels
on $\rA \rC$.
\end{lemma}
\Proof Take a purification of $R$, say $\Psi_R \in \Stset_1 ( \rC \rB
\tilde \rA )$ for some purifying system $\rC$.  One has
\begin{equation}
  \begin{aligned}  \Qcircuit @C=1em @R=.7em @! R {\multiprepareC{2}{\Psi_R} & \qw \poloFantasmaCn{\rC} &\measureD e\\
      \pureghost{\Psi_R} &  \qw \poloFantasmaCn{\rB} &\measureD e\\
      \pureghost{\Psi_R}& \qw \poloFantasmaCn{\tilde \rA} &\qw}
\end{aligned} ~=~
\begin{aligned}  \Qcircuit @C=1em @R=.7em @! R {\multiprepareC{1}{R} & \qw \poloFantasmaCn{\rB} &\measureD e\\
    \pureghost{R}& \qw \poloFantasmaCn{\tilde \rA} &\qw}
\end{aligned} 
~=~
\begin{aligned} \Qcircuit @C=1em @R=.7em @! R {\multiprepareC{1}{\Psi^{(\rA)}} & \qw \poloFantasmaCn{\rA} &\measureD{e}\\
    \pureghost{\Psi^{(\rA)}}& \qw \poloFantasmaCn{\tilde \rA} &\qw}
\end{aligned}
\end{equation}
that is, the pure states $\Psi_R$ and $\Psi^{(\rA)}$ have the same marginal on system $\tilde \rA$.  Applying the uniqueness of purification as expressed by Lemma \ref{lem:purichan} one then obtains
\begin{equation}
  \begin{aligned}  \Qcircuit @C=1em @R=.7em @! R {\multiprepareC{2}{\Psi_R} & \qw \poloFantasmaCn{\rC} &\qw\\
      \pureghost{\Psi_R} &  \qw \poloFantasmaCn{\rB} &\qw\\
      \pureghost{\Psi_R}& \qw \poloFantasmaCn{\tilde \rA} &\qw}
  \end{aligned}
  ~=~ 
  \begin{aligned}  \Qcircuit @C=1em @R=.7em @! R {\prepareC{\varphi_0}&\qw&\qw\poloFantasmaCn{\rB  \rC}&\qw &
      \multigate{1}{\tU}&\qw&\qw\poloFantasmaCn{\rm A }&\measureD e \\
      \multiprepareC{1}{\Psi^{(\rA)}}&\qw &\qw \poloFantasmaCn{\rA}&\qw&\ghost{\tU} &
      \qw&\qw\poloFantasmaCn{\rB\rC}&\qw \\
      \pureghost{\Psi^{(\rA)}}&\qw& \qw& \qw& \qw& \qw&\qw\poloFantasmaCn{\tilde \rA}&\qw}
  \end{aligned}
\end{equation}
Applying the deterministic effect on system $\rC$ on both sides, one then proves Eq.
(\ref{eq:detchoi}). Moreover, if $\tV' := \tU' \K{\varphi_0'}_{\rB
  \rC}$ is channel such that Eq. (\ref{eq:detchoi}) holds, then the pure
states $\tV \K{\Psi^{(\rA)}}_{\rA \tilde \rA}$ and $\tV'
\K{\Psi^{(\rA)}}_{\rA \tilde \rA}$ have the same marginal on system $\rB \tilde \rA$. Uniqueness of purification then implies
\begin{equation}
\begin{aligned}  \Qcircuit @C=1em @R=.7em @! R {& & \pureghost{\tV'} &\qw \poloFantasmaCn {\rA \rC}&\qw\\ 
    \multiprepareC{1}{\Psi^{(\rA)}} & \qw \poloFantasmaCn{\rA} &\multigate{-1}{\tV'} &\qw \poloFantasmaCn {\rB} &\qw \\
    \pureghost{\Psi^{(\rA)}} & \qw & \qw&\qw \poloFantasmaCn {\tilde \rA}&\qw }
\end{aligned}
~=~ 
\begin{aligned}
  \Qcircuit @C=1em @R=.7em @! R {& & \pureghost{\tV} &\qw \poloFantasmaCn {\rA \rC}&\gate{\tW}&\qw \poloFantasmaCn{\rA \rC}&\qw\\ 
    \multiprepareC{1}{\Psi^{(\rA)}} & \qw \poloFantasmaCn{\rA} &\multigate{-1}{\tV} &\qw \poloFantasmaCn {\rB} &\qw&\qw&\qw \\
    \pureghost{\Psi^{(\rA)}} & \qw & \qw&\qw \poloFantasmaCn {\tilde \rA}&\qw &\qw&\qw}
\end{aligned}
\end{equation} 
for some reversible channel $\tW \in \Trnset(\rA \rC)$. Since $\Psi^{(\rA)}$ is dynamically faithful for $\rA$, this implies $\tV' = \tW \tV$. \qed
\medskip 

We now give the definitions of dilation, environment, and reversible dilation:
\begin{definition}[Dilation of a channel]
A \emph{dilation} of channel $\tC \in \Trnset (\rA, \rB)$ is a channel $\tV \in \Trnset(\rA, \rB\rE)$ such that  
\begin{equation}
\begin{aligned} \Qcircuit @C=1em @R=.7em @! R { 
&\qw  \poloFantasmaCn{\rA}&\gate{\tC}&\qw\poloFantasmaCn{\rB}&\qw}
\end{aligned} ~=~
\begin{aligned} \Qcircuit @C=1em @R=.7em @! R {   & & \pureghost{\tV} & \qw\poloFantasmaCn{\rE}&\measureD e \\ 
 &\qw\poloFantasmaCn{\rA}& \multigate{-1}{\tV}&\qw\poloFantasmaCn{\rB}&\qw }
\end{aligned}
\end{equation}  
We refer to system $\rE$ as to the \emph{environment}.
\end{definition}
\begin{definition}[Reversible dilation] A dilation $\tV \in \Trnset
  (\rA, \rB\rE)$ is called \emph{reversible} if there exists a system
  $\rE_0$ such that $\rA \rE_0 \simeq \rB \rE$ and
\begin{equation}
\begin{aligned}  \Qcircuit @C=1em @R=.7em @! R {
    & & \pureghost{\tV} & \qw\poloFantasmaCn{\rE}&\qw \\ 
    &\qw\poloFantasmaCn{\rA}& \multigate{-1}{\tV}&\qw\poloFantasmaCn{\rB}&\qw  }   
  \end{aligned} ~=~ \begin{aligned} \Qcircuit @C=1em @R=.7em @! R {\prepareC{\varphi_0}&\qw\poloFantasmaCn{\rE_0}& \multigate{1}{\tU}&\qw\poloFantasmaCn{\rE}&\qw  \\
    &\qw \poloFantasmaCn{\rA}& \ghost{\tU} & \qw\poloFantasmaCn{\rB}&\qw } \end{aligned}
\end{equation}
for some pure state $\varphi_0\in \Stset_1(\rE_0 )$ and some reversible
channel $\tU\in\Trnset(\rA \rE_0,\rB \rE )$.
\end{definition}

According to the above definitions, we have the following dilation theorem:
 \begin{theorem}[Reversible dilation of channels]\label{theo:stine} Every channel
   $\tC\in\Trnset(\rm{A,B})$ has a reversible dilation $\tV \in
   \Trnset (\rA, \rB\rE)$.  If $\tV, \tV'\in \Trnset(\rA, \rB\rE)$ are two reversible  dilations of the same channel, then they are connected by a reversible transformation on the environment,
  namely
\begin{align}
\begin{aligned} \Qcircuit @C=1em @R=.7em @! R {
    & & \pureghost{\tV'} & \qw\poloFantasmaCn{\rE}&\measureD e \\ 
    &\qw\poloFantasmaCn{\rA}& \multigate{-1}{\tV'}&\qw\poloFantasmaCn{\rB}&\qw  }\end{aligned}
&~=~
\begin{aligned} \Qcircuit @C=1em @R=.7em @! R {
    & & \pureghost{\tV} & \qw\poloFantasmaCn{\rE}&\measureD e \\ 
    &\qw\poloFantasmaCn{\rA}& \multigate{-1}{\tV}&\qw\poloFantasmaCn{\rB}&\qw  }\end{aligned} 
\nonumber\\
  ~\Longrightarrow~ 
\begin{aligned} \Qcircuit @C=1em @R=.7em @! R {
    & & \pureghost{\tV'} & \qw\poloFantasmaCn{\rE}&\qw \\ 
    &\qw\poloFantasmaCn{\rA}& \multigate{-1}{\tV'}&\qw\poloFantasmaCn{\rB}&\qw  } \end{aligned}
&~=~
\begin{aligned} \Qcircuit @C=1em @R=.7em @! R {
    & & \pureghost{\tV} & \qw\poloFantasmaCn{\rE}& \gate \tW &\qw\poloFantasmaCn \rE &\qw \\ 
    &\qw\poloFantasmaCn{\rA}& \multigate{-1}{\tV}&\qw\poloFantasmaCn{\rB}&\qw &\qw&\qw  } 
\end{aligned}
\end{align}   
for some reversible channel $\tW\in \grp G_\rE$.
\end{theorem}
\Proof Let us store the channel $\tC$ in the faithful state $\Psi^{(\rA)}\in\Stset_1(\rA \tilde \rA)$, thus getting the state $R_\tC$, as in Eq. (\ref{stor}).  
Since  $\tC$ is a channel, it satisfies the normalization condition  
\begin{equation}\label{eq:Ce}
\begin{aligned}  \Qcircuit @C=1em @R=.7em @! R {
    &\poloFantasmaCn{\rA}\qw&\gate{\tC}&\poloFantasmaCn{\rB}\qw&\measureD{e}}\end{aligned} ~=~ \begin{aligned} \Qcircuit
  @C=1em @R=.7em @! R {
   & \poloFantasmaCn{\rA}\qw&\measureD{e}}
\end{aligned}
\end{equation}
which implies
\begin{equation}
\begin{split}
\begin{aligned} \Qcircuit @C=1em @R=.7em @! R {
\multiprepareC{1}{R_\tC}&\qw\poloFantasmaCn{\rB}&\measureD{e}\\
\pureghost{R_\tC}&\qw\poloFantasmaCn{\tilde \rA}&\qw }\end{aligned} &~=~ 
\begin{aligned} \Qcircuit @C=1em @R=.7em @! R {
\multiprepareC{1}{\Psi^{(\rA)}}&\qw\poloFantasmaCn{\rA}&\gate{\tC} &\qw\poloFantasmaCn{\rB}&\measureD{e} \\
\pureghost{\Psi^{(\rA)}}&\qw\poloFantasmaCn{\tilde \rA}&\qw &\qw &\qw} \end{aligned}\\
& ~=~ \begin{aligned} \Qcircuit @C=1em @R=.7em @! R {
\multiprepareC{1}{\Psi^{(\rA)}}&\qw\poloFantasmaCn{\rA}&\measureD{e}\\
\pureghost{\Psi^{(\rA)}}&\qw\poloFantasmaCn{\tilde \rA}&\qw } \end{aligned}
\end{split}
\end{equation} 
Now, applying Lemma \ref{lem:fordilation} we  obtain
\begin{equation}
  \begin{aligned} \Qcircuit @C=1em @R=.7em @! R {\multiprepareC{1}{R_\tC} & \qw \poloFantasmaCn{\rB} &\qw  \\
    \pureghost{R_\tC}& \qw \poloFantasmaCn{\tilde \rA} &\qw}\end{aligned} ~=~ 
  \begin{aligned}\Qcircuit @C=1em @R=.7em @! R {\prepareC{\varphi_0}&\qw&\qw\poloFantasmaCn{\rB  \rC}&\qw &
    \multigate{1}{\tU}&\qw&\qw\poloFantasmaCn{\rm A \rC}&\measureD{e} \\
    \multiprepareC{1}{\Psi^{(\rA)}}&\qw &\qw \poloFantasmaCn{\rA}&\qw&\ghost{\tU} &
    \qw&\qw\poloFantasmaCn{\rB}&\qw \\
    \pureghost{\Psi^{(\rA)}}&\qw& \qw& \qw& \qw& \qw&\qw\poloFantasmaCn{\tilde \rA}&\qw}
\end{aligned}
\end{equation} 
Since $\Psi^{(\rA)}$ is dynamically faithful for system $\rA$, this implies 
\begin{equation}
\begin{aligned}  \Qcircuit @C=1em @R=.7em @! R { & \qw\poloFantasmaCn{\rA}&\gate \tC & \qw \poloFantasmaCn \rB &\qw}
\end{aligned}   
  ~=~ \begin{aligned}  \Qcircuit @C=1em @R=.7em @! R {\prepareC{\varphi_0}&\qw\poloFantasmaCn{\rB \rC}& \multigate{1}{\tU}&\qw\poloFantasmaCn{\rA\rC}&\measureD e  \\
    &\qw \poloFantasmaCn{\rA}& \ghost{\tU} & \qw\poloFantasmaCn{\rB}&\qw }
\end{aligned}
\end{equation}
Therefore, $\tV:= \tU \K{\varphi_0}_{\rB \rC}$ is a reversible
dilation of $\tC$, with $\rE_0 := \rB\rC$ and $\rE := \rA\rC$.
Finally, the uniqueness clause in Lemma \ref{lem:fordilation}
implies uniqueness of the dilation.\qed
\medskip 

Moreover, two reversible dilations of the same channel with different
environments are related as follows:

\begin{lemma}\label{lem:uniquenessdifferentE}
  Let $\tV \in \Trnset (\rA, \rB \rE)$ and $\tV' \in \Trnset (\rA,
  \rB\rE')$ be two reversible dilations of the same channel $\tC$,
  with generally different environments $\rE$ and $\rE'$. Then there
  is a  channel $\tZ$ from $\rE$ to $\rE\rE'$ such that
\begin{equation}
\begin{aligned}  \Qcircuit @C=1em @R=.7em @! R { &  &  \pureghost{\tV'} & \qw \poloFantasmaCn \rE' &\qw \\
& \qw \poloFantasmaCn \rA & \multigate{-1}{\tV'} &\qw\poloFantasmaCn \rB &\qw} \end{aligned}  ~=~ \begin{aligned}  \Qcircuit @C=1em @R=.7em @! R {&&&&\pureghost{\tZ}&\qw\poloFantasmaCn{\rE}&\measureD{e}\\
 &  &  \pureghost{\tV} & \qw \poloFantasmaCn \rE &  \multigate{-1}{\tZ} & \qw\poloFantasmaCn {\rE'} &\qw\\
& \qw \poloFantasmaCn \rA & \multigate{-1}{\tV} &\qw\poloFantasmaCn \rB &\qw &\qw&\qw } \end{aligned}
\end{equation}
The channel $\tZ$ has the form 
\begin{equation}\label{zeta}
\begin{aligned}  \Qcircuit @C=1em @R=.7em @! R {&&\pureghost{\tZ}&\qw\poloFantasmaCn{\rE}&\qw\\&\poloFantasmaCn{\rE}\qw &\multigate{-1}{\tZ}&\qw\poloFantasmaCn{\rE'}&\qw}
\end{aligned}\;=\; \begin{aligned} \Qcircuit @C=1em @R=.7em @! R {\prepareC{\eta_0}&\qw&\qw\poloFantasmaCn{\rE'}&\qw &
    \multigate{1}{\tU}&\qw&\qw\poloFantasmaCn{\rE}&\qw \\
     &\qw &\qw \poloFantasmaCn{\rE}&\qw&\ghost{\tU} &
    \qw&\qw\poloFantasmaCn{\rE'}&\qw }
\end{aligned}
\end{equation} 
for some pure state $\eta_0\in \Stset_1 (\rE')$ and some reversible transformation $\tU \in \Trnset (\rE\rE')$. 
\end{lemma}

\Proof Apply $\tV$ and $\tV'$ to the faithful state $\Phi^{(\rA)}$, and then use the uniqueness of purification stated in Lemma \ref{lem:purichan}. \qed

The above results represent  the general version---holding in all probabilistic
theories with purification---of the dilation scheme implied by
Stinespring's Theorem \cite{stine} in quantum theory.  However, differently from
the proof of Stinespring's Theorem, the present proof does not require
any C*-algebraic structure, being based just on the purification
postulate. In fact, it is easy to see that the purification of states and
the  reversible dilation of channels are equivalent features, in the following sense:
\begin{corollary}{\bf (Equivalence between purification and reversible dilation)}
Existence and uniqueness (up to reversible channels on the purifying system) of the purification of states is equivalent to existence and uniqueness (up to reversible channels on the environment) of the reversible dilation of channels. 
\end{corollary}
\Proof The direction ``purification $\Rightarrow$ dilation'' has been
just proved by the dilation theorem. The converse is obvious, since a
normalized state $\rho \in \Stset_1 (\rB)$ is a special case of channel from the
trivial system $\rI$ to $\rB$, and in this special case purification
coincides with dilation. \qed
\medskip

Finally, the reversible dilation of a channel allows one to define the \emph{complementary channel} as follows
\begin{definition}[Complementary channel] Let $\tV \in \Trnset (\rA, \rB \rE)$ be a reversible dilation of channel $\tC \in \Trnset (\rA, \rB)$, as in Theorem \ref{theo:stine}. The \emph{complementary channel} of $\tC$ is the channel $\widetilde{\tC} \in \Trnset (\rA, \rE)$ defined by 
\begin{equation}
 \begin{aligned} \Qcircuit @C=1em @R=.7em @! R {&\poloFantasmaCn{\rA}\qw &\gate{\widetilde{\tC}}&\qw\poloFantasmaCn{\rE}&\qw}
\end{aligned} ~=~ \begin{aligned} \Qcircuit @C=1em @R=.7em @! R {   & & \pureghost{\tV} & \qw\poloFantasmaCn{\rE}&\qw \\ 
 &\qw\poloFantasmaCn{\rA}& \multigate{-1}{\tV}&\qw\poloFantasmaCn{\rB}&\measureD e }
\end{aligned}
\end{equation}
\end{definition}
Note that the complementary channel $\widetilde{\tC}$ is unique up to reversible transformations on the environment $\rE$. 

The notion of complementary channel has played a crucial role in the
research about capacity of quantum information channels (see e.g.
\cite{comple1,comple2,comple3}) and we expect that having the same
definition in general probabilistic theories will be very fruitful (in
fact, a number of consequences is already presented in the Section \ref{sec:errcorr}).
\subsection{Reversible dilation of tests}  
We now generalize the dilation of channels  (i.e. single-outcome tests) to the case of arbitrary tests. For this purpose, we need the analogue of Lemma \ref{lem:fordilation}:
\begin{lemma}\label{lem:fordilation2}
Let $\{R_i\}_{i \in \rX}$ be a preparation-test for system $\rB \tilde \rA$ with the property 
\begin{equation}
\sum_{i \in \rX} \begin{aligned}  \Qcircuit @C=1em @R=.7em @! R {\multiprepareC{1}{R_i} & \qw \poloFantasmaCn{\rB} &\measureD{e}\\
\pureghost{R_i}& \qw \poloFantasmaCn{\tilde \rA} &\qw}\end{aligned} ~=~\begin{aligned} \Qcircuit @C=1em @R=.7em @! R {\multiprepareC{1}{\Psi^{(\rA)}} & \qw \poloFantasmaCn{\rA} &\measureD{e}\\
\pureghost{\Psi^{(\rA)}}& \qw \poloFantasmaCn{\tilde \rA} &\qw} \end{aligned}
\end{equation}
where $\Psi^{(\rA)}$ is the purification of an internal state of
system $\rA$.   Then, there exists a system $\rC$, a pure state
$\varphi_0 \in \Stset_1 (\rB \rC)$, a reversible channel $\tU \in
\Trnset (\rA\rB \rC)$, and an observation-test $\{c_i\}_{i \in \rX}$
on $\rC$ such that
\begin{equation}  
\begin{aligned} \Qcircuit @C=1em @R=.7em @! R {\multiprepareC{1}{R_i} & \qw \poloFantasmaCn{\rB} &\qw  \\
    \pureghost{R_i}& \qw \poloFantasmaCn{\tilde \rA} &\qw} \end{aligned}~=~
\begin{aligned}  \Qcircuit @C=1em @R=.7em @! R
  {\multiprepareC{1}{\varphi_0}&\qw&\qw\poloFantasmaCn{\rC}&\qw &
    \multigate{2}{\tU}&\qw&\qw\poloFantasmaCn{\rC}&\measureD{c_i} \\
    \pureghost{\varphi_0}&\qw&\qw\poloFantasmaCn{\rB}&\qw &
    \ghost{\tU}&\qw&\qw\poloFantasmaCn{\rA}&\measureD{e} \\
    \multiprepareC{1}{\Psi^{(\rA)}}&\qw &\qw
    \poloFantasmaCn{\rA}&\qw&\ghost{\tU} &
    \qw&\qw\poloFantasmaCn{\rB}&\qw \\
    \pureghost{\Psi^{(\rA)}}&\qw& \qw& \qw& \qw& \qw&\qw\poloFantasmaCn{\tilde
      \rA}&\qw} \end{aligned}
\end{equation}
for any outcome $i \in \rX$. By suitably choosing system $\rC$, the
observation-test $\{c_i\}_{i \in \rX}$ can be taken to be a
discriminating test.
\end{lemma}
\Proof Take a purification of the coarse-grained state $R = \sum_i R_i$, say $\Psi_R \in \Stset_1 (\rC\rB \tilde \rA )$ for some purifying system $\rC$.  According to Theorem \ref{theo:purificami-ens}, there is an observation-test $\{c_i\}_{i \in \rX}$ on $\rC$ such that
\begin{equation}
  \K{R_i}_{\rB\tilde \rA} =  \B{c_i}_\rC \K{\Psi_R}_{\rC\rB \tilde \rA } \qquad \forall i \in \rX,
\end{equation}  
and, by suitably choosing $\rC$, $\{c_i\}$ can be chosen to be a
discriminating test. Following the same line of Lemma
\ref{lem:fordilation} we then obtain the thesis. \qed

\medskip

Following the proof of the reversible dilation of channels given in Theorem \ref{theo:stine}  we have the following
\begin{theorem}[Reversible dilation of tests] For every test
  $\{\tC_i\}_{i \in \rX}$ from system $\rA$ to system $\rB$ there
  exist a system $\rC$, a pure state $\varphi_0\in \Stset_1(\rB \rC)$, a
  reversible channel $\tU\in\Trnset(\rA \rB \rC)$, and an
  observation-test $\{c_i\}_{i\in \rX}$ on $\rC$ such that for all outcomes $i \in \rX$
  \begin{equation}
\begin{aligned}    \Qcircuit @C=1em @R=.7em @! R {&\poloFantasmaCn{\rA}\qw &\gate{\tC_i}&\qw\poloFantasmaCn{\rB}&\qw}\end{aligned}\;=\; \begin{aligned} \Qcircuit @C=1em @R=.7em @! R {\multiprepareC{1}{\varphi_0}&\qw &\qw\poloFantasmaCn{\rC}&\qw &
      \multigate{2}{\tU}&\qw&\qw\poloFantasmaCn{\rC}&\measureD{c_i} \\ \pureghost {\varphi_0}&\qw &\qw\poloFantasmaCn{\rB}&\qw &
      \ghost{\tU}&\qw&\qw\poloFantasmaCn{\rm A }&\measureD{e} \\ 
      &\qw &\qw \poloFantasmaCn{\rA}&\qw&\ghost{\tU} &
      \qw&\qw\poloFantasmaCn{\rB}&\qw } \end{aligned}
\end{equation}
   By suitably choosing  system $\rC$, the observation-test $\{c_i\}_{i \in \rX}$ can be taken to be a discriminating test.
\end{theorem}

In the case we choose the observation-test $\{c_i\}_{i \in \rX}$ to be
discriminating, the above Theorem yields a (simplified) version of
Ozawa's Theorem in quantum theory \cite{ozawa}. Here the simplification comes
from the fact that we consider finite dimensional state spaces and
tests with finite outcomes, whereas the challenging part of Ozawa's
Theorem is  the rigorous treatment of infinite dimension and
continuous spectrum. 

Moreover, we can apply the dilation theorem to tests with trivial output $\rB \equiv \rI$, thus obtaining the operational version of Naimark's Theorem \cite{naimark} in the finite-outcome case: 
\begin{corollary}{\bf (Discriminating dilation of observation-tests)}  For
  every observation-test $\{a_i\}_{i \in \rX}$ on $\rA$ there exists a
  system $\rC$, a pure state $\varphi_0\in \Stset_1(\rC)$, a
  reversible channel $\tU\in\Trnset(\rA  \rC)$, and a discriminating
  test $\{c_i\}_{i\in \rX}$ on $\rC$ such that
\begin{equation}\label{eq:thesis}
\begin{aligned}  \Qcircuit @C=1em @R=.7em @! R {&\poloFantasmaCn{\rA}\qw &\measureD{a_i}} \end{aligned}\;=\; \begin{aligned} \Qcircuit @C=1em @R=.7em @! R {\prepareC{\varphi_0}&\qw &\qw\poloFantasmaCn{\rC}&\qw &
    \multigate{1}{\tU}&\qw&\qw\poloFantasmaCn{ \rC}&\measureD{c_i} \\
    &\qw &\qw \poloFantasmaCn{\rA}&\qw&\ghost{\tU} & \qw &\qw \poloFantasmaCn{\rA} &\measureD{e}} \end{aligned} 
\end{equation}
for all outcomes $i \in \rX$. 
\end{corollary}

Another corollary is the following:
\begin{corollary}{\bf (Characterization of theories with purification)}
  In a theory with purification every test can be realized using
  only pure states, reversible transformations, and discriminating
  tests.
\end{corollary}
In fact, only one pure state for each system is enough, since due to
Corollary \ref{lem:transitivity} all pure states can be obtained from
a fixed one by acting with reversible channels.

\section{States-transformations isomorphism}\label{sec:iso}

The results of the previous Section allow a complete identification of
transformations with bipartite states, thus providing the general
version of the Choi-Jamio\l kowski isomorphism \cite{choi,jamiolkowski} in quantum theory.
The correspondence is summarized in the following
\begin{theorem}\label{theo:iso} {\bf (States-transformations isomorphism)}
  The storing map $\tC \mapsto R_{\tC} := \tC \K{\Psi^{(\rA)}}_{\rA
    \tilde \rA}$, where $\K{\Psi^{(\rA)}}_{\rA \tilde \rA}$ is a pure dynamically faithful state for system $\rA$, has the following properties:
\begin{enumerate}
\item it defines a bijective correspondence between tests $\{\tC_i \}_{i\in \rX}$ from $\rA$ to $\rB$ and preparation-tests $\{R_i\}_{i\in \rX}$ for $\rB\tilde \rA$ satisfying
 \begin{equation}\label{eq:choiinstr}\sum_{i \in \rX} \B{e}_{\rB} \K{R_{i}}_{\rB
      \tilde \rA} =  \B{e}_{\rA} \K{\Psi^{(\rA)}}_{\rA\tilde \rA}.
 \end{equation}
\item a transformation $\tC$ is atomic (according to Definition \ref{def:atomic})  if and only if the
 corresponding state $R_\tC$ is pure.
\item in  convex theory the map $\tC \mapsto R_\tC$ defines a bijective correspondence between transformations $\tC \in
  \Trnset (\rA, \rB)$ and bipartite states $R \in \Stset (\rB \tilde
  \rA)$ satisfying the property
  \begin{equation}\label{dominance2}
    \B{e}_{\rB} \K{R}_{\rB \tilde \rA} \in D_{\tilde \omega}\qquad \K{\tilde \omega}_{\tilde \rA} =
   \B{e}_{\rA} \K{\Psi^{(\rA)}}_{\rA\tilde \rA}. 
\end{equation}  
\end{enumerate}
\end{theorem}  

\Proof Let us start from the proof of item 1. One direction is obvious: if  $\{\tC_i\}_{i \in \rX}$ is a test from $\rA$ to $\rB$, it must satisfy the normalization condition $\sum_{i \in \rX} \B e_\rB  \tC_i = \B  e_\rA$ (see  Eq. (\ref{eq:testnorm})).   The preparation-test $\{R_{\tC_i}\}_{i \in \rX}$ defined by $\K{R_{\tC_i}}_{\rB\tilde\rA} = \tC_i \K{\Psi^{(\rA)}}_{\rA\tilde\rA}$ satisfies  the property $\sum_{i\in \rX} \B e_\rB \K{ R_{\tC_i}}_{\rB\tilde \rA}  = \sum_{i \in \rX}  \B e_\rB   \tC_i  \K {\Phi^{(\rA)}}_{\rA\tilde \rA}  = \B e_\rA \K{\Psi^{(\rA)}}_{\rA\tilde\rA}  $, that is, it satisfies Eq. (\ref{eq:choiinstr}).  Moreover, if two tests $\{\tC_i\}_{i\in\rX}$ and $\{\tC_i'\}_{i\in \rX}$ satisfy $R_{\tC_i} = R_{\tC_i'}$ for all $i\in \rX$, then by injectivity of the map $\tC \mapsto R_\tC$ (proved in Corollary \ref{cor:stor}), one has $\tC_i = \tC_i'$ for all $i\in \rX$. Conversely, suppose that $\{R_i\}_{i \in \rX}$ is a preparation-test satisfying Eq. (\ref{eq:choiinstr}). Then, by  Lemma \ref{lem:fordilation2} there is a a system $\rC$, a pure state
$\varphi_0 \in \Stset_1 (\rB \rC)$, a reversible channel $\tU \in
\Trnset (\rA\rB \rC)$, and an observation-test $\{c_i\}_{i \in \rX}$
on $\rC$ such that for every outcome $i\in \rX$ one has
\begin{equation} 
\begin{aligned} \Qcircuit @C=1em @R=.7em @! R {\multiprepareC{1}{R_i} & \qw \poloFantasmaCn{\rB} &\qw  \\
    \pureghost{R_i}& \qw \poloFantasmaCn{\tilde \rA} &\qw} \end{aligned} ~=~
\begin{aligned}  \Qcircuit @C=1em @R=.7em @! R
  {\multiprepareC{1}{\varphi_0}&\qw&\qw\poloFantasmaCn{\rC}&\qw &
    \multigate{2}{\tU}&\qw&\qw\poloFantasmaCn{\rC}&\measureD{c_i} \\
    \pureghost{\varphi_0}&\qw&\qw\poloFantasmaCn{\rB}&\qw &
    \ghost{\tU}&\qw&\qw\poloFantasmaCn{\rA}&\measureD{e} \\
    \multiprepareC{1}{\Psi^{(\rA)}}&\qw &\qw
    \poloFantasmaCn{\rA}&\qw&\ghost{\tU} &
    \qw&\qw\poloFantasmaCn{\rB}&\qw \\
    \pureghost{\Psi^{(\rA)}}&\qw& \qw& \qw& \qw& \qw&\qw\poloFantasmaCn{\tilde
      \rA}&\qw} \end{aligned}   \end{equation}  
Defining the test $\{\tC_i\}_{i\in \rX}$ by $\tC_i := \B{c_i}_\rC \B e_\rA \tU \K{\varphi_0}_{\rB\rC}$, we then obtain
\begin{equation} 
 \begin{aligned} \Qcircuit @C=1em @R=.7em @! R {\multiprepareC{1}{R_i} & \qw \poloFantasmaCn{\rB} &\qw  \\
    \pureghost{R_i}& \qw \poloFantasmaCn{\tilde \rA} &\qw} \end{aligned} ~=~ \begin{aligned}  \Qcircuit @C=1em @R=.7em @! R
  { \multiprepareC{1}{\Psi^{(\rA)}} &\qw
    \poloFantasmaCn{\rA}&\gate{\tC_i}  & \qw\poloFantasmaCn{\rB}&\qw \\
    \pureghost{\Psi^{(\rA)}}&\qw&\qw&\qw\poloFantasmaCn{\tilde
      \rA}&\qw} \end{aligned} 
\end{equation}
 This completes the proof of item 1. Item 2 is an immediate
consequence of the item 1: If $\tC$ is atomic, then
$R_{\tC}$ must be pure, otherwise we would have a non-trivial
decomposition of $\tC$. Vice-versa, if $R_\tC$ is pure, then $\tC$ must
be atomic, otherwise we would have a non-trivial decomposition of
$R_{\tC}$. 
 Regarding item 3, injectivity was already established in Corollary \ref{cor:stor}. To prove surjectivity, suppose that $R\in\Stset(\rB\tilde \rA)$ is such that $\B e_\rB \K R_{\rB \tilde \rA} $ is in the refinement set of $\K{\tilde \omega}_{\tilde \rA}$.  This means that there is a preparation-test $\{\tilde \omega_i\}_{i\in \rX}$ such that $\tilde \omega = \sum_{i \in \rX} \tilde \omega_i$ and $\B e_\rB \K R_{\rB \tilde \rA} =\K{\tilde \omega_{i_0}}_{\tilde \rA}$ for some outcome $i_0$.  Now choose an arbitrary set of normalized states  $\{\rho_i\}_{i \in \rX} \subset \Stset_1 (\rB)$ and consider the collection of states $\{R_i\}_{i\in \rX}$ defined as follows:  $R_{i_0} = R$, $R_i = \rho_i\otimes \tilde \omega_i$ for $i\not=i_0$.  Because the theory is convex the collection of states $\{R_i\}_{i\in \rX}$ is a preparation-test (it can be obtained by randomization of the normalized states $\bar R_i =R_i/\SC e {R_i}_{\rB \tilde \rA}$ with probabilities $p_i=\SC e {R_i}_{\rB\tilde \rA}$).  Moreover, it clearly satisfies Eq. (\ref{eq:choiinstr}). Therefore, using item 1 we see that there exists a test $\{\tC_i\}_{i\in \rX}$ from $\rA$ to $\rB$ such that $R_{i} = R_{\tC_i}$. In particular, $R = R_{i_0} = R_{\tC_{i_0}}$, thus proving surjectivity.  \qed
\medskip

Clearly, the correspondence $\tC \mapsto R_{\tC}$ can be extended via
 linear combinations to an injective linear map between the vector
spaces $\Trnset_\Reals (\rA, \rB)$ and $\Stset_\Reals (\rB \tilde
\rA)$.

An immediate consequence of the states-transformations isomorphism is the following
\begin{corollary}[Existence of an ultimate refinement] 
In a convex theory with purification, every test $\{\tC_i\}_{i \in \rX}$ from $\rA$ to $\rB$ admits an ultimate refinement $\{\tD_j\}_{j \in \rY}$ where every transformation $\tD_j$ is atomic.
\end{corollary}

\Proof Consider the preparation-test $\{R_{\tC_i}\}_{i \in \rX}$ and
take the normalized states $\overline{R}_{\tC_i} = R_{\tC_i} / \SC e
{R_{\tC_i}}_{\rB \tilde \rA}$.  Since the states form a finite-dimensional compact convex
set, each state $\overline{R}_{\tC_i}$ has a convex decomposition on a finite number of
pure states. Collecting together all these decompositions yields a
preparation-test $\{R_j\}_{j \in \rY}$, containing only pure states,
that refines $\{R_{\tC_i}\}_{i \in \rX}$.  By the
states-transformations isomorphism, one has $R_j = R_{\tD_j}$, for a
test $\{\tD_j\}_{j \in \rY}$ that refines $\{\tC_i\}_{i \in \rX}$ and
contains only atomic transformations.\qed

\subsection{First consequences of the isomorphism}

Two simple consequences of the states-transformations isomorphism are the following:
\begin{corollary}\label{cor:invertiblereversible}
  A channel $\tV$ from $\rA$ to $\rA \rB$  is atomic if and only
  if it is of the form
\begin{equation}
 \begin{aligned} \Qcircuit @C=1em @R=.7em @! R {&&\pureghost{\tV}&\qw\poloFantasmaCn{\rB}&\qw\\&\poloFantasmaCn{\rA}\qw &\multigate{-1}{\tV}&\qw\poloFantasmaCn{\rA}&\qw}\end{aligned}\;=\; \begin{aligned} \Qcircuit @C=1em @R=.7em @! R {\prepareC{\varphi_0}&\qw&\qw\poloFantasmaCn{\rB}&\qw &
    \multigate{1}{\tU}&\qw&\qw\poloFantasmaCn{\rB}&\qw \\
     &\qw &\qw \poloFantasmaCn{\rA}&\qw&\ghost{\tU} &
    \qw&\qw\poloFantasmaCn{\rA}&\qw } \end{aligned}
\end{equation}  
for some pure state $\varphi_0 \in \Stset_1 (\rB)$ and some reversible channel $\tU \in \grp G_{\rA\rB}$.
\end{corollary}

\Proof Clearly a channel of the form $\tV = \tU \K{\varphi_0}_{\rB}$
is atomic, since the corresponding state $R_\tC = \tU \K{\Psi^{(\rA)}}_{\rA
  \tilde \rA} \K{\varphi_0}_{\rB}$ is pure.  Conversely, if $\tV$ is
atomic, then $R_\tV$ is a purification of the state $\K{\tilde \omega}_{\tilde \rA} := \B e_\rA \K{ \Psi^{(A)}}_{\rA \tilde \rA}$. Since $R_\tV$ and $\Psi^{(\rA)}$ are both purifications
of the same state, by the uniqueness of purification stated by Lemma \ref{lem:purichan} we have $R_\tV =\tU
\K{\Psi^{(\rA)}}_{\rA \tilde \rA} \K{\varphi_0}_{\rB}$ for some pure state
$\varphi_0 \in \Stset_1 (\rB)$ and some reversible channel $\tU \in
\grp G_{\rA\rB}$. Since $\Psi^{(\rA)}$ is dynamically faithful for system $\rA$, this implies
$\tV = \tU \K{\varphi_0}_\rB $.  \qed
\medskip

When system $\rB$ is trivial, we have the more specific result:
\begin{corollary}\label{cor:atomicreversible}
A channel from $\rA$ to $\rA$ is atomic if and only if it is
reversible. 
\end{corollary}
\Proof Special case of Corollary \ref{cor:invertiblereversible} with $\rB \equiv \rI$. \qed 

\medskip

The states-transformations isomorphism also allows one to prove that the sets of transformations, channels, reversible channels, and pure states are compact with respect to the operational norm induced by optimal discrimination:

\begin{corollary} The set of physical transformations $\Trnset (\rA, \rB)$ is compact in the operational norm. 
\end{corollary}
\Proof 
By Theorem \ref{theo:iso}, we have
$\dim (\Trnset_\Reals (\rA, \rB)) \le \dim (\Stset_\Reals (\rB \tilde
\rA))$, namely transformations span a finite-dimensional vector space.
Since we are in finite dimensions, to prove compactness it is enough
to prove that the set of transformations is closed. To see this, suppose that
$\{\tC_n\}$ is a Cauchy sequence of transformations.  By definition, each transformation $\tC_n$ arises in some test, which can be taken to be binary without loss of generality. Let $\{\tC_n,\tD_n\}$ be such a binary test, and let $\{R_{\tC_n},R_{\tD_n}\}$
be the corresponding preparation-test.  Since the set of all states
$\Stset (\rB \tilde \rA)$ is compact (by hypothesis it is finite dimensional and closed), there is a subsequence $\{R_{\tC_{n_k}},R_{\tD_{n_k}}\}$
converging to a binary preparation-test $\{R_0,R_1\}$. Now, since each test $\{R_{\tC_{n_k}},R_{\tD_{n_k}}\}$ satisfies
Eq. (\ref{eq:choiinstr}), also $\{R_0,R_1\}$ satisfies it. By the
states-transformations isomorphism, this implies that there is a binary
test $\{\tC,\tD\}$ such that $R_0= R_{\tC}$ and $R_1 = R_\tD$. Finally, using the bound
of Eq. (\ref{bounddist}) we see that $\tC_{n_k}$ (and hence $\tC_n$) converges to $\tC$ in the
operational norm. \qed

\begin{corollary}
The set of channels from $\rA$ to $\rB$ is compact in the operational norm.
\end{corollary}

\Proof Again, since we are in finite dimension, it is enough to prove that the set of channels is closed. Let
$\{\tC_n\}$ be a Cauchy sequence of channels. Since the set of
transformations is closed, the sequence converges to some
transformation $\tC$. Moreover, $\tC$ is a channel. Indeed, since $\tC_n$ is a channel we have $\B e_\rB \tC_n =\B e_\rA$, and, for every state $\rho$,  $\B e_\rB \tC \K\rho_\rA = \lim_{n \to \infty} \B e_\rB  \tC_n \K{\rho}_\rA = \SC  e \rho_\rA $, which implies $\B e_\rB \tC = \B e_\rA$.  \qed

\begin{corollary} The group $\grp G_\rA$ of all reversible
  transformations of system $\rA$ is a compact Lie group.
\end{corollary}

\Proof Let $\{\tU_n\}$ be a sequence of reversible channels converging
to some channel $\tC$.  We now show that $\tC$ is reversible. Indeed,
consider the sequence $\{\tU_n^{-1}\}$. Since the set of channels is
compact, it is possible to choose a subsequence $\{\tU^{-1}_{n_k}\}$ that
converges to some channel $\tD$. But now we have $\tC\tD = \lim_{k\to\infty}  \tU_{n_k} \tU_{n_k}^{-1} = \tI_\rA$, and, $\tD\tC =
\lim_{k \to \infty} \tU_{n_k}^{-1} \tU_{n_k} = \tI_\rA$ \cite{limprod}, that is,
$\tC$ is reversible and $\tD = \tC^{-1}$. This proves that $\grp
G_\rA$ is closed, and, therefore, compact.  Finally, since $\grp
G_\rA$ is compact and has a faithful finite-dimensional matrix
representation, it is a Lie group (see e.g.  Theorem 5.13 of Ref.
\cite{Folland}).  \qed

\begin{corollary}
The set of pure states of system $\rA$ is compact.
\end{corollary}

\Proof Let $\{\varphi_n\}$ be a sequence of pure states converging to some state $\rho$. We now prove that $\rho$ is pure. To see this, let us fix a pure state $\varphi_0$. By Lemma \ref{lem:transitivity} for every $n$ there is a reversible channel $\tU_n$ such that $\varphi_n = \tU_n \varphi_0$. Since the group $\grp G_\rA$ is compact, we can take a subsequence $\{\tU_{n_k}\}$ that converges to a reversible channel $\tU$.  Therefore we have $\rho = \lim_{n \to \infty} \varphi_{n} = \lim_{k \to \infty} \tU_{n_k} \varphi_0 = \tU \varphi_0$. Since $\rho$ is connected to a pure state by a reversible channel it must be pure. \qed 

\medskip

We conclude this Subsection with two results that will be useful in the construction of deterministic teleportation:

\begin{corollary}[Existence of a twirling test]\label{l:tw} In a (convex) theory with purification there always exists a
  twirling test $\{p_i \tU_i\}_{i \in \rX}$ (according to Def. \ref{def:twirling}), where $\{p_i\}$ are probabilities and $\{\tU_i\}$ are reversible channels. In particular, one of the channels $\{\tU_i\}$ can be always chosen to be the identity.
\end{corollary}
\Proof Let ${\rm d} \tW$ be the normalized Haar measure over the
compact group $\grp G_\rA$, and define the channel $\tT := \int {\rm
  d} \tW ~ \tW$, which is clearly a twirling channel, since by
invariance of the Haar measure one has $\tU \tT = \tT$ for every $\tU
\in \grp G_\rA$.  Since the reversible channels span a
finite-dimensional space, their convex hull is a finite-dimensional
 convex set.  Then by Caratheodory's theorem the integral can be written as
a finite convex combination of reversible transformations, i.e. $\tT =
\sum_{i\in \rX} p_i \tU_i$.  Since $\tU \tT = \tT$, we can pick an outcome
$i_0$, and apply $\tU_{i_0}^{-1}$, thus obtaining a new twirling test where
one channel is the identity.  \qed

\begin{corollary}[Uniqueness of the  invariant state]\label{lem:uniquenessinvariant} 
  For every system $\rA$, there is a unique state $\chi_\rA$ invariant
  under all reversible transformations in $\grp G_\rA$. Moreover, $\chi_\rA$ is internal.
\end{corollary}

\Proof Let $\tT$ be the twirling channel defined in the previous Corollary.  Since for two arbitrary pure states $\psi,
\psi'$ there is a reversible channel $\tU$ such that $\psi '= \tU
\psi$ (Lemma \ref{lem:transitivity}), this implies
\begin{equation}
\tT (\psi') = \int {\rm d} \tW ~ \tW\tU \psi = \int {\rm d} \tW' ~ \tW' \psi  = \tT (\psi):= \chi,
\end{equation}
having used the invariance of the Haar measure. Now, since the
twirling channel is constant on pure states, it is constant on every
state, namely $\tT (\rho) = \chi$ for every $\rho$. In particular, if
$\rho$ is an invariant state, then we have $\rho = \tT (\rho) = \chi$.  This proves that the invariant state is unique. 
Finally, Corollary \ref{l:tw} implies that the integral $\tT (\rho)$ can be
written as the sum of the transformations of a twirling test containing the identity, namely
\begin{equation}
\chi=\tT (\rho)=\sum_jp_j\tU_j\rho=p_{i_0}\rho+\sum_{j\not = i_0}\tU_0^{-1}\tU_j\rho
\end{equation}
whence $p_{i_0}\rho$ belongs to the refinement set $D_\chi$ of $\chi$
for every state $\rho$. This proves  that $\chi$ is internal.  \qed

\subsection{Entanglement breaking channels}

An interesting consequence of the states-transformations isomorphism regards the identification of  \emph{measure-and-prepare channels} and \emph{entanglement breaking channels} , the latter defined as follows
\begin{definition}[Entanglement-breaking channel] A channel $\tC \in
  \Trnset (\rA, \rB)$ is \emph{entanglement breaking} if the output
  state $\tC  \K{\sigma}_{\rA \rC}$ is separable for every state
  $\sigma \in \Stset_1 (\rA \rC)$, namely
\begin{equation}
\tC  \K{\sigma}_{\rA \rC} = \sum_{i \in \rX} p_i  \K{\beta_i}_\rB  \K{\tilde \rho_i}_\rC ,
\end{equation}
for some separable preparation-test $\{p_i \rho_i \otimes \tilde \rho_i\}_{i \in \rX}$, $\beta_i\in\Stset_1 (\rB), \tilde \rho_i\in\Stset_1 (\tilde \rA)$.
\end{definition}

The following Theorem extends to arbitrary theories with purification
the characterization of entanglement breaking channels presented in
quantum theory by Horodecki, Shor, and Ruskai in Ref.
\cite{EBchannels}:

\begin{corollary}{\bf (Structure of entanglement-breaking channels)}
In a theory with purification, the following are equivalent
\begin{enumerate}
\item $\tC$ is entanglement-breaking
\item $R_\tC$ is separable
\item $\tC$ is measure-and-prepare 
\end{enumerate}
\end{corollary}
\Proof \emph{(1) $\Rightarrow$ (2)} If $\tC$ is entanglement-breaking, then in particular $\K{R_\tC}_{\rB \tilde \rA} =
\tC  \K{\Psi^{(\rA)}}_{\rA \tilde \rA}$ is
separable. \emph{(2) $\Rightarrow$  (3)} Suppose that $R_\tC $ is separable, namely $R_\tC =
\sum_{i \in \rX} p_i \beta_i \otimes \tilde \rho_i$ for some separable
preparation-test $\{p_i \beta_i \otimes \tilde \rho_i\}_{i \in \rX}$ (with $\beta_i \in \Stset_1 (\rB)$ and $\tilde \rho_i \in \Stset_1 (\tilde \rA)$).
Now, the preparation-test $\{p_i\tilde \rho_i\}_{i \in \rX}$ has the property 
\begin{equation}
\sum_i p_i \tilde \rho_i = \B e_\rB \K{R_\tC}_{\rB \tilde \rA} = \B e_\rA \K{\Psi^{(\rA)}} := \K{\tilde \chi}_{\tilde \rA},
\end{equation}
having used that $\K{R_\tC}_{\rB \tilde \rA} := \tC
\K{\Psi^{(\rA)}}_{\rA \tilde \rA}$, and the fact that $\tC$ is a
channel. Applying the first item of Theorem \ref{theo:iso} with $\rB
\equiv \rI$, we then deduce that $p_i \tilde \rho_i = R_{a_i}$ for
some suitable observation-test $\{a_i\}$ on $\rA$. Considering the
measure-and-prepare channel $\tD := \sum_{i \in \rX} \K{\beta_i}_{\rB}
\B{a_i}_{\rA}$ we then obtain $R_\tD = R_\tC$, which implies $\tC =
\tD$. Hence, $\tC$ is measure-and-prepare. \emph{ (3) $\Rightarrow$
  (1)} If $\tC$ is measure-and-prepare, it is easily seen that it is
entanglement-breaking. \qed

\subsection{Completeness  of theories with purification}\label{subsect:completeness}

As a consequence of the states-transformations isomorphism, in a
theory with purification we cannot enlarge the set of transformations
without enlarging the set of states. Indeed, we can compare
different theories that have the same set of systems in the following
way:
\begin{definition}[Inclusion of theories] The theory $\Theta'$ is
  \emph{larger} than the theory $\Theta$ if for every couple of
  systems $\rA, \rB$ one has $\Trnset (\rA, \rB) \subseteq \Trnset'
  (\rA, \rB)$, where $\Trnset' (\rA, \rB)$ denotes the set of all
  transformations from $\rA$ to $\rB$ allowed by $\Theta'$.
\end{definition}

Then we have the following
\begin{lemma}{\bf (Maximality of theories with
    purification)}\label{lem:maximality}
Let $\Theta$ be a convex theory with purification, and $\Theta'$ be a convex theory with the same sets of normalized states of $\Theta$, \emph{i.e.}  $\Stset_1 (\rA) = \Stset_1' (\rA)$ for every $\rA$. If $\Theta'$ is larger than $\Theta$, then $\Theta' = \Theta$. 
\end{lemma} 
\Proof First of all, note that the deterministic effect, uniquely
defined by the condition $\SC e \rho_{\rA} = 1, \forall \rho \in
\Stset_1 (\rA)$ is the same in both theories. Now suppose that
$\{\tC_i'\}_{i \in \rX}$ is one of the tests from $\rA$ to $\rB$
allowed by theory $\Theta'$. Let $\{R_{\tC_i'}\}_{i \in \rX}$ be the
corresponding preparation-test for system $\rB \tilde \rA$, as defined
by the state-transformations isomorphism of Theorem \ref{theo:iso}. Since the theories $\Theta'$ and $\Theta$
have the same states, each $R_{\tC_i'}$ is also a state in $\Theta$.
Now, convexity of the set of states implies that $\{R_{\tC_i'}\}_{i
  \in \rX}$ is a legitimate preparation-test in $\Theta$. Moreover, we
have $\sum_{i \in \rX} \B e_\rB \K{R_{\tC'_i}}_{\rB \tilde \rA} =\B
e_\rB \K{\Phi^{(\rA)}}_{\rB \tilde \rA} := \K {\tilde \omega}_{\tilde
  \rA}$. Then, by Theorem \ref{theo:iso} there must be a test
$\{\tC_i\}_{i \in \rX}$ from $\rA$ to $\rB$, allowed by theory
$\Theta$, such that $R_{\tC_i'} = R_{\tC_i}:= \tC_i
\K{\Psi^{(\rA)}}_{\rA \tilde \rA}$. Since $\K{\Psi^{(\rA)}}_{\rA
  \tilde \rA}$ is dynamically faithful for system $\rA$, this implies
$\tC_i' = \tC_i$ for every $i \in \rX$. Therefore, $\Theta'$ and
$\Theta$ have exactly the same tests. \qed

\medskip

The states-transformations isomorphism has also the very strong
consequence that any transformation that is ``mathematically
admissible'' can be actually realized as a test.  To make this statement precise,
let us give the following definitions:

\begin{definition}{\bf (Positive transformation)}
  A transformation $\tC \in \Trnset_{\Reals} (\rA, \rB)$ is
  \emph{positive} if  for every $\rho \in \Stset_+ (\rA)$ one has $\tC \K{\rho}_{\rA}
  \in \Stset_+ (\rB) $.
\end{definition}

\begin{definition}{\bf ($\rS$-positive transformation)}
Given a system $\rS$,  a transformation $\tC \in \Trnset_{\Reals} (\rA, \rB)$ is
  \emph{$\rS$-positive} if $\tC \otimes \tI_\rS$ is positive.
\end{definition}
\begin{definition}{\bf (Completely positive transformation)}
  A transformation $\tC \in \Trnset_{\Reals} (\rA, \rB)$ is
  \emph{completely positive} (CP) if it is $\rS$-positive
  for every system $\rS$.
\end{definition}
\begin{definition}[Admissible instrument]
  An \emph{admissible instrument} with input $\rA$ and output $\rB$ is a
  collection of CP transformations $\{\tC_i\}_{i \in \rX}$ such that
\begin{equation}\label{inst}
\sum_{i \in \rX} \B{e}_\rB \tC_i = \B{e}_{\rA}.
\end{equation}
\end{definition}

The following Theorem establishes that every admissible instrument
must be feasible in a convex theory with purification:
\begin{theorem}{\bf (Completeness of theories with purification)}\label{completeness}
  In a convex theory with purification every admissible instrument
  from $\rA$ to $\rB$ is a test. In particular, every admissible instrument from $\rA$ to $\rI$ is an observation-test. 
\end{theorem} 
\Proof Call $\Theta$ the theory under consideration, and consider the
set of all admissible instruments that are conceivable in $\Theta$.
This set is closed under parallel/sequential composition and under coarse-graining and conditioning.  Therefore this set defines a new theory $\Theta'$ that is larger
than $\Theta$. Moreover, by construction  $\Theta'$ and $\Theta$ have the same states. By Lemma
\ref{lem:maximality}, this implies $\Theta' = \Theta$.  \qed

\begin{corollary}\label{cor:phystrans}{\bf (Characterization of physical transformations)}   In a convex theory with purification the following are equivalent
\begin{enumerate}
\item $\tC$ is a physical transformation from $\rA$ to $\rB$ 
\item $\tC $ is a CP transformation from $\rA$ to $\rB$
  and $\B e_\rA -\B e_\rB \tC$
  is $CP$.
\end{enumerate}
\end{corollary}
\Proof The direction $1 \Rightarrow 2$ is obvious. Conversely, suppose that condition $2$ is satisfied, and define the CP transformations $\B a_\rA :=\B e_\rA - \B e_\tB \tC  $ and $\tD := \K \beta_\rB  \B a_\rA$ where $\K\beta_\rB$ is some normalized state of system $\rB$. Then the collection of CP transformations $\{\tC,\tD\}$ is an admissible instrument. By the completeness of Theorem \ref{completeness} this implies that $\{\tC,\tD\}$ is a test allowed by the theory. Hence, $\tC$ is a physical transformation. \qed
\medskip

We are now in position to prove a stronger result than Lemma
\ref{lem:maximality}, namely the fact that a theory with purification
is completely specified once we have declared the states for every
system: 
\begin{theorem}{\bf (States specify the theory)}\label{statesspecify}
  Let $\Theta, \Theta'$ be two convex theories with purification. If $\Theta$ and $\Theta'$ have the same sets of normalized states, then $\Theta ' = \Theta$. 
\end{theorem} 

\Proof Given two theories $\Theta, \Theta'$ with the same set of
states we can take the new theory $\Theta \cup \Theta'$ that is
generated by $\Theta$ and $\Theta'$ by taking sequential and parallel
composition of the corresponding CP transformations.  Since by
construction $\Theta \cup \Theta'$ contains $\Theta$ and $\Theta'$ and
has the same sets of states by Lemma \ref{lem:maximality} we have
$\Theta = \Theta \cup \Theta' = \Theta'$. \qed
\medskip

We conclude this Subsection by discussing the implication of the
no-restriction hypothesis of Def. \ref{def:no-restriction} and of Lemma \ref{lem:no-restrictioncausal}, which states that every element in the dual cone of states is proportional to a possible effect.  In
this case, we have the following characterization:

\begin{lemma}
  In theory satisfying the no-restriction hypothesis of Def.
  \ref{def:no-restriction} the following are equivalent:
\begin{enumerate} 
\item $a \in \Trnset_\Reals (\rA, \rI)$ is  CP
\item $a$ is an element of the dual cone $\Stset_+ (\rA)^*$
\item $a$ is an element of the cone $\Cntset_+(\rA)$
\end{enumerate}
\end{lemma}
\Proof 1 $\Rightarrow $ 2. Any CP transformation $\tC$ from $\rA$ to
$\rI$ defines a unique element $a$ of the dual cone $\Stset_+
(\rA)^*$, via the relation $a (\rho) := \tC \K{\rho}_\rA$.  In fact,
$\tC$ and $a$ are identified: if two CP transformations $\tC$ and
$\tC'$ define the same effect, then we also have $(\tC \otimes
\tI_\rC) \K{ \sigma}_{\rA\rC} = (\tC' \otimes \tI_\rC) \K{
  \sigma}_{\rA\rC}$ for every system $\rC$ and for every state $\rho
\in \Stset (\rA \rC)$.  Therefore $\tC \equiv \tC'$, and we can
identify $\tC$ with $a$.  2 $\Rightarrow$ 3.  By the consequence of the no-restriction
hypothesis stated by Lemma \ref{lem:no-restrictioncausal}, if $a$ is in the dual $\Stset_+ (\rA)^*$ then $a$ is in
$\Cntset_+ (\rA)$.  3 $\Rightarrow$ 1 By definition, an element of
$\Cntset_+ (\rA)$ is proportional to an effect (with a positive
proportionality constant). Now every effect is a physical
transformation from $\rA$ to $\tI$, and physical transformations are
by definition CP. \qed

\begin{definition}[Effect-valued measures]
An admissible instrument from $\rA$ to $\rI$ is an \emph{effect-valued measure (EVM)}, that is, a collection of effects $\{a_i\}_{i \in \rX}$ such that $\sum_{i \in \rX} \B{a_i}_\rA = \B e_\rA$.  
\end{definition}

The completeness Theorem \ref{completeness} now implies:
 
\begin{corollary}[Characterization of observation-tests]\label{cor:evm}
  In a convex theory with purification every effect-valued measure is an observation-test. If the no-restriction hypothesis of Def. \ref{def:no-restriction} holds, 
  every probability rule  (collection of positive functionals that sum to the deterministic effect) is an observation-test.
\end{corollary}

Finally, the characterization of Corollary \ref{cor:phystrans}  becomes:

\begin{corollary} In a convex theory with purification satisfying the
  no-restriction hypothesis of Def. \ref{def:no-restriction} the following are
  equivalent
\begin{enumerate}
\item $\tC$ is a physical transformation from $\rA$ to $\rB$ 
\item $\tC $ is a CP transformation from $\rA$ to $\rB$ and is \emph{normalization non-increasing}, i.e., $\B e_\rB \tC \K{\rho}_\rA \le \SC e \rho_\rA$ for every $\rho \in \Stset (\rA)$.
\end{enumerate}
\end{corollary}

\section{Error correction}\label{sec:errcorr}

\subsection{Basic definitions}

Here we give some basic definitions that will be used in the next Subsections.

\begin{definition}[Correctable channels]\label{def:correctable} A channel $\tC \in \Trnset (\rA, \rB)$ is \emph{correctable} upon input of $\rho\in \Stset_1 (\rA)$ if there is a \emph{recovery channel} $\tR \in \Trnset (\rB , \rA)$ such that $\tR \circ \tC =_\rho \tI_\rA$. If $\rho$ is an internal state, we simply say that $\tC$ is \emph{correctable}.
\end{definition}

\begin{definition}[Deletion channels]\label{def:deletion} A channel $\tC \in \Trnset (\rA, \rB)$ is a \emph{deletion channel} upon input of $\rho\in \Stset_1 (\rA)$ if there is a fixed state $\sigma \in \Stset_1 (\rB)$ such that $ \tC =_\rho \K{\sigma}_{\rB} \B{e}_\rA$.
\end{definition}

\begin{definition}[Purification-preserving channels]\label{def:puripres} A channel $\tC \in \Trnset (\rA, \rB)$ is \emph{purification-preserving} for $\rho\in \Stset (\rA)$ if there  is a recovery channel $\tR \in \Trnset (\rB , \rA)$ such that $\tR \tC  \K{\Psi_\rho}_{\rA \rR} = \K{\Psi_\rho}_{\rA \rR}$, with  $\Psi_\rho\in\Stset_1 (\rA\rR)$ arbitrary purification of $\rho$. 
\end{definition}

In the context of error correction, the purifying system $\rR$ will be referred to as the \emph{reference}.

\begin{definition}[Correlation-erasing channels]\label{def:correras} A channel $\tC \in
  \Trnset (\rA, \rB)$ is \emph{correlation-erasing} for $\rho\in
  \Stset (\rA)$ if there is a state $\sigma \in \Stset (\rB)$ such
  that $\tC \K{\Psi_\rho}_{\rA \rR} = \K{\sigma}_\rB \K{\tilde
    \rho}_{\rR}$, where $\Psi_\rho\in\Stset_1 (\rA\rR)$ is an arbitrary purification of $\rho$,
  and $\tilde \rho$ is the complementary state $\K{\tilde \rho}_\rR :=
  \B{e}_{\rA}\K{\Psi_{\rho}}_{\rA\rR}$.
\end{definition}

In a theory with purification, the interplay between these four definitions is the basic underlying structure of error correction.  The simplest relations can be immediately recovered from  Theorem \ref{theo:uponinput}, which related the equality upon input of $\rho$ to the equality on a purification of $\rho$.
\begin{corollary}\label{cor:correct-correl}
  A channel is correctable upon input of $\rho$ if and only if it is
  purification-preserving for $\rho$.
\end{corollary}
\begin{corollary}\label{cor:erasing}
If a channel is correlation-erasing for $\rho$, then it is a deletion channel upon input of $\rho$.  If local discriminability holds, the converse is also true. 
\end{corollary}

Another simple fact about error correction, which holds in all theories with purification, is the following
\begin{lemma}\label{lem:correctrefinement} If a channel
  $\tC \in \Trnset (\rA, \rB)$ is correctable upon input of $\rho \in \Stset_1 (\rA)$ with recovery channel $\tR$,
  and  $\tD \in
  D_{\tC}$ is a transformation in the refinement set of $\tC$ (Def. \ref{def:refset}), then 
  $\tD $ is correctable upon input of $\rho$, with recovery channel $\tR$, \emph{i.e.}  $\tR \tD = _\rho ~ p \tI_\rA$ for some probability $p >0$.
\end{lemma}  

\Proof By definition, since $\tD$ is in the refinement set of $\tC$, there is a test
$\{\tD_i\}_{i \in \rX}$ such that $\tD \equiv \tD_{i_0}$ and $\tC =
\sum_{i \in \rX} \tD_i$. Since $\tC$ is correctable with recovery
channel $\tR$, one has $\tI_\rA =_\rho \tR \tC = \sum_{i \in \rX} \tR
\tD_i$. This means that the test $\{\tR \tD_i\}_{i \in \rX}$ is
non-disturbing upon input of $\rho$. By the ``no-information without
disturbance'' Theorem \ref{theo:no-info} one then has $\tR \tD_i
=_{\rho}~ p_i \tI_\rA$ for every $i \in \rX$.  \qed

\subsection{Error correction and the complementarity between correctable and deletion channels}

We now discuss some necessary and sufficient conditions for the correctability of channels. The simplest case is that of channels from a system to itself: 

\begin{theorem}\label{theo:corrAA} A channel $\tC$ from $\rA$ to $\rA$
  is correctable if and only if it is reversible.
\end{theorem}
\Proof Clearly, if $\tC = \tU \in \grp G_\rA$ one can correct $\tC$ by applying
$\tU^{-1}$. Conversely, suppose that $\tC$ is correctable with some
recovery channel $\tR$.  Let $\tC= \sum_{i \in\rX} \tC_i$ be a
refinement of $\tC$ where each $\tC_i$ is an atomic transformation.
Then, the composition $\{\tR \tC_i\}_{i \in \rX}$ is a non-disturbing
test, and  Theorem \ref{theo:no-info} implies $\tR \tC_i = p_i \tI_\rA$.
Since $\tR$ is a channel, applying the deterministic effect we obtain $\B e_\rA \tR \tC_i = \B e_\rA \tC_i = p_i \B e_\rA$, that is,
$\tC_i$ is proportional to an atomic channel $\tU_i$. By Corollary
\ref{cor:atomicreversible}, an atomic channel from $\rA$ to $\rA$ is reversible. 
Therefore, we have $\tR \tU_i = \tI_\rA$, which implies $\tR = \tU_i^{-1}$ for every $i$.  Hence, all channels $\tU_i$ must be equal, and one has $\tC = \tU$ for some reversible channel $\tU\in \grp G_\rA $. \qed 

\medskip

We now give necessary and sufficient conditions for error correction
in the general case of channels from $\rA$ to $\rB$. The following condition
was presented in the quantum case in Refs.
\cite{schumacher,nielschumacher,barnumnielschumacher}.

\begin{theorem}\label{theo:facto}{\bf (Factorization of reference and environment)} 
  A channel $\tC \in \Trnset (\rA ,\rB)$ is correctable upon input of
  $\rho$ if and only if there are a reversible dilation $\tV \in
  \Trnset (\rA, \rB \rE)$ of $\tC$ and a purification
  $\K{\Psi_\rho}_{\rA \rR}$ of $\rho$ such that systems  $\rE$
  and $\rR$ remain uncorrelated. Diagrammatically,
\begin{equation}\label{ERfact}
  \begin{aligned}\Qcircuit @C=1em @R=.7em @! R {& & \pureghost{\tV} &\qw \poloFantasmaCn {\rE}&\qw\\ 
    \multiprepareC{1}{\Psi_\rho} & \qw \poloFantasmaCn{\rA} &\multigate{-1}{\tV} &\qw \poloFantasmaCn {\rB} &\measureD{e} \\
    \pureghost{\Psi_\rho} & \qw & \qw&\qw \poloFantasmaCn {\rR}&\qw } \end{aligned} ~=~ \begin{aligned}
  \Qcircuit @C=1em @R=.7em @! R {\prepareC{\sigma} & \qw \poloFantasmaCn{\rE} &\qw\\
\prepareC{\tilde \rho}& \qw \poloFantasmaCn{\rR} &\qw } \end{aligned}
\end{equation} 
where $\sigma$ is some state of  $\rE$ and $\tilde \rho$ is the complementary state of $\rho$ on system $\rR$.
\end{theorem}

\Proof Suppose that $\tC$ is correctable upon input of $\rho$ with some
recovery channel $\tR$.  Then, by Theorem \ref{theo:uponinput} we have
\begin{equation}
\begin{aligned} \Qcircuit @C=1em @R=.7em @! R {\multiprepareC{1}{\Psi_{\rho}} & \qw \poloFantasmaCn \rA  & \gate{\tC} & \qw \poloFantasmaCn \rB & \gate{\tR} & \qw \poloFantasmaCn \rA &\qw\\
\pureghost{\Psi_\rho} & \qw  & \qw  & \qw &\qw &\qw  \poloFantasmaCn \rR &\qw}\end{aligned} ~=~ 
\begin{aligned} \Qcircuit @C=1em @R=.7em @! R{\multiprepareC{1}{\Psi_\rho} & \qw \poloFantasmaCn {\rA}&\qw \\
\pureghost{\Psi_\rho} & \qw \poloFantasmaCn \rR&\qw}
\end{aligned} 
\end{equation} 
and, inserting two reversible dilations for $\tC$ and $\tR$, respectively,
\begin{equation}
 \begin{aligned}
\Qcircuit @C=1em @R=.7em @! R {
& & \pureghost{\tV} &  \qw  & \qw& \qw \poloFantasmaCn \rE & \measureD{e} \\ 
& & \pureghost{\tV} & &  \pureghost{\tW}  &\qw \poloFantasmaCn \rF &\measureD{e} \\
\multiprepareC{1}{\Psi_{\rho}} & \qw \poloFantasmaCn \rA  & \multigate{-2}{\tV} & \qw \poloFantasmaCn \rB & \multigate{-1}{\tW} & \qw \poloFantasmaCn \rA &\qw \\
\pureghost{\Psi_\rho} & \qw   & \qw  & \qw &\qw &\qw  \poloFantasmaCn \rR &\qw } \end{aligned}~=~  
\begin{aligned} \Qcircuit @C=1em @R=.7em @! R{\multiprepareC{1}{\Psi_\rho} & \qw \poloFantasmaCn {\rA}&\qw \\
\pureghost{\Psi_\rho} & \qw \poloFantasmaCn \rR&\qw} \end{aligned} 
\end{equation} 
This means that $ \tW \tV \K{\Psi_{\rho}}_{\rA\rR}$ is a purification of $\K{\Psi_\rho}_{\rA \rR}$. Then, Lemma \ref{lem:purestatepur} ensures that  $ \tW \tV \K{\Psi_{\rho}}_{\rA\rR}$ is of the form
\begin{equation}
\begin{aligned} \Qcircuit @C=1em @R=.7em @! R {
& & \pureghost{\tV} &  \qw  & \qw& \qw \poloFantasmaCn \rE &\qw \\ 
& & \pureghost{\tV} & &  \pureghost{\tW}  &\qw \poloFantasmaCn \rF &\qw \\
\multiprepareC{1}{\Psi_{\rho}} & \qw \poloFantasmaCn \rA  & \multigate{-2}{\tV} & \qw \poloFantasmaCn \rB & \multigate{-1}{\tW} & \qw \poloFantasmaCn \rA &\qw \\
\pureghost{\Psi_\rho} & \qw   & \qw  & \qw &\qw &\qw  \poloFantasmaCn \rR &\qw } \end{aligned} ~=~ \begin{aligned}
 \Qcircuit @C=1em @R=.7em @! R{
\multiprepareC{1}{\widetilde{\Psi}_{\phantom{\rho}}} & \qw \poloFantasmaCn \rE &\qw\\
\pureghost{\Psi_{\rho}} & \qw \poloFantasmaCn \rF &\qw \\ 
\multiprepareC{1}{\Psi_\rho} & \qw \poloFantasmaCn {\rA}&\qw \\
\pureghost{\Psi_\rho} & \qw \poloFantasmaCn \rR&\qw} \end{aligned}
\end{equation}  
where $\widetilde {\Psi}$ is some pure state on $\rE \rF$. Applying the deterministic effect on $\rF \rA$ and using the fact that $\tW$ is a channel, we then obtain Eq. (\ref{ERfact}). Conversely, suppose that  Eq. (\ref{ERfact}) holds for some dilation $\tV$ and some purification $\K{\Psi_\rho}_{\rA \rR}$. Then take a purification of $\sigma$, say $\Psi_{\sigma} \in \Stset_1 (\rE \rF)$. Since $\tV \K{\Psi_{\rho}}_{\rA \rR}$ and $\K{\Psi_{\rho}}_{\rA \rR} \K{\Psi_\sigma}_{\rE \rF}$ are both purifications of $\K{\sigma}_{\rE}\K{\tilde \rho}_{\rR}$, by Lemma \ref{lem:purichan} we have 
\begin{equation}
\begin{aligned} \Qcircuit @C=1em @R=.7em @! R {
& & \pureghost{\tV} &  \qw  & \qw& \qw \poloFantasmaCn \rE &\qw \\ 
& & \pureghost{\tV} & &  \pureghost{\tD}  &\qw \poloFantasmaCn \rF &\qw \\
\multiprepareC{1}{\Psi_{\rho}} & \qw \poloFantasmaCn \rA  & \multigate{-2}{\tV} & \qw \poloFantasmaCn \rB & \multigate{-1}{\tD} & \qw \poloFantasmaCn \rA &\qw \\
\pureghost{\Psi_\rho} & \qw   & \qw  & \qw &\qw &\qw  \poloFantasmaCn \rR &\qw} \end{aligned} ~=~\begin{aligned} 
 \Qcircuit @C=1em @R=.7em @! R{
\multiprepareC{1}{{\Psi}_{\sigma}} & \qw \poloFantasmaCn \rE &\qw\\
\pureghost{\Psi_{\sigma}} & \qw \poloFantasmaCn \rF &\qw \\ 
\multiprepareC{1}{\Psi_\rho} & \qw \poloFantasmaCn {\rA}&\qw \\
\pureghost{\Psi_\rho} & \qw \poloFantasmaCn \rR&\qw} \end{aligned} 
\end{equation}  
for some channel $\tD \in \Trnset (\rB, \rF \rA)$. 
Applying the deterministic effect on $\rE$ and $\rF$ and defining $\tR := \B{e}_{\rF}\tD$ we then obtain 
\begin{equation}
\begin{aligned} \Qcircuit @C=1em @R=.7em @! R {\multiprepareC{1}{\Psi_{\rho}} & \qw \poloFantasmaCn \rA  & \gate{\tC} & \qw \poloFantasmaCn \rB & \gate{\tR} & \qw \poloFantasmaCn \rA &\qw\\
\pureghost{\Psi_\rho} & \qw  & \qw  & \qw &\qw &\qw  \poloFantasmaCn \rR &\qw}  \end{aligned}
~=~ \begin{aligned} \Qcircuit @C=1em @R=.7em @! R{\multiprepareC{1}{\Psi_\rho} & \qw \poloFantasmaCn {\rA}&\qw \\
\pureghost{\Psi_\rho} & \qw \poloFantasmaCn \rR&\qw} \end{aligned} 
\end{equation}  By Theorem \ref{theo:uponinput}, this implies $\tR \circ \tC =_\rho \tI_\rA$, namely $\tC$ is correctable upon input of $\rho$. \qed
\medskip

An immediate consequence of the factorization Theorem \ref{theo:facto} is:

\begin{corollary}\label{cor:corrCNES}{\bf (Complementarity of purification-preserving and correlation-erasing channels)}
A channel $\tC\in\Trnset (\rA, \rB)$ is purification-preserving for $\rho\in \Stset_1 (\rA)$ (according to Def. \ref{def:puripres}) if and only if its complementary channel $\widetilde\tC \in \Trnset (\rA , \rE)$ is correlation-erasing for $\rho$ (according to Def. \ref{def:correras}).
\end{corollary}
\Proof By corollary \ref{cor:correct-correl}, $\tC$ is
purification-preserving for $\rho$ iff it is correctable upon input of
$\rho$ and, by the previous Theorem, iff Eq. (\ref{ERfact}) holds.
But Eq. (\ref{ERfact}) is the definition of $\widetilde \tC$ being a
correlation-erasing channel for $\rho$. \qed \medskip

In a theory with purification, since the global evolution of system and environment is reversible, it would be natural to expect that if no information goes to the environment, then the whole information about the input state is contained in the system.  While this intuition is correct in  theories with local discriminability (see  Ref. \cite{private-correctable} for the quantum case), in general theories this situation is trickier. Indeed, as we will see in the following, in a  theory without local discriminability some information can remain ``locked'' in the global state, in a way that makes it inaccessible both from the system and from the environment separately.

\begin{corollary}\label{cor:partialduality}{\bf (Complementarity of correctable and deletion channels)}
  If a channel $\tC \in \Trnset (\rA,\rB)$ is correctable upon input
  of $\rho\in \Stset_1 (\rA)$  (according to Definition \ref{def:correctable}), then its complementary channel
  $\widetilde \tC \in \Trnset (\rA, \rE)$ is a deletion channel upon input of $\rho$ (according to Definition \ref{def:deletion}).  If local discriminability holds, the converse
  is also true.
\end{corollary}
\Proof Direct consequence of corollaries  \ref{cor:correct-correl}, \ref{cor:corrCNES}, and \ref{cor:erasing}. \qed 
\medskip 
{\bf Counterexample.}  We show that in a theory without local
discriminability the complementarity between correctable and
deletion channels does not hold. Consider the case of quantum
mechanics on real Hilbert spaces, and consider the isometry $V$ from a
real qubit to two real qubits defined by
\begin{equation}
V = | \Phi_+ \> \< 0 | + |\Psi_-\> \< 1| 
\end{equation} 
with $ |\Phi_+\>:= \frac {|0\>|0\> + |1\>|1\>}{\sqrt 2}$, and $ |\Psi_-\> := \frac {|0\>|1\> - |1\>|0\>}{\sqrt 2}$. 
In this case the complementary channels $\tC (\rho):=\Tr_1 [V \rho V^\tau]$  and  $\widetilde \tC (\rho):=\Tr_2 [V \rho V^\tau]$  are both deletion channels: indeed, one has
\begin{equation}
\tC (\rho) = \frac{ I_1}2 \qquad \widetilde \tC (\rho) = \frac{I_2} 2~,
\end{equation}
for any real density matrix $\rho$.

\subsection{Error correction with one-way classical communication from the environment}

Here we briefly discuss a more general kind of correction, in which the environment is not completely inaccessible, but rather some operations on it are allowed. Particularly interesting is the case of LOCC operations, which do not require the exchange of systems from the environment, but only communication of outcomes and conditioned operations. In particular, we will focus here on the case of a single round of forward classical communication from the environment to the output system. With the term ``classical''  we mean that only outcomes are communicated.   

\begin{definition}[One-way correctable channels]
A channel $\tC \in \Trnset (\rA, \rB)$ is \emph{one-way correctable} upon input of $\rho$ if for every dilation $\tV \in \Trnset (\rA, \rB \rE)$ there is an observation-test $\{a_i\}_{i \in \rX}$ on $\rE$ and  a collection of recovery channels $\{\tR_i\}_{i \in \rX} \subset \Trnset (\rB, \rA)$ such that 
\begin{equation}
\sum_{i \in \rX} 
\begin{aligned}  
\Qcircuit @C=1em @R=.7em @! R {
&& \pureghost{\tV} &\qw \poloFantasmaCn {\rE}&\measureD{a_i}\\ 
& \qw \poloFantasmaCn{\rA} &\multigate{-1}{\tV} & \qw \poloFantasmaCn {\rB} &\gate{\tR_i}& \qw \poloFantasmaCn \rA &\qw } 
\end{aligned} 
~=_\rho~ 
\begin{aligned}  
\Qcircuit @C=1em @R=.7em @! R {
& \qw\poloFantasmaCn\rA &\gate\tI & \qw \poloFantasmaCn \rA&\qw }
\end{aligned}
\end{equation} 
If $\rho$ is an internal state, we simply say that $\tC$ is \emph{one-way
correctable}.
\end{definition}

The following theorem states that one-way correctable channels are
nothing but randomizations of correctable channels.  The quantum
version of it was given by Gregoratti and Werner in Ref.
\cite{gregwern}.
\begin{theorem}\label{theo:onewaycorr}{\bf (Characterization of one-way correctable
    channels)} A channel $\tC \in \Trnset (\rA, \rB)$ is one-way
  correctable upon input of $\rho\in \Stset_1 (\rA)$ if and only if
  $\tC$ is a the coarse-graining of a test $\{\tC_i\}_{i \in \rX}$  where each transformation $\tC_i$ is correctable upon input of $\rho$.  In particular, if $\rho$ is internal, then $\tC$ is a randomization of correctable channels.
\end{theorem}

\Proof Suppose that $\tC$ is one-way correctable upon input of $\rho$. 
Defining the test $\{\tC_i\}_{i \in \rX}$ by $\tC_i:= \B{a_i}_{\rE} \tV$, and  using Theorem \ref{theo:no-info},  we then obtain $ \tR_i \tC_i =_{\rho} p_i \tI_\rA$.  Therefore, $\tC$ is the coarse-graining of a test where each transformation is correctable upon input of $\rho$. Moreover,  if $\rho$ is internal, using the fact that  each $\tR_i$ is a channel, we obtain 
\begin{equation}
\B{e}_{\rA} \tR_i \tC_i = \B{e}_{\rB}  \tC_i = p_i \B{e}_{\rA}, 
\end{equation}
namely each $\tC_i$ must be proportional to a channel, say $\tC_i =
p_i \tD_i$, with channel $\tD_i$ correctable upon input of $\rho$.   Conversely, suppose that $\tC = \sum_{i \in \rX} \tC_i$ for some test $\{\tC_i\}$ where each transformation $\tC_i$ is correctable
upon input of $\rho$. Dilating such a test, we then obtain a channel
$\tV \in \Trnset(\rA, \rB \rE)$ and an observation-test $\{a_i\}_{i
  \in \rX}$ on $\rE$ such that
\begin{equation}
  \begin{aligned}\Qcircuit @C=1em @R=.7em @! R {&& \pureghost{\tV} &\qw \poloFantasmaCn {\rE}&\measureD{a_i}\\ 
     & \qw \poloFantasmaCn{\rA} &\multigate{-1}{\tV} & \qw \poloFantasmaCn {\rB} &\qw  } \end{aligned}  ~=~
 \begin{aligned} \Qcircuit @C=1em @R=.7em @! R {& \qw\poloFantasmaCn\rA &\gate{\tC_i} & \qw \poloFantasmaCn \rB&\qw }
\end{aligned}
\end{equation}    
for every outcome $i \in \rX$. Since each $\tC_i$ is correctable upon input of $\rho$, knowing the outcome $i\in \rX$, we can perform the recovery channel for $\tC_i$, thus correcting channel $\tC$. \qed

In the case of channels from $\rA$ to itself, the above theorem takes the simple form
\begin{corollary}
A channel $\tC \in \Trnset (\rA)$ is one-way correctable if and only if it is a randomization of reversible channels.
\end{corollary}

\Proof  Just combine Theorem \ref{theo:onewaycorr} with the characterization of correctable channels from $\rA$ to $\rA$ (Theorem \ref{theo:corrAA}). \qed  
\section{Causally ordered channels and channels with memory}

In Ref. \cite{bgnp} Beckman, Gottesmann, Nielsen, and Preskill
introduced the notions of {\em semicausal} and {\em semilocalizable}
quantum channel for the purpose of studying the constraints on quantum
dynamics of bipartite systems imposed by relativistic causality.
Subsequently, Eggeling, Schlingemann, and Werner \cite{eggeshliwe}
proved the equivalence between semicausality and semilocalizability
(see also Ref. \cite{pianihoro} for an extensive discussion on the topic).
The same notions were generalized to the case of multipartite channels
by Kretschmann and Werner in Ref.  \cite{kretschwe}.  From different
points of view Refs.  \cite{kretschwe,watgut,comblong} studied the
structure of multipartite causal channels, showing that they can
always be realized as sequences of channels with memory.  In this
Section we show that all these results, originally obtained in quantum
mechanics, actually hold in any causal theory with purification.

Unfortunately, the nomenclature used in the literature is not fully
consistent if we go from bipartite to multipartite channels
\cite{notgood}.  In order to have a consistent nomenclature, instead of ``semicausal'' and ``semilocalizable channel'' we use here the plain expressions  \emph{causally ordered bipartite channel}  and \emph{sequence of two channels with memory}, respectively.

\begin{definition}{\bf (Causally ordered bipartite channel)} A bipartite channel $\tC$ from $\rA_1 \rA_2$ to $\rB_1 \rB_2$ is \emph{causally ordered} if there is a channel $\tD$ from $\rA_1$ to $\rB_1$ such that
\begin{equation}
\B{e}_{\rB_2} \tC = \tD \otimes \B{e}_{\rA_2}. 
\end{equation} 
Diagrammatically,
\begin{equation}\label{semicausal} 
  \begin{aligned}
\Qcircuit @C=1em @R=.7em @! R {& \qw\poloFantasmaCn{\rA_1} & \multigate{1}{\tC} & \qw \poloFantasmaCn {\rB_1}& \qw\\
     &\qw\poloFantasmaCn{\rA_{2}}  & \ghost{\tC} &\qw \poloFantasmaCn {\rB_2} &\measureD e} \end{aligned} 
  ~=~
\begin{aligned}  \Qcircuit @C=1em @R=.7em @! R {
    & \qw\poloFantasmaCn{\rA_1} & \gate \tD  & \qw \poloFantasmaCn {\rB_1}&\qw \\
  & \qw &\qw & \qw\poloFantasmaCn{\rA_2} & \measureD e } 
\end{aligned}
\end{equation}
\end{definition}
Eq. (\ref{semicausal}) means that the channel $\tC$ does not allow for
signaling from the input system $\rA_2$ to the output system $\rB_1$.
In a relativistic context, this can be interpreted as $\rB_1$ being
outside the causal future of $\rA_2$.
 
\begin{definition}\label{def:sequencetwochannels}{\bf (Sequence of two channels with memory)} A
  bipartite channel $\tC$ from $\rA_1 \rA_2$ to $\rB_1 \rB_2$ can be realized as a
  \emph{sequence of two channels with memory} if there exist two
  systems $\rE_1, \rE_2$, called \emph{memory systems}, and two channels $\tC_1 \in
  \Trnset (\rA_1, \rB_1 \rE_1)$ and $\tC_2 \in\Trnset(\rA_2 \rE_1,
  \rB_2 \rE_2 )$ such that
\begin{equation}
  \tC=\B e_{\rE_2}(\tC_2\otimes\tI_{\rB_1})(\tI_{\rA_2}\otimes\tC_1).
\end{equation}
Diagrammatically,
\begin{equation}\label{dilationsemicausal}
  \begin{aligned}
\Qcircuit @C=1em @R=.7em @! R {
    & \qw\poloFantasmaCn{\rA_1} & \multigate{1}{\tC}  & \qw \poloFantasmaCn {\rB_1}&\qw \\
    & \qw\poloFantasmaCn{\rA_2} & \ghost{\tC}  & \qw \poloFantasmaCn {\rB_2} &\qw}\end{aligned} ~=~
\begin{aligned}  \Qcircuit @C=1em @R=.7em @! R { 
    & \qw \poloFantasmaCn{\rA_1} & \multigate{1}{\tC_1} & \qw \poloFantasmaCn{\rB_1} & \qw && \qw \poloFantasmaCn{\rA_2} & \multigate{1}{\tC_2} & \qw \poloFantasmaCn{\rB_2} & \qw  \\ 
    &             & \pureghost{\tC_1} & \qw \poloFantasmaCn{\rE_1} & \qw & \qw & \qw &\ghost{\tC_2} & \qw \poloFantasmaCn {\rE_2} & \measureD e}
\end{aligned}
\end{equation}
\end{definition}

\subsection{Dilation of causally ordered channels}

For causally ordered bipartite channels the dilation theorem implies the following result:

\begin{theorem}\label{theo:decompsemicausal}{\bf (Causal ordering is memory)} A bipartite channel
  $\tC$ from $\rA_1 \rA_2$ to $\rB_1 \rB_2$ is causally ordered if and
  only if it can be realized as a sequence of two channels with memory. Moreover, the channels $\tC_1, \tC_2$ in Eq. (\ref{dilationsemicausal}) can be always chosen such that $\tC_2 \tC_1$ is a reversible dilation of $\tC$.
 \end{theorem}

\Proof If Eq. (\ref{dilationsemicausal}) holds, the channel $\tC$ is
clearly causally ordered, with the channel $\tD$ given by $\tD := \B
e_{\rE_1} \tC_1$. Conversely, suppose that $\tC$ is causally ordered.  Take
a reversible dilation of $\tC$, say $\tV \in \Trnset (\rA_1 \rA_2, \rB_1 \rB_2
\rE)$, and a reversible dilation of $\tD$, say $\tV_1 \in \Trnset (\rA_1, \rB_1
\rE_1)$.  Now, by definition of causally ordered channel (Eq. (\ref{semicausal}) )we have
\begin{equation}
 \begin{aligned} \Qcircuit @C=1em @R=.7em @! R {& \qw\poloFantasmaCn{\rA_1} & \multigate{2}{\tV} & \qw \poloFantasmaCn {\rB_1}& \qw\\
     &\qw\poloFantasmaCn{\rA_{2}}  & \ghost{\tV} &\qw \poloFantasmaCn {\rB_2} &\measureD e\\
& & \pureghost{\tV}  & \qw \poloFantasmaCn{\rE} &\measureD e }
\end{aligned} 
  ~=~
\begin{aligned}  \Qcircuit @C=1em @R=.7em @! R {
    & \qw\poloFantasmaCn{\rA_1} & \multigate{1} {\tV_1}  & \qw \poloFantasmaCn {\rB_1}&\qw \\
    &   & \pureghost{ \tV_1 } & \qw \poloFantasmaCn {\rE_1}&\measureD e \\
  & \qw &\qw & \qw\poloFantasmaCn{\rA_2} & \measureD e }  
\end{aligned} 
\end{equation}
This means that $\tV$ and $\tV_1 \otimes \tI_{\rA_2}$ are two reversible dilations
of the same channel. By the uniqueness of the reversible dilation expressed by Lemma \ref{lem:uniquenessdifferentE} we then
obtain
\begin{equation}
\begin{aligned}  \Qcircuit @C=1em @R=.7em @! R {& \qw\poloFantasmaCn{\rA_1} & \multigate{2}{\tV} & \qw \poloFantasmaCn {\rB_1}& \qw\\
    &\qw\poloFantasmaCn{\rA_{2}}  & \ghost{\tV} &\qw \poloFantasmaCn {\rB_2} &\qw\\
    & & \pureghost{\tV}  & \qw \poloFantasmaCn{\rE} &\qw } \end{aligned} ~=~ 
\begin{aligned}  \Qcircuit @C=1em @R=.7em @! R {
    & \qw\poloFantasmaCn{\rA_1} & \multigate{1} {\tV_1}  & \qw \poloFantasmaCn {\rB_1}&\qw &\qw &\qw &\qw \\
    &   & \pureghost{ \tV_1 } & \qw \poloFantasmaCn {\rE_1}&\multigate{2}{\tZ} & \qw &\qw \poloFantasmaCn{\rB_2} &\qw \\
    & \qw &\qw & \qw\poloFantasmaCn{\rA_2} & \ghost{\tZ} &\qw &\qw \poloFantasmaCn{\rE} &\qw \\
    &&&&\pureghost{\tZ}&\qw&\qw\poloFantasmaCn{\rE_1\rA_2}&\measureD{e}} \end{aligned}  \end{equation} 
Once we have defined $\rE_2 := \rE\rE_1 \rA_2 $ it only remains to
observe that the above diagram is nothing but the thesis, with $\tC_1
= \tV_1$ and $\tC_2 =
\tZ$. By construction, $\tC_2 \tC_1$ is a reversible dilation of $\tC$. 
\qed

\medskip

The definition of causally ordered bipartite channel is easily
extended to the multipartite case as follows:

\begin{definition}[Causally ordered channel] An $N$-partite channel $\tC^{(N)}$ from $ \rA_1  \dots \rA_N$ to $\rB_1 \dots \rB_N $ is \emph{causally ordered} if for every $k\le N$ there is a channel $\tC^{(k)}$ from $\rA_1 \dots \rA_k$ to $\rB_1 \dots \rB_k$ such that
\begin{equation} 
 \begin{aligned}\Qcircuit @C=1em @R=.7em @! R {
  & \qw\poloFantasmaCn{\rA_1} & \multigate{5}{\tC^{(N)}}  & \qw \poloFantasmaCn {\rB_1}&\qw \\
  & \poloFantasmaCn{\vdots} & \pureghost{\tC^{(N)}}  &  \poloFantasmaCn {\vdots} &  \\
  & \qw\poloFantasmaCn{\rA_k} & \ghost{\tC^{(N)}}  & \qw \poloFantasmaCn {\rB_k} &\qw\\ 
 & \qw\poloFantasmaCn{\rA_{k+1}} & \ghost{\tC^{(N)}}  & \qw \poloFantasmaCn {\rB_{k+1}} &\measureD e\\
  & \poloFantasmaCn{\vdots} & \pureghost{\tC^{(N)}}  &  \poloFantasmaCn {\vdots} & \\
  & \qw\poloFantasmaCn{\rA_N} & \ghost{\tC^{(N)}}  & \qw \poloFantasmaCn {\rB_N} &\measureD e} \end{aligned}
~=~
\begin{aligned}\Qcircuit @C=1em @R=.7em @! R {
  & \qw\poloFantasmaCn{\rA_1} & \multigate{2}{\tC^{(k)}}  & \qw \poloFantasmaCn {\rB_1}&\qw \\
  & \poloFantasmaCn{\vdots} & \pureghost{\tC^{(k)}}  &  \poloFantasmaCn {\vdots} &  \\
  & \qw\poloFantasmaCn{\rA_k} & \ghost{\tC^{(k)}}  & \qw \poloFantasmaCn {\rB_k} &\qw\\ 
  & \qw & \qw & \qw\poloFantasmaCn{\rA_{k+1}} & \measureD e\\
  & \poloFantasmaCn{\vdots} & & \poloFantasmaCn {\vdots} & \\
  & \qw & \qw & \qw\poloFantasmaCn{\rA_N} &\measureD e}\end{aligned} 
\end{equation}
\end{definition}
The definition means that the output systems $\rB_1 \dots \rB_k$ are outside the causal future of any input system $\rA_l$ with $l > k$.

Causally ordered channels can be characterized as follows:
\begin{theorem}\label{theo:memorychan}{\bf (Causal ordering is memory for general $N$)}  An $N$-partite channel  $\tC^{(N)}$  from $ \rA_1  \dots \rA_N$ to $\rB_1 \dots \rB_N $ is causally ordered if and only if there exists a sequence of memory systems $\{\rE_k\}_{k=0}^{N}$ with $\rE_0 = \rI$ and a sequence of  channels $\{\tV_k\}_{k=1}^{N}$, with $\tV_k \in \Trnset (\rA_k \rE_{k-1},\rB_k \rE_{k})$ such that
\begin{eqnarray}\label{decompcausalchannel}
  &&\begin{aligned} \Qcircuit @C=1em @R=.7em @! R {
    & \qw\poloFantasmaCn{\rA_1} & \multigate{2}{\tC^{(N)}}  & \qw \poloFantasmaCn {\rB_1}&\qw \\
    & \poloFantasmaCn{\vdots} & \pureghost{\tC^{(N)}}  &  \poloFantasmaCn {\vdots} &  \\
    & \qw\poloFantasmaCn{\rA_N} & \ghost{\tC^{(N)}}  & \qw \poloFantasmaCn {\rB_N} &\qw} \end{aligned}~=~ \\
  && \nonumber\\
  &&  =~\begin{aligned}\Qcircuit @C=1em @R=.7em @! R { 
    & \qw \poloFantasmaCn{\rA_1} & \multigate{1}{\tV_1} & \qw \poloFantasmaCn{\rB_1} & \qw && \qw \poloFantasmaCn{\rA_2} & \multigate{1}{\tV_2} & \qw \poloFantasmaCn{\rB_2} & \qw  & \dots & & \qw \poloFantasmaCn{\rA_N} & \multigate{1}{\tV_N} &\qw \poloFantasmaCn{\rB_N} &\qw \\ 
    &             & \pureghost{\tV_1} & \qw \poloFantasmaCn{\rE_1} & \qw & \qw & \qw &\ghost{\tV_2} & \qw \poloFantasmaCn {\rE_2} & \qw    & \dots&  &\qw \poloFantasmaCn{\rE_{N-1}} &\ghost{\tV_N} & \qw \poloFantasmaCn{\rE_N} &\measureD e} \end{aligned}\nonumber
\end{eqnarray}
Moreover, $\tV_N \dots \tV_1$ is a reversible dilation of $\tC$.
\end{theorem} 

\Proof It is trivial to see that if $\tC^{(N)}$ is a sequence of channels with memory, it is  a causally ordered channel. Here we prove the converse. For $N=1$ the thesis is just the dilation theorem for channels.
We now show that if the thesis holds for $N$, then it has to hold also
for $N+1$. Since $\tC^{(N+1)}$ is a causal channel, we have in particular 
\begin{equation}\label{questa}
\B{e}_{\rB_{N+1}} \tC^{(N+1)}  = \tC^{(N)}  \otimes \B e_{\rA_{N+1}}~.
\end{equation}
This means that $\tC^{(N+1)}$ can be viewed as a bipartite causally ordered  channel from $\rC_1 \rC_2$ to $\rD_1 \rD_2$, where $\rC_1 := \rA_1 \dots \rA_N$, $\rC_2:=\rA_{N+1}$, $\rD_1 :=\rB_1 \dots
\rB_N$, and $\rD_2 := \rB_{N+1}$.  Then Theorem \ref{theo:decompsemicausal} yields two  channels $\tW_1 \in \Trnset (\rC_1, \rD_1 \rF_1)$ and $\tW_2 \in \Trnset (\rC_2 \rF_1, \rD_2 \rF_2)$ such that 
\begin{equation}
  \begin{aligned} \Qcircuit @C=1em @R=.7em @! R {
    & \qw\poloFantasmaCn{\rC_1} & \multigate{1}{\tC}  & \qw \poloFantasmaCn {\rD_1}&\qw \\
    & \qw\poloFantasmaCn{\rC_2} & \ghost{\tC}  & \qw \poloFantasmaCn {\rD_2} &\qw}\end{aligned} ~=~
\begin{aligned}  \Qcircuit @C=1em @R=.7em @! R { 
    & \qw \poloFantasmaCn{\rC_1} & \multigate{1}{\tW_1} & \qw \poloFantasmaCn{\rD_1} & \qw && \qw \poloFantasmaCn{\rC_2} & \multigate{1}{\tW_2} & \qw \poloFantasmaCn{\rD_2} & \qw  \\ 
    &             & \pureghost{\tW_1} & \qw \poloFantasmaCn{\rF_1} & \qw & \qw & \qw &\ghost{\tW_2} & \qw \poloFantasmaCn {\rF_2} & \measureD e} \end{aligned}
\end{equation} 
Now, applying the deterministic effect on $\rD_2$, and using Eq.
(\ref{questa}) the above diagram implies also that $\tW_1$ is a
dilation of $\tC^{(N)}$. On the other hand, by the induction
hypothesis $\tC^{(N)}$ has a reversible dilation $\tV^{(N)}$ of the form of Eq.
(\ref{decompcausalchannel}), namely 
\begin{equation}
\tV^{(N)} = \tT_N \dots \tT_1,
\end{equation}
for some sequence of  channels $(\tT_k)_{k=1}^N \in \Trnset
(\rA_k \rG_{k-1}, \rB_k \rG_k)$ and some sequence of memory systems
$(\rG_k)_{k=0}^N$, with $\rG_0 = \rI$. Since $\tW_1$ and $\tV^{(N)}$
are reversible dilations of the same channel, the uniqueness of the reversible dilation of Lemma
\ref{lem:uniquenessdifferentE} implies $\tW_1 = \B{e}_{\rG_N} \tZ
\tV^{(N)}$, with $\tZ \in \Trnset (\rG_N, \rG_N\rF_1)$ of the form of
Eq.~\eqref{zeta}.  Then, the thesis follows by defining the memory systems
as
\begin{equation}
  \rE_k := \left\{ 
\begin{array}{ll}
  \rG_k \qquad &  k <N\\ 
  \rG_N \rF_1  \qquad & k=N\\
\rG_N  \rF_2 \qquad   & k=N+1. 
\end{array}
\right.
\end{equation}
 and by defining the channels as  
\begin{equation}
  \tV_k := \left\{ 
\begin{array}{ll}
  \tT_k \qquad &  k <N\\ 
  \tZ \tT_N  \qquad & k=N\\
\tI_{\rG_N} \otimes \tW_2 \qquad   & k=N+1. 
\end{array}
\right.
\end{equation}
By construction, the channel $\tV_{N+1} \tV_N \dots \tV_1$ is a reversible dilation of the channel $\tC^{(N+1)}$.
\qed

\medskip

Moreover, since the realization of the previous Theorem is just the reversible dilation of $\tC^{(N)}$, we have the uniqueness result: 
\begin{corollary}{\bf (Uniqueness of the reversible dilation)}
  Let $\{\tV_k\}_{k=1}^{N}$, $\tV_k \in \Trnset (\rA_k \rE_{k-1},\rB_k
  \rE_{k})$ be a reversible realization of the causally ordered channel
  $\tC^{(N)}$ as a sequence of channels with memory, as in Theorem
  \ref{theo:memorychan}.  Suppose that $\{\tV'_k\}_{k=1}^{N}$, $\tV'_k
  \in \Trnset (\rA_k \rE'_{k-1},\rB_k \rE'_{k})$ is another reversible
  realization of $\tC^{(N)}$ as a sequence of channels with memory. Then there exists a channel $\tR$ from
  $\rE_N$ to $\rE_N'$ such that
\begin{widetext}
  \begin{equation}
  \begin{aligned}\Qcircuit @C=1em @R=.7em @! R { 
    & \qw \poloFantasmaCn{\rA_1} & \multigate{1}{\tV'_1} & \qw \poloFantasmaCn{\rB_1} & \qw && \qw \poloFantasmaCn{\rA_2} & \multigate{1}{\tV'_2} & \qw \poloFantasmaCn{\rB_2} & \qw  & \dots & & \qw \poloFantasmaCn{\rA_N} & \multigate{1}{\tV'_N} &\qw \poloFantasmaCn{\rB_N} &\qw \\ 
    &             & \pureghost{\tV'_1} & \qw \poloFantasmaCn{\rE'_1} & \qw & \qw & \qw &\ghost{\tV'_2} & \qw \poloFantasmaCn {\rE'_2} & \qw    & \dots&  &\qw \poloFantasmaCn{\rE'_{N-1}} &\ghost{\tV'_N} & \qw \poloFantasmaCn{\rE'_N} &\qw} \end{aligned} 
  ~=~ \begin{aligned}
 \Qcircuit @C=1em @R=.7em @! R { 
    & \qw \poloFantasmaCn{\rA_1} & \multigate{1}{\tV_1} & \qw \poloFantasmaCn{\rB_1} & \qw && \qw \poloFantasmaCn{\rA_2} & \multigate{1}{\tV_2} & \qw \poloFantasmaCn{\rB_2} & \qw  & \dots & & \qw \poloFantasmaCn{\rA_N} & \multigate{1}{\tV_N} &\qw \poloFantasmaCn{\rB_N} &\qw &\qw &\qw\\ 
    &             & \pureghost{\tV_1} & \qw \poloFantasmaCn{\rE_1} & \qw & \qw & \qw &\ghost{\tV_2} & \qw \poloFantasmaCn {\rE_2} & \qw    & \dots&  &\qw \poloFantasmaCn{\rE_{N-1}} &\ghost{\tV_N} & \qw \poloFantasmaCn{\rE_N} &\gate \tR & \qw \poloFantasmaCn {\rE_N'} &\qw} \end{aligned}
\end{equation} 
\end{widetext}
\end{corollary}

\Proof The channels $\tV:= \tV_N \dots \tV_1 \in
\Trnset (\rA_1 \dots \rA_N, \rB_1 \dots \rB_N \rE_N)$ and $\tV':= \tV'_N
\dots \tV'_1 \in \Trnset (\rA_1 \dots \rA_N, \rB_1 \dots \rB_N \rE'_N)$
are two reversible dilations of the channel $\tC^{(N)}$. The
statement is the direct application of the uniqueness of the dilation stated by Lemma
\ref{lem:uniquenessdifferentE}. \qed

\subsection{No bit commitment}

Sequences of channels with memory can be used to describe sequences of
moves of a given party in a cryptographic protocol or in a multiparty
game (see Ref. \cite{watgut} for the case of quantum games). In this scenario,
the memory systems are the private systems available to a party, while
the other input-output systems are the systems exchanged in the communication with other
parties.  In this context, the uniqueness of the realization of a
causal channel directly implies the impossibility of tasks like
unconditionally secure bit commitment (see Refs.
\cite{bitcomm,bitcomb} and references therein for the definition of
the problem). A proof in the general case is given by the following:
\begin{corollary}{\bf (No perfectly secure bit commitment)} In a theory with purification, if an $N$-round
  protocol is perfectly concealing, then there is a perfect cheating.
\end{corollary}     
\Proof We first prove the impossibility for protocols that do not involve the exchange of classical information. Let $\tA_0, \tA_1 \in \Trnset (\rA_1 \dots, \rA_N, \rB_1 \dots \rB_{N-1} 
\rB_N \rF_N)$ be two causally ordered $N$-partite channels (here the last output system of the causally-ordered channels is the bipartite system $\rB_N \rF_N$),
representing Alice's moves to encode the bit value $b =0,1$,
respectively. The system $\rF_N$  is the system sent from Alice to
Bob at the final phase of the protocol (called the \emph{opening}) in order to unveil the value of the bit. If the protocol is
perfectly concealing, then the reduced channels before the opening phase
must be indistinguishable, namely $\B e_{\rF_N} \tA_0 = \B e_{\rF_N}
\tA_1 := \tC$. Now, take two reversible dilations $\tV_0 \in \Trnset
(\rA_1 \dots, \rA_N, \rB_1 \dots \rB_N \rF_N \rG_0)$ and $\tV_1 \in
\Trnset (\rA_1 \dots, \rA_N, \rB_1 \dots \rB_N \rF_N \rG_1)$ for
$\tA_0$ and $\tA_1$, respectively. Since $\tV_0$ and $\tV_1$ are also
two dilations of the channel $\tC$, there is a channel $\tR $ from $\rF_N
\rG_0$ to $\rF_N \rG_1$ such that $\tV_1 = \tR \tV_0$.  Applying this
channel to her private systems, Alice can switch from $\tV_0$ to
$\tV_1$ just before the opening.  Discarding the auxiliary system
$\rG_1$, this yields channel $\tA_1$. The cheating is perfect, since
Alice can play the strategy $\tV_0$ until the end of the commitment,
and decide the bit value before the opening without being detected by
Bob.  The above reasoning can be extended to $N$-round protocols
involving the exchange of classical information. Indeed, classical messages
can be modelled by perfectly distinguishable states, while classical
channels can be modelled by measure-and-prepare channels where the
observation-test is discriminating, and the prepared states are
perfectly distinguishable. The fact that some systems can only be prepared in perfectly distinguishable states will be referred to as the ``communication interface" of the protocol \cite{bitcomm,bitcomb}.   In this case, to construct Alice's cheating strategy we
can first take the reversible dilations $\tV_0, \tV_1$  and the channel $\tR$ such that $\tV_1 = \tR \tV_0$. In order to comply with the communication
interface of the protocol, one can compose $\tV_0$ and $\tV_1$
 with classical channels on all systems that must be ``classical"  before the opening, thus obtaining two channels $\tD_0$ and $\tD_1$ that are no longer reversible, but still satisfy $\tD_1 = \tR \tD_0$.     Discarding the auxiliary system $\rG_1$ and, if required by the communication interface, applying a classical channel on $\rF_N$, Alice then obtains channel $\tA_1$.  Again, this strategy allows Alice to decide the value of the bit just before the opening without being detected. \qed
\medskip

\section{Deterministic programming of reversible transformations}

In Section \ref{sec:probtele} we saw that transformations can be
stored into states, in such a way that they can be retrieved at later
time with non-zero probability of success.  This provides an instance
of \emph{probabilistic programming}, in which a state plays the role
of program for a transformation, and a suitable machine is able to
read out the program and to reproduce (with some probability) the
correct transformation.  Of course, one would like also to have
deterministic programmable machines, which correctly retrieve the
transformations with unit probability.  We now show that such machines
are much more demanding in terms of resources: indeed to program a
certain number of reversible transformations one needs to have an
equal number of perfectly distinguishable program states. This theorem is the general version of the quantum no-programming theorem by Nielsen and Chuang \cite{no-prog}.
 
\begin{theorem}{\bf (No perfect deterministic programming of reversible channels
    without distinguishable program states)} Let $\{\tU_i\}_{i \in
      \rX}$ be a set of reversible channels on $\rA$, and $\{
    \eta_i\}_{i \in \rX}$ be a set of pure states of $\rB$. If there
    exists a channel $\tR \in \Trnset (\rA \rB, \rA)$ such that
  \begin{equation}
   \begin{aligned} \Qcircuit @C=1em @R=.7em @! R {&&\qw \poloFantasmaCn \rA & \multigate{1}{\tR} & \qw \poloFantasmaCn \rA &\qw \\
&\prepareC{\eta_i} & \qw \poloFantasmaCn \rB  & \ghost{\tR} & & }\end{aligned} ~=~ 
\begin{aligned}   \Qcircuit @C=1em @R=.7em @! R {&\qw \poloFantasmaCn \rA &\gate {\tU_i} & \qw \poloFantasmaCn \rA &\qw}
\end{aligned}
\end{equation}
then the states $\{\eta_i\}_{i \in \rX}$ are perfectly distinguishable. 
\end{theorem}

\Proof Take a dilation of $\tR$, with pure state $\varphi_0 \in \Stset_1
(\rC)$ and reversible channel $\tU \in \Trnset (\rA \rB \rC)$.  Upon
defining the pure states $\varphi_i := \eta_i \otimes\varphi_0$ we
have
\begin{equation}
\begin{aligned}    \Qcircuit @C=1em @R=.7em @! R {&&\qw \poloFantasmaCn \rA & \multigate{1}{\tU} & \qw \poloFantasmaCn \rA &\qw \\
&\prepareC{\varphi_i} & \qw \poloFantasmaCn {\rB\rC}  & \ghost{\tU} & \qw \poloFantasmaCn {\rB \rC}& \measureD e } \end{aligned} ~=~ \begin{aligned} 
   \Qcircuit @C=1em @R=.7em @! R {&\qw \poloFantasmaCn \rA &\gate {\tU_i} & \qw \poloFantasmaCn \rA &\qw}\end{aligned}
\end{equation}
Since this is a dilation of the reversible transformation $\tU_i$, by the uniqueness of the  reversible dilation stated by Theorem \ref{theo:stine} there must be a pure state $\psi_i \in \Stset_1 (\rB \rC)$ such that
\begin{equation}\label{equaa}
\begin{aligned}    \Qcircuit @C=1em @R=.7em @! R {&&\qw \poloFantasmaCn \rA & \multigate{1}{\tU} & \qw \poloFantasmaCn \rA &\qw \\
&\prepareC{\varphi_i} & \qw \poloFantasmaCn {\rB\rC}  & \ghost{\tU} & \qw \poloFantasmaCn {\rB \rC}& \qw } \end{aligned}  ~=~ \begin{aligned}
   \Qcircuit @C=1em @R=.7em @! R {&\qw \poloFantasmaCn \rA &\gate {\tU_i} & \qw \poloFantasmaCn \rA &\qw 
\\
&\prepareC{\psi_i} &\qw& \qw \poloFantasmaCn{\rB \rC} &\qw& } \end{aligned}
\end{equation}
By  applying $\tU_i^{-1}$  on both sides of Eq. (\ref{equaa}), one has 
\begin{equation}
\begin{aligned}    \Qcircuit @C=1em @R=.7em @! R {&\qw \poloFantasmaCn \rA & \gate{\tU_i^{-1}} &\qw \poloFantasmaCn \rA & \multigate{1}{\tU} & \qw \poloFantasmaCn \rA &\qw \\
&&\prepareC{\varphi_i} & \qw \poloFantasmaCn {\rB\rC}  & \ghost{\tU} & \qw \poloFantasmaCn {\rB \rC}& \qw } \end{aligned}  ~=~ \begin{aligned}
   \Qcircuit @C=1em @R=.7em @! R {&  &\qw& \qw \poloFantasmaCn \rA &\qw 
\\
&\prepareC{\psi_i} &\qw& \qw \poloFantasmaCn{\rB \rC} &\qw& } \end{aligned}
\end{equation}
and, applying $\tU^{-1}$,
\begin{equation}\label{equab}
\begin{aligned}    
 \Qcircuit @C=1em @R=.7em @! R {&\qw \poloFantasmaCn \rA &\gate {\tU^{-1}_i} & \qw \poloFantasmaCn \rA &\qw 
\\
\prepareC{\varphi_i} &\qw&\qw & \qw \poloFantasmaCn{\rB \rC} &\qw } \end{aligned} ~=~\begin{aligned}\Qcircuit @C=1em @R=.7em @! R {&&\qw \poloFantasmaCn \rA & \multigate{1}{\tU^{-1}} & \qw \poloFantasmaCn \rA &\qw \\
&\prepareC{\psi_i} & \qw \poloFantasmaCn {\rB\rC}  & \ghost{\tU^{-1}} & \qw \poloFantasmaCn {\rB \rC}& \qw }\end{aligned} 
\end{equation}
Composing Eqs. (\ref{equaa}) and (\ref{equab}) we then obtain  
\begin{widetext}
\begin{equation}
\begin{aligned}  \Qcircuit @C=1em @R=.7em @! R {
&  &\qw \poloFantasmaCn {\rA} & \multigate{1}{\tU} & \qw \poloFantasmaCn {\rA} &\qw && \qw \poloFantasmaCn{\rA} & \multigate{1}{\tU^{-1}} & \qw \poloFantasmaCn {\rA} &\qw\\
    &\prepareC{\varphi_i} & \qw \poloFantasmaCn {\rB\rC}  & \ghost{\tU}  & \qw \poloFantasmaCn {\rB\rC}  & \qw& \qw& \qw &  \ghost{\tU^{-1}} & \qw \poloFantasmaCn {\rB \rC}& \qw  } \end{aligned} 
~=~
\begin{aligned}  \Qcircuit @C=1em @R=.7em @! R {&\qw \poloFantasmaCn {\rA} &\gate {\tU_i} & \qw \poloFantasmaCn {\rA} &\qw  &&\qw \poloFantasmaCn {\rA} &\gate {\tU^{-1}_i} & \qw \poloFantasmaCn {\rA} &\qw
    \\
    \prepareC{\varphi_i} &\qw& \qw  &\qw &\qw  \poloFantasmaCn{\rB \rC} &\qw &\qw &\qw &\qw &\qw}\end{aligned}
\end{equation} 
\end{widetext}
This means that we can obtain an unbounded number of copies of
$\tU_i$ and $\tU_i^{-1}$ by iterating the application of $\tU$ and
$\tU^{-1}$.  Now, if $\tU_i$ and $\tU_j$ are different, the
probability of error in discriminating between them using $N$ copies
should go to zero as $N$ goes to infinity (this can be seen by
repeating $N$ times the optimal test and using majority voting, as in the
proof of Theorem \ref{lemma:clon-disc}).  Since programming the transformations $\{(\tU_i \otimes \tU_i^{-1} )^{\otimes N}\}$ and discriminating among them is a particular way of discriminating between the program states $\{\varphi_i\}$, the latter must be perfectly distinguishable.   Finally, since the states
$\varphi_i = \eta_i \otimes \varphi_0$ are perfectly distinguishable,
also the program states $\eta_i$ must be so. \qed

\medskip 

Note that trying to use mixed program states $\{\rho_i\}$ cannot help
in reducing the number of perfectly distinguishable states needed in
the program system $\rB$. Indeed, suppose that $\rho_i$ is the
following mixture $\rho_i = \sum_j p_j^{(i)} \psi^{(i)}_j$.  Since
reversible transformations are atomic, this means that each
pure state $\psi^{(i)}_j$ must work as a program for $\tU_i$. But the
above theorem implies that, whichever choice we make, the pure states
$\{\varphi^{(i)}_{j_i}\}_{i \in \rX}$ must be perfectly
distinguishable.

\section{Purification with conjugate systems}

\subsection{Conjugate purifying systems}
 
All the results derived so far were consequence of the sole fact that
every state has a purification, unique up to reversible
transformations of the purifying system. We now add more
structure, by introducing the notion of conjugate purifying systems:

\begin{postulate}[Conjugate purifying systems]\label{post:conjugate} For every system $\rA$ there
  exists a \emph{conjugate purifying system} $\tilde \rA$ such that
\begin{enumerate}
\item  for every state $\rho \in \Stset_1( \rA)$ there is a purification $\Psi_\rho$ in $\Stset_1 (\rA \tilde \rA)$ \emph{(completeness for purification)}
\item  $\tilde{\tilde \rA} = \rA$ \emph{(symmetry)}
\item $\widetilde {\rA \rB} = \tilde \rA \tilde \rB$ \emph{(regularity under composition)} 
\end{enumerate}
\end{postulate}

The above postulate could be derived from more basic assumptions.
However, we will not discuss this issue here, and, for the moment, the
existence of conjugate systems will be taken as a Postulate.

Conjugate purifying systems have particularly nice properties, some of
which are given in the following:\begin{lemma} Let $\tilde {\rA}$ be the
  conjugate system of $\rA$. Then, $\dim \Stset_\Reals  (\tilde
  \rA) =\dim \Stset_\Reals  ( \rA) $.
\end{lemma} 

\Proof Trivial consequence of the bound on dimensions given in Eq.  (\ref{eq:dimpurifierinternal}) and of the symmetry condition $\tilde {\tilde \rA} = \rA$.\qed 
\medskip

In a theory with conjugate purifying systems, the dynamically faithful pure states considered in Subsection \ref{ssect:dynfaith} enjoy the following symmetry property:
\begin{lemma}\label{lem:internaltilde}
If the pure state $\Psi \in \Stset_1 (\rA \tilde \rA)$ is dynamically faithful for system $\rA$, then it is dynamically faithful for system $\tilde \rA$. 
\end{lemma}

\Proof Let $\tilde \omega$ be the marginal of $\Psi$ on system $\tilde \rA$, namely $\K {\tilde \omega}_{\tilde \rA} = \B e_\rA \K\Psi_{\rA\tilde\rA}$. Since $\Psi$ is dynamically faithful for system $\rA$, the map $\tau : \Cntset_\Reals  (\rA) \to
\Span (D_{\tilde \omega})$ defined by $\B{a}_\rA \mapsto
\K{\tau_a}_{\tilde \rA} = \B{a}_\rA \K \Psi_{\rA \tilde \rA}$ is
injective (and surjective, by definition). This implies $\dim \Span (D_{\tilde \omega}) = \dim \Cntset_\Reals  (\rA) $. On the other hand, using the previous Lemma one has $\dim \Cntset_\Reals  (\rA) \equiv \dim \Stset_\Reals  (\rA) =\dim \Stset_\Reals  (\tilde \rA)$.  This proves that $\tilde \omega$ is internal in $\Stset (\tilde \rA)$. Since $\Psi$ is the purification of an internal state, by Theorem \ref{theo:dynfaith} it is faithful for system $\tilde \rA$.  \qed 

\medskip

Using the previous Lemma it is quite simple to show that conjugate systems are unique up to operational equivalence:
\begin{lemma}[Uniqueness of the conjugate system] For any system $\rA$ the conjugate system $\tilde \rA$ is unique up to operational equivalence (see Def. \ref{def:opeq}).  
\end{lemma}
\Proof Suppose that $\tilde \rA'$ is another conjugate system of
$\rA$. Then take an internal state $\omega \in \Stset_1(\rA)$ and consider
its purifications $\Psi \in \Stset_1 (\rA \tilde \rA)$ and $\Psi' \in
\Stset_1 (\rA \tilde \rA')$. By the uniqueness of purification expressed by Lemma \ref{lem:purichan}, since $\Psi$
and $\Psi'$ are purifications of the same state, there are two
channels $\tC \in \Trnset (\tilde \rA, \tilde \rA')$ and $\tD \in
\Trnset (\tilde \rA', \tilde \rA)$ such that
\begin{eqnarray}
\K{\Psi'}_{\rA \tilde \rA'} & = &\tC  \K{\Psi}_{\rA \tilde \rA} \\
\K{\Psi}_{\rA \tilde \rA} & = &\tD  \K{\Psi'}_{\rA \tilde \rA'}.
\end{eqnarray}
Clearly, this implies that 
\begin{eqnarray}
\K{\Psi}_{\rA \tilde \rA} & =&  \tD\tC  \K{\Psi}_{\rA \tilde \rA} \\
\K{\Psi'}_{\rA \tilde \rA'} & =& \tC \tD  \K{\Psi'}_{\rA \tilde \rA'}.
\end{eqnarray}
On the other hand, by the previous Lemma the states $\Psi$ and $\Psi'$
are dynamically faithful for systems $\tilde \rA$ and $\tilde \rA'$,
respectively. Hence, one has $\tD \tC = \tI_{\tilde \rA}$ and $\tC \tD
= \tI_{\tilde \rA'}$, namely the channels $\tC$ and $\tD$ are reversible. By Definition \ref{def:opeq}, this means that
$\rA$ and $\rA'$ are operationally equivalent.  \qed

\subsection{States-transformations isomorphism for conjugate purifying systems}

If we use conjugate purifying systems to build up dynamically faithful states some of the
results derived so far become simpler and more elegant. First of all, according to Lemma \ref{lem:internaltilde},  if a pure state $\Psi_{\rA\tilde \rA}$ is dynamically faithful for system $\rA$, then it is also dynamically faithful for system $\tilde \rA$.  This means that we can simply use the expression ``dynamically faithful pure state $\K \Psi_{\rA\tilde\rA}$" without further specifications.  Accordingly, we will drop the superscript $\rA$ in the state $\K{\Psi^{(\rA)}}$.  We now show that we can also drop the condition Eq. (\ref{dominance2}) in the
isomorphism between transformations and bipartite states:

\begin{theorem}\label{theo:iso2}{\bf (Strong version of the states-transformations isomorphism)}
  The storing map $\tC \mapsto \K{R_{\tC}}_{\rB\tilde \rA} := \tC \K{\Psi}_{\rA \tilde
    \rA}$, with $\Psi$ dynamically faithful pure state, has the following properties:
\begin{enumerate}
\item it defines a bijective correspondence between tests $\{\tC_i \}_{i\in \rX}$ from $\rA$ to $\rB$ and preparation-tests $\{R_i\}_{i\in \rX}$ for $\rB\tilde \rA$ satisfying
  \begin{equation}\label{eq:choiinstr2}\sum_{i \in \rX} \B{e}_{\rB} \K{R_{i}}_{\rB
      \tilde \rA} =  \B{e}_{\rA} \K{\Psi}_{\rA\tilde \rA}.
  \end{equation}
\item a transformation $\tC$ is atomic (according to Definition \ref{def:atomic})  if and only if the
  corresponding state $R_\tC$ is pure.
\item in  convex theory the map $\tC \mapsto R_\tC$ defines a bijective correspondence between the cones $
  \Trnset_+ (\rA, \rB)$ and  $ \Stset_+ (\rB \tilde
  \rA)$.
  \end{enumerate}
\end{theorem}  

We  now have the following remarkable fact:

\begin{theorem}\label{theo:atomiceffect}
For every effect $a \in \Cntset (\rA)$ there is an atomic transformation $\tC_a \in \Trnset (\rA)$ such that 
\begin{equation}\label{atomiceffect}
  \begin{aligned}\Qcircuit @C=1em @R=.7em @! R {
    &\qw \poloFantasmaCn \rA &\measureD a}\end{aligned}  \quad =\quad \begin{aligned}   
  \Qcircuit @C=1em @R=.7em @! R {& \qw \poloFantasmaCn \rA & \gate {\tC_a} & \qw \poloFantasmaCn \rA & \measureD e} \end{aligned}.  
\end{equation}
Moreover, the transformation $\tC_a$ is unique up to reversible channels on the output. 
\end{theorem}

\Proof Let  $p_0$ and $p_1$ be the probabilities defined by $p_0 :=  \B a_\rA  \B e_{\tilde \rA}\K \Psi_{\rA
  \tilde \rA}$ and $p_1 :=  \B {e-a}_\rA  \B e_{\tilde \rA}\K \Psi_{\rA
  \tilde \rA}$.  Let $\K{\Psi_0}_{\rA \tilde \rA} $ and $\K{\Psi_{1}}_{\rA \tilde \rA}$ be  purifications of the normalized states $\K{\rho_0}_{\tilde \rA} := \B a_\rA \K \Psi_{\rA
  \tilde \rA}/p_0$ and $\K{\rho_{1}}_{\tilde \rA} := \B {e-a}_\rA \K \Psi_{\rA
  \tilde \rA}/p_1$, respectively. Now, the collection of states $\{p_0\Psi_0,p_1 \Psi_{1}\}$ is a preparation-test (it can be prepared via randomization).  Moreover, such a preparation-test has the property
\begin{equation}
p_0\B e_\rA \K{\Psi_0}_{\rA \tilde \rA} + p_1\B e_\rA \K{\Psi_{1}}_{\rA \tilde \rA}  = \B e_\rA \K{\Psi}_{\rA\tilde \rA},
\end{equation}    
namely it satisfies Eq. (\ref{eq:choiinstr2}). 
By the states-transformations isomorphism, it must correspond to a
test $\{\tC_0, \tC_{1}\}$ from $\rA$ to
$\rA$: in particular we must have  
\begin{equation}
p_0 ~ \begin{aligned}  \Qcircuit @C=1em @R=.7em @! R {
    \multiprepareC{1}{\Psi_0}&\qw\poloFantasmaCn{\rA}&\qw \\
    \pureghost{\Psi_0}&\qw\poloFantasmaCn{\tilde \rA}&\qw }
  \end{aligned}~ =~
\begin{aligned}
  \Qcircuit @C=1em @R=.7em @! R {
    \multiprepareC{1}{\Psi}&\qw\poloFantasmaCn{\rA}&\gate{\tC_0} &\qw\poloFantasmaCn{ \rA}&\qw \\
    \pureghost{\Psi}&\qw\poloFantasmaCn{\tilde \rA}&\qw &\qw &\qw}
\end{aligned}
\end{equation}
Applying the deterministic effect on $\rA$ we then obtain 
\begin{equation}
\begin{split}
 \begin{aligned} \Qcircuit @C=1em @R=.7em @! R { \multiprepareC{1} \Psi & \qw \poloFantasmaCn \rA & \measureD a \\
    \pureghost \Psi & \qw \poloFantasmaCn {\tilde \rA} & \qw }\end{aligned} & =~
 p_0 ~ \begin{aligned}\Qcircuit @C=1em @R=.7em @! R { \prepareC {\rho_0} &\qw \poloFantasmaCn {\tilde \rA} & \qw } \end{aligned}~ =~p_0 ~\begin{aligned} \Qcircuit @C=1em @R=.7em @! R {
    \multiprepareC{1}{\Psi_0}&\qw\poloFantasmaCn{\rA}&\measureD e \\
    \pureghost{\Psi_0}&\qw\poloFantasmaCn{\tilde \rA}&\qw } \end{aligned}\\ 
&\\
 & = ~\begin{aligned} \Qcircuit @C=1em @R=.7em @! R {
    \multiprepareC{1}{\Psi}&\qw\poloFantasmaCn{\rA}&\gate{\tC_0} &\qw\poloFantasmaCn{ \rA}&\measureD e \\
    \pureghost{\Psi}&\qw\poloFantasmaCn{\tilde \rA}&\qw &\qw &\qw} \end{aligned}
\end{split}
\end{equation}
Since $\Psi$ is dynamically faithful, this implies Eq.
(\ref{atomiceffect}) with $\tC_a := \tC_0$. Moreover,  the states-transformation isomorphism states that $\tC_0$ is atomic since
$p_0 \K{\Psi_0}_{\rA \tilde \rA} = \tC_0 \K{\Psi}_{\rA \tilde \rA}$ is pure.  Finally, suppose
that $\tC_0' \in \Trnset (\rA)$ is another atomic transformation such that
Eq. (\ref{atomiceffect}) holds, and define the pure state $\K{\Psi'_0} := \tC_0' \K{\Psi}_{\rA \tilde \rA}/p_0$. Since $\Psi_0$ and $\Psi_0'$ are purifications of the same state $\K{\rho_0}_{\tilde \rA}$, then they are connected by a reversible channel $\tU$ on $\rA$. Using the fact that $\Psi$ is dynamically faithful, this implies $\tC_0' = \tU \tC_0$. \qed 

\medskip

Moreover, having conjugate purifying systems allows for a more elegant
description of the composition of transformations in terms of
composition of states. We recall that to treat the composition of states we need a system of purifications, as defined in Subsect. \ref{subsect:link}. The nice thing now is that we can take the system of purifications to be symmetric: 

\begin{definition}[Symmetric system of purifications] A \emph{symmetric system of purification} is a choice of dynamically faithful pure states $\K{\Psi}_{\rA \tilde \rA}$ and teleportation effects $\B E_{\tilde \rA \rA}$ that satisfies the properties
\begin{equation}
\begin{split}
\K{\Psi}_{\rA \rB \tilde \rA \tilde \rB} & = \K {\Psi}_{\rA \tilde \rA} \K{\Psi}_{\rB \tilde \rB}\\
\B E_{\tilde \rA\tilde\rB \rA\rB} &= \B E_{\tilde \rA\rA} \B E_{\tilde \rB \rB}.
\end{split}
\end{equation}
\end{definition}
Regarding the probabilities of conclusive teleportation, we now have
$p_\rA = p_{\tilde \rA}$ (compare Eqs.  (\ref{telerho})  and (\ref{telerhotilde}) in the teleportation protocol of Corollary \ref{cor:probtele}).

In the next Subsection we will see that
there is a canonical choice of internal states, namely choosing
$\K{\omega}_\rA = \K \chi_\rA$, where $\K{\chi}_\rA$ is the unique
invariant state of system $\rA$ (for the uniqueness, see Lemma
\ref{lem:uniquenessinvariant}).  We will choose a fixed purification
of $\K\chi_\rA$ and refer to it as to the \emph{canonical faithful
  state}, denoted by $\K{\Phi}_{\rA\tilde \rA}$. In Corollary
\ref{cor:welldone} we will show that this  notation is consistent, since
$\K\Phi_{\rA\tilde \rA}$ is also a purification of the unique
invariant state of $\tilde \rA$.

\subsection{Conjugated transformations}

The most important consequence of the existence of conjugate purifying systems is the possibility of defining a one-to-one correspondence between the reversible transformations of one system $\rA$ and the reversible transformations of its conjugate system $\tilde \rA$. As we will see, this implies in particular the possibility of deterministic teleportation. The correspondence is set by the following Lemma:

\begin{lemma}[Transposition of reversible channels]\label{lem:revchanconj}
  Let $\Phi\in \Stset_1 (\rA \tilde \rA)$ be a purification of the
  unique invariant state $\chi\in \Stset_1 (\rA)$. Then, for every
  reversible channel $\tU \in \grp G_\rA$ there exists a unique
  reversible channel $\tU^\tau \in \grp G_{\tilde \rA}$, here called
  the \emph{transpose} of $\tU$ with respect to $\Phi$, such that
  \begin{equation}\label{transpose}
  \begin{aligned}\Qcircuit @C=1em @R=.7em @! R {
    \multiprepareC{1}{\Phi}&\qw\poloFantasmaCn{\rA}&\gate{\tU} &\qw\poloFantasmaCn{ \rA}&\qw \\
    \pureghost{\Phi}&\qw\poloFantasmaCn{\tilde \rA}&\qw &\qw &\qw} \end{aligned}
  ~=~
\begin{aligned} \Qcircuit @C=1em @R=.7em @! R {
    \multiprepareC{1}{\Phi}&\qw\poloFantasmaCn{\rA}&\qw &\qw&\qw \\
    \pureghost{\Phi}&\qw\poloFantasmaCn{\tilde \rA}&\gate{\tU^\tau} &\qw \poloFantasmaCn {\tilde \rA}&\qw}\end{aligned}
\end{equation}
Transposition is an injective map satisfying the properties
\begin{eqnarray}
\label{Itrans}\tI_{\rA}^\tau & = & \tI_{\tilde \rA}\\
\label{TUtrans}\left( \tU_1 \tU_2 \right)^\tau & = & \tU_2^\tau \tU_1^\tau.  
\end{eqnarray} 
\end{lemma}  

\Proof Since $\K{\chi}_{\rA}$ is invariant, the states $\K{\Phi}_{\rA
  \tilde \rA}$ and $\tU\K{ \Phi}_{\rA \tilde \rA}$ are both
purifications of it.  Then, there must be a reversible transformation
$\tU^\tau \in \grp G_{\tilde \rA}$ such that Eq. (\ref{transpose})
holds.  Moreover, since the invariant state $\K{\chi}_\rA$ is
internal, its purification $\Phi$ is dynamically faithful, both for
system $\rA$ and for system $\tilde \rA$. Dynamical
faithfulness on system $\tilde \rA$ implies that the transformation $\tU^\tau$ is uniquely defined, while dynamical faithfulness on system $\rA$ implies that transposition is injective.  Finally, Eq. (\ref{Itrans})  is obvious, while Eq.  (\ref{TUtrans}) is easily proved by repeated application of Eq. (\ref{transpose}):
\begin{equation}
\begin{split}
\left(\tI_\rA \otimes \left(\tU_1 \tU_2\right)^\tau \right)  \K{\Phi}_{\rA \tilde \rA}& =  (\tU_1 \tU_2 \otimes \tI_{\tilde \rA} )  \K{\Phi}_{\rA \tilde \rA}\\
 & = (\tU_1  \otimes \tU_2^\tau) \K{\Phi}_{\rA \tilde \rA}\\
& = \left(\tI_\rA \otimes  \tU_2^\tau \tU_1^\tau \right) \K{\Phi}_{\rA \tilde \rA},
\end{split}
\end{equation} 
using the fact that $\Phi$ is dynamically faithful for system $\tilde \rA$. \qed

\begin{lemma}[Continuity of transposition]\label{lem:transcont} Transposition is continuous with respect to the operational norm. Moreover, if $C\subseteq \grp G_\rA$ is closed, then $\tau (C) \subseteq \grp G_{\tilde \rA}$ is closed.
\end{lemma}
\Proof  Let $p_\rA$ be the probability of  teleportation for the canonical faithful state $\K\Phi_{\rA\tilde \rA}$. Define $\K{R_{\tU}}_{\rA \tilde \rA} : = (\tU \otimes \tI_{\tilde \rA} )\K\Phi_{\rA \tilde \rA}$.    For every $\epsilon >0$, if     $\tU, \tV \in \grp G_\rA$ are such that $|\!| \tU - \tV |\!|_{\rA, \rA} < \epsilon$, then using Eq. (\ref{bounddist}) one has  $|\!|  \tU^\tau - \tV^{\tau} |\!|_{\tilde \rA, \tilde \rA} \le |\!|  R_\tU - R_{\tV}  |\!|_{\rA \tilde \rA} /p_\rA < \epsilon/p_\rA$.  This proves continuity. Now, suppose that $C\subseteq \grp G_\rA$ is a closed set, and suppose that $\{\tU^{\tau}_n\}$ is a sequence in $\tau (C)$ converging to some reversible transformation $\tV\in\grp G_{\tilde \rA}$. It is easy to see that $\tV$ must be in $\tau (C)$. Indeed, consider the sequence $\{\tU_n\} \subset \grp G_\rA$. Since $\grp G_\rA$ is compact, there must be a subsequence $\tU_{n_k}$ such that  $\tU_{n_k} \to \tU$ for some  $\tU\in\grp G_\rA$. Moreover, since $C$ is closed, one has $\tU \in C$.  Now, using continuity we obtain $\tU_{n_k}^\tau \to \tU^\tau$. This implies that $\tV = \lim_{n\to \infty}  \tU_n^\tau =\tU^\tau$, that is, the limit point is in $\tau (C)$.  Hence, $\tau (C)$ is closed.  \qed

\begin{lemma}
The transposition map $\tau: \tU \mapsto \tU^\tau$ defined in Eq. (\ref{transpose}) is surjective on $\grp G_{\tilde \rA}$.
\end{lemma}

\Proof Take the invariant state $\K{\chi}_{\tilde \rA}$, a purification of
it, say $\K{\Phi^{(\tilde \rA)}}_{\rA \tilde \rA}$, and define the
transpose $\tilde {\tau}$ with respect to $\Phi^{(\tilde \rA)}$.
Since $\tau$ and $\tilde \tau$ are both injective transformations,
their composition $\iota := \tau \tilde \tau: \grp G_{\tilde \rA} \to
\grp G_{\tilde \rA}$ is injective too.  Moreover, $\iota$ is a
homomorphism, since $\iota (\tI_{\tilde \rA}) = \tI_{\tilde \rA}$ and
$\iota (\tV \tW) = \iota(\tV) \iota(\tW)$ for every $\tV , \tW$ in
$\grp G_{\tilde \rA}$. We now claim that $\iota$ is surjective. Of course, since $\iota := \tau \tilde \tau$, this will also prove that $\tau$ is surjective.   Consider the sequence $\{\grp H_n\}$ defined by
$\grp H_n := \iota^{n}(\grp G_{\tilde \rA})$.  By the previous Lemma \ref{lem:transcont}, each $\grp H_n$ is a closed subgroup of $\grp G_{\tilde \rA}$, and one has
\begin{equation}
\grp G_\rA:= \grp H_0 \supseteq \grp H_1 \supseteq \dots \supseteq  \grp H_n \supseteq \grp H_{n+1},
\end{equation}
namely $\{\grp H_n\}$ is a descending chain of subgroups of $\grp G_{\tilde \rA}$.
Since $\grp G_{\tilde \rA}$ is a compact Lie group, every descending chain of
closed subgroups must be eventually  constant (see e.g.  p. 136 of \cite{fordescending}), i.e. there exists a finite $\bar n$ such that 
\begin{equation}
\grp H_{n}= \grp H_{n+1} \qquad n \ge \bar n.
\end{equation} 
Applying $\iota^{-n}$ on both sides, this implies $\grp H_0 = \grp
H_1$, namely $\grp G_{\tilde \rA} = \iota (\grp G_{\tilde \rA})$.
Therefore, $\iota$ is surjective.  \qed
\medskip 

The first consequences of the properties of transposition are given by the following corollary
\begin{corollary}\label{cor:welldone}
  Let $\Phi \in \Stset_1 (\rA \tilde \rA)$ be a purification of the
 unique  invariant state $\chi_\rA \in \Stset_1 (\rA)$. Then the complementary state $\K{\tilde \chi}_{\tilde
    \rA} := \B e _\rA \K{\Phi}_{\rA \tilde \rA}$ is the unique invariant state of $\tilde \rA$.  
\end{corollary}

\Proof For every $\tU \in \grp G_\rA$ we have
 \begin{equation}
 \begin{split}
   \begin{aligned} \Qcircuit @C=1em @R=.7em @! R { \prepareC{\tilde \chi} & \qw
     \poloFantasmaCn {\tilde \rA} &\qw}\end{aligned} &=~\begin{aligned}  \Qcircuit @C=1em @R=.7em @! R {
     \multiprepareC{1}{\Phi}&\qw\poloFantasmaCn{\rA} &\gate{\tU} &\qw \poloFantasmaCn{\tilde \rA}&\measureD e \\
     \pureghost{\Phi}&\qw \poloFantasmaCn{\tilde \rA}&\qw &\qw&\qw } \end{aligned}
   \\
   & =~\begin{aligned} \Qcircuit @C=1em @R=.7em @! R {
     \multiprepareC{1}{\Phi}&\qw\poloFantasmaCn{\rA}  &\measureD e & & \\
     \pureghost{\Phi}&\qw\poloFantasmaCn{\tilde \rA} &\gate{\tU^\tau}
     &\qw\poloFantasmaCn{ \rA}&\qw} \end{aligned} \\
&= ~ \begin{aligned}\Qcircuit @C=1em @R=.7em @! R { \prepareC{\tilde \chi} & \qw
     \poloFantasmaCn {\tilde \rA} &\gate {\tU^\tau} & \qw \poloFantasmaCn {\tilde \rA} & \qw} \end{aligned}
\end{split}
\end{equation}
Since $\tau$ is surjective, $\tU^\tau$ is an arbitrary element of $\grp G_{\tilde \rA}$, hence $\tilde \chi$ is invariant. \qed

\begin{definition}[Conjugate of a reversible channel]
  The \emph{conjugate} of the reversible channel $\tU\in \Trnset
  (\rA)$ with respect to the state $\Phi \in \Stset (\rA \tilde \rA)$ is the reversible channel $\tU^* \in \Trnset(\tilde \rA)$
  defined by $\tU^* := (\tU^\tau)^{-1}$, where the transpose is defined with respect to $\Phi$.
\end{definition}

Note that with this definition the canonical faithful state $\K{\Phi}_{\rA \tilde \rA}$ is \emph{isotropic}, i.e. it is invariant under combined reversible channels on the conjugate systems $\rA$ and $\tilde \rA$: 
  \begin{equation}\label{isotropic}
\begin{aligned}    \Qcircuit @C=1em @R=.7em @! R {
      \multiprepareC{1}{\Phi}&\qw\poloFantasmaCn{\rA}&\qw  \\
      \pureghost{\Phi}&\qw\poloFantasmaCn{\tilde \rA}&\qw } \end{aligned} ~=~ \begin{aligned}
      \Qcircuit @C=1em @R=.7em @! R {
      \multiprepareC{1}{\Phi}&\qw\poloFantasmaCn{\rA}&\gate{\tU}  &\qw \poloFantasmaCn \rA &\qw \\
      \pureghost{\Phi}&\qw\poloFantasmaCn{\tilde \rA}&\gate{\tU^*} &\qw \poloFantasmaCn {\tilde \rA}&\qw}\end{aligned} \qquad \forall \tU \in \grp G_\rA.
\end{equation}
Moreover, we have also the converse:
\begin{corollary}[Isotropic states] A pure state $\Psi\in \Stset_1 (\rA
  \tilde \rA)$ is isotropic if and only if $\K{\Psi}_{\rA \tilde \rA} = (\tV\otimes \tI_{\tilde \rA}) \K{\Phi}_{\rA \tilde \rA}$ for some
  reversible $\tV \in \grp G_\rA$ such that
\begin{equation}\label{center}
  \tU \tV = \tV \tU \qquad \forall \tU \in \grp G_\rA.
\end{equation}  
\end{corollary}
\Proof Clearly, a state of the above form is isotropic. Conversely, if
$\Psi$ is isotropic, it satisfies Eq. (\ref{isotropic}), and,
therefore, its marginal state on system $\tilde \rA$ is the invariant state
$\K{\chi}_{\tilde \rA}$. Since $\Psi$ and $\Phi$ are purifications of the same
state, there must exist a reversible channel $\tV \in \grp
G_{ \rA}$ such that $\K{\Psi}_{\rA \tilde \rA}  = \tV
\K{\Phi}_{\rA \tilde \rA}$. The isotropy condition then gives 
\begin{equation}
\begin{split}
(\tV \otimes \tI_{\tilde \rA})\K{\Phi}_{\rA \tilde \rA} & = \K{\Psi}_{\rA \tilde \rA} \\ 
&=(\tU \otimes \tU^*) \K{\Psi}_{\rA \tilde \rA}\\ &= (\tU\tV \otimes \tU^*) \K{\Phi}_{\rA \tilde \rA}\\
 &= (\tU\tV \tU^{-1}\otimes \tI_{\tilde \rA}) \K{\Phi}_{\rA \tilde \rA}\\
\end{split}
\end{equation}
Dynamical faithfulness of $\Phi$ then implies $\tV = \tU \tV \tU^{-1}$, namely Eq. (\ref{center}). \qed 
\medskip

Recalling that the \emph{center} of the group $\grp G_\rA$ is the set of all elements $\tV\in\grp G_\rA$ such that $\tU\tV = \tV\tU$ for every $\tU\in \grp G_\rA$, it is immediate to state the following
\begin{corollary} The canonical faithful state $\K{\Phi}_{\rA \tilde \rA}$ is the unique isotropic state of system $\rA \tilde \rA$ if and only if the compact Lie group $\grp G_A$ has trivial center. 
\end{corollary}

The conclusion of this Subsection is summarized by the following theorem:

\begin{theorem}[Isomorphism of groups]
  The reversible channels on $\rA$ and $\tilde \rA$ form two
  isomorphic Lie groups, with the isomorphism given by the conjugation
  map $* : \grp G_{\rA} \to \grp G_{\tilde \rA}, \tU \mapsto \tU^*$.
\end{theorem}
\Proof Clearly, $*$ is a homomorphism, namely $\tI_{\rA}^* =
\tI_{\tilde \rA}$ and $ (\tU_1 \tU_2)^* = \left((\tU_1 \tU_2 )^\tau
\right)^{-1} = \tU_1^* \tU_2^* $.  Moreover, $*$ is injective and
surjective, since it is the composition of two injective and
surjective maps, namely transposition and inversion. \qed

\subsection{Deterministic teleportation}

\begin{lemma} Let $\tT_\rA $ and $\tT_{\tilde \rA}$ be the twirling
  channels on $\rA$ and $\tilde \rA$, respectively, and let $\Phi \in
  \Stset_1 (\rA \tilde \rA)$ be the canonical faithful state. Then, one
  has
\begin{equation}\label{twirltwirl}
  \begin{aligned}\Qcircuit @C=1em @R=.7em @! R {
    \multiprepareC{1}{\Phi}&\qw\poloFantasmaCn{\rA}&\gate{\tT} &\qw\poloFantasmaCn{ \rA}&\qw \\
    \pureghost{\Phi}&\qw\poloFantasmaCn{\tilde \rA}&\qw &\qw &\qw} \end{aligned}~=~ \begin{aligned}
  \Qcircuit @C=1em @R=.7em @! R {
    \multiprepareC{1}{\Phi}&\qw &\qw &\qw&\qw\poloFantasmaCn{\rA} \\
    \pureghost{\Phi}&\qw\poloFantasmaCn{\tilde \rA}&\gate \tT &\qw \poloFantasmaCn {\tilde \rA} &\qw} \end{aligned} 
\end{equation}
and
  \begin{equation}
 \begin{aligned} \Qcircuit @C=1em @R=.7em @! R {
    \multiprepareC{1}{\Phi}&\qw\poloFantasmaCn{\rA}&\gate{\tT} &\qw\poloFantasmaCn{ \rA}&\qw \\
    \pureghost{\Phi}&\qw\poloFantasmaCn{\tilde \rA}&\qw &\qw &\qw} \end{aligned} 
~=~\begin{aligned} \Qcircuit @C=1em @R=.7em @! R { \prepareC{\chi} &\qw \poloFantasmaCn \rA &\qw \\
\prepareC{\chi} &\qw \poloFantasmaCn {\tilde \rA} &\qw} \end{aligned}
\end{equation}
\end{lemma}
\Proof We have
\begin{equation}
\begin{split}
  \tT_{\rA} \K{\Phi}_{\rA \tilde \rA} & = \int_{\grp G_{\rA}} {\rm d} \tU ~  \tU   \K{\Phi}_{\rA \tilde \rA} \\
  &=  \int_{\grp G_{\rA}} {\rm d} \tU ~  (\tU^*)^{-1}  \K{\Phi}_{\rA \tilde \rA}\\
  &= \tT_{\tilde \rA} \K{\Phi}_{\rA \tilde \rA},
\end{split}
\end{equation}
having used the fact that $\grp G_{\rA}$ and $\grp G_{\tilde \rA}$ are isomorphic, and, therefore, have the same Haar measure.  Moreover, since the output of the twirling channel is an invariant state, we have that $\K\sigma_{\rA \tilde \rA} :=\tT_{\rA}  \K{\Phi}_{\rA \tilde \rA}$ is invariant under local reversible transformations, i.e.
\begin{equation}
\K{\sigma}_{\rA \tilde \rA} = (\tU \otimes \tV)  \K{\sigma}_{\rA \tilde \rA} \qquad \forall \tU \in \grp G_{\rA}, \forall \tV \in \grp G_{\tilde \rA}.  
\end{equation}
Finally, we invoke Theorem \ref{theo:uniqueLI}, which states that the
unique state invariant under local reversible transformations is
$\chi_\rA \otimes \chi_{\tilde \rA}$. \qed

\begin{theorem}[Deterministic teleportation] Let $\rA$ and $\rA'$ be
  two operationally equivalent systems, and let $\{p_i \tU_i\}_{ i \in
    \rX}$ be a twirling test, where each $\tU_i$ is a reversible
  channel on $\rA$. Then there exists an observation-test $\{B_i\}_{i
    \in \rX}$ on $ \tilde \rA\rA'$ such that for every outcome $i$ one
  has
\begin{equation}\label{teleforalli}
\begin{aligned} 
 \Qcircuit @C=1em @R=.7em @! R { 
\multiprepareC{1}{\Phi} &\qw \poloFantasmaCn{\rA} &\gate{\tU_i^{-1}} & \qw \poloFantasmaCn \rA & \qw  \\
 \pureghost{\Phi} &\qw \poloFantasmaCn{\tilde \rA} & \multimeasureD{1}{B_i}&& \\
     &  \qw  \poloFantasmaCn{\rA'}  & \ghost{B_i} & &}
\end{aligned}
~= ~ p_i ~
\begin{aligned}
  \Qcircuit @C=1em @R=.7em @! R { & \poloFantasmaCn{\rA'} \qw & \gate{
      \tI} &\qw \poloFantasmaCn{\rA}&\qw }
\end{aligned}
\end{equation}
Moreover, each effect $B_i$ must be atomic.        
\end{theorem}

\Proof Define the preparation-test $\{p_i \Phi_i\}_{i \in \rX}$ with $\K{\Phi_i}_{\rA \tilde \rA} := \tU_i \K{\Phi}_{\rA \tilde \rA} $. By the previous Lemma, we have $\sum_i p_i \Phi_i = \chi_{\rA} \otimes \chi_{\tilde \rA}$, namely  coarse-graining of the preparation-test $\{p_i \Phi_i\}$  yields the invariant state of $\rA \tilde \rA$.  By the states-transformations isomorphism of Theorem \ref{theo:iso2}, there exists an observation-test on $\tilde \rA' \rA'$, say $\{B_i\}_{i \in \rX}$, such that 
\begin{equation}
\begin{aligned}  \Qcircuit @C=1em @R=.7em @! R { \multiprepareC{1}{\Phi} &\qw \poloFantasmaCn{\rA} &\qw \\
    \pureghost{\Phi} &\qw \poloFantasmaCn{\tilde \rA'} & \multimeasureD{1}{B_i} \\
    \multiprepareC{1}{\Phi} &\qw \poloFantasmaCn{\rA'} &\ghost{B_i}  \\
    \pureghost{\Phi} &\qw \poloFantasmaCn{\tilde \rA} &\qw  } 
\end{aligned}
~ = ~ p_i ~ 
\begin{aligned}\Qcircuit @C=1em @R=.7em @! R {\multiprepareC{1}{\Phi_i} &\qw \poloFantasmaCn{\rA} &\qw \\
    \pureghost{\Phi_i} &\qw \poloFantasmaCn{\tilde \rA} &\qw  }
\end{aligned}
\end{equation}
Clearly, the states-transformations isomorphism implies that each
effect $B_i$ must be atomic (indeed, the corresponding state is pure).  Applying $\tU_i^{-1}$ on system
$\rA$  we obtain
\begin{equation}
\begin{aligned}
\Qcircuit @C=1em @R=.7em @! R { 
\multiprepareC{1}{\Phi} &\qw \poloFantasmaCn{\rA} &\gate {\tU_i^{-1}} &\qw \poloFantasmaCn{\rA}&\qw \\
    \pureghost{\Phi} &\qw \poloFantasmaCn{\tilde \rA'} & \multimeasureD{1}{B_i} && \\
    \multiprepareC{1}{\Phi} &\qw \poloFantasmaCn{\rA'} &\ghost{B_i} &&  \\
    \pureghost{\Phi} &\qw \poloFantasmaCn{\tilde \rA} &\qw &\qw&\qw }
\end{aligned} ~=~p_i 
\begin{aligned} \Qcircuit @C=1em @R=.7em @! R {\multiprepareC{1}{\Phi} &\qw \poloFantasmaCn{\rA} &\qw \\
    \pureghost{\Phi} &\qw \poloFantasmaCn{\tilde \rA} &\qw  }
\end{aligned}
\end{equation}       
The thesis follows from the fact that $\Phi$ is dynamically faithful. \qed

\medskip

In theories with local discriminability we have the additional result:
\begin{corollary}
  Let $\{p_i\tU_i\}_{i \in \rX}$ be a twirling test where each $\tU_i$ is a
  reversible channel. In a theory with local
  discriminability the number of outcomes $|\rX|$ cannot be smaller
  than $\dim \Stset_\Reals (\rA)$. 
\end{corollary}
\Proof  By Eq. (\ref{teleforalli}) the state $\Psi_i := (\tU_i^{-1} \otimes \tI_\rA) \Phi$ and the effect $B_i$ achieve teleportation with probability $p_i$. In a theory with local discriminability the bound of Eq. (\ref{maxprob}) gives $p_i \le 1/\dim \Stset_\Reals (\rA)$. We then have $1 =\sum_{i\in \rX} p_i \le |\rX|/\dim \Stset_\Reals (\rA)$.  \qed 

\medskip

If two parties share the pure state $\K{\Phi}_{\rA \tilde \rA}$, then
by the teleportation protocol they can convert it in an arbitrary
state $\Psi \in \Stset_1 (\rA \tilde \rA)$ using only local operations
and one round of classical communication (one-way LOCC). We now show
that the state $\K{\Phi}_{\rA \tilde \rA}$ is the \emph{maximally entangled} state of
$\Stset_1 (\rA \tilde \rA)$, that is, if  we can convert another state
$\Psi$ to $\Phi$ by one-way LOCC, then $\Psi = (\tU \otimes \tI_\rA ) \Phi$ for some
local reversible channel $\tU \in \Trnset (\rA)$.  To see that, we
show that if $\Psi$ allows for deterministic teleportation, then
$\Psi= (\tU \otimes \tI_\rA ) \Phi$.

\begin{theorem}\label{theo:uniquetele}{\bf (Unique structure of deterministic teleportation)} Let
  $\Psi \in \Stset_1 (\rA \tilde \rA)$ be a pure state, $\{\tR_i\}_{i
    \in \rX}$ be a collection of channels on $\rA$, $\{p_i\}_{i \in \rX}$ a set of probabilities, and $\{M_i\}_{i \in
    \rX}$ be an observation-test on $\tilde \rA \rA'$, with $\rA'$ and
  $\rA$ operationally equivalent systems. If for every outcome $i$ one has
\begin{equation}
\begin{aligned}
 \Qcircuit @C=1em @R=.7em @! R { 
\multiprepareC{1}{\Psi} &\qw \poloFantasmaCn{\rA} &\gate{\tR_i} & \qw \poloFantasmaCn \rA & \qw  \\
    \pureghost{\Psi} &\qw \poloFantasmaCn{\tilde \rA} & \multimeasureD{1}{M_i} && \\
     &  \qw  \poloFantasmaCn{\rA'}  &\ghost{M_i}&&}
 \end{aligned}
~= ~ p_i ~
\begin{aligned}
\Qcircuit @C=1em @R=.7em @! R {
 & \poloFantasmaCn{\rA'} \qw &  \gate{ \tI} &\qw \poloFantasmaCn{\rA}&\qw }
\end{aligned}
\end{equation}       
then 
\begin{enumerate}
\item each channel $\tR_i$ is reversible, namely $\tR_i = \tU_i^{-1}$ for some $\tU_i \in \grp G_\rA$
\item there is a reversible channel $\tU \in \grp G_{\rA}$ such that $\Psi = \tU \Phi$
\item each effect $M_i$ has the property $\B{M_i}_{\tilde \rA \rA'} \K{\chi}_{\tilde \rA} = p_i \B e_{\rA'}$ 
\item $\sum_{i \in \rX} p_i \tU_i = \tT$, where $\tT$ is the twirling
  channel
\end{enumerate} 
\end{theorem} 

\Proof Define the transformation $\tA_i$ as  
\begin{equation}
\begin{aligned}  \Qcircuit @C=1em @R=.7em @! R {&\qw \poloFantasmaCn{\rA} &\gate{\tA_i} & \qw \poloFantasmaCn \rA &\qw}
\end{aligned} ~:=~ 
  \Qcircuit @C=1em @R=.7em @! R { \multiprepareC{1}{\Psi} &\qw \poloFantasmaCn{\rA} & \qw  \\
    \pureghost{\Psi} &\qw \poloFantasmaCn{\tilde \rA} & \multimeasureD{1}{M_i} \\
    &  \qw  \poloFantasmaCn{\rA}  &\ghost{M_i}}
\end{equation}
With this definition we have $\tR_i \tA_i = p_i \tI_\rA$ for every outcome $i$. Moreover, applying
the deterministic effect on both sides of the equality we obtain 
\begin{equation}\B e_{\rA} \tA_i = \B{e}_\rA \tR_i
\tA_i = p_i \B e_\rA,
\end{equation}
that is, each $\tA_i$ is proportional to a channel $\tC_i$, i.e. $\tA_i
= p_i \tC_i$.  We now have $\tR_i \circ \tC_i = \tI_\rA$, that means that the
channel $\tC_i$ is invertible. By corollary
\ref{cor:invertiblereversible}, this implies that $\tC_i$ is
reversible, namely $\tC_i = \tU_i$ for some $\tU_i \in \grp G_\rA$.
Clearly, this requires $\tR_i = \tU_i^{-1}$.  Now consider the marginal of $\Psi$ on system $\rA$: one has
\begin{equation}
\begin{split}
\begin{aligned}  \Qcircuit @C=1em @R=.7em @! R {\multiprepareC{1} {\Psi} &\qw\poloFantasmaCn{\rA} &\qw \\
    \pureghost{\Psi} &\qw \poloFantasmaCn {\tilde \rA} &\measureD e}
\end{aligned}
& =~
\begin{aligned}  \Qcircuit @C=1em @R=.7em @! R { \multiprepareC{1}{\Psi} &\qw \poloFantasmaCn{\rA} &\qw  \\
    \pureghost{\Psi} &\qw \poloFantasmaCn{\tilde \rA} & \measureD{e} \\
    \prepareC{\chi} &  \qw  \poloFantasmaCn{\rA}  &\measureD e} 
\end{aligned}\\
  & = \sum_{i \in \rX} ~
\begin{aligned}  \Qcircuit @C=1em @R=.7em @! R { \multiprepareC{1}{\Psi} &\qw \poloFantasmaCn{\rA} &\qw  \\
    \pureghost{\Psi} &\qw \poloFantasmaCn{\tilde \rA} & \multimeasureD{1}{M_i} \\
    \prepareC{\chi} &  \qw  \poloFantasmaCn{\rA}  &\ghost{M_i}} 
\end{aligned}\\
  & = \sum_{i \in \rX}  ~\begin{aligned}
  \Qcircuit @C=1em @R=.7em @! R {\prepareC \chi & \poloFantasmaCn{\rA} \qw &  \gate{ \tA_i} &\qw \poloFantasmaCn{\rA}&\qw }
\end{aligned}\\
  &  = \sum_{i \in \rX} p_i ~\begin{aligned}  \Qcircuit @C=1em @R=.7em @! R {\prepareC \chi & \poloFantasmaCn{\rA} \qw &  \gate{ \tU_i} &\qw \poloFantasmaCn{\rA}&\qw }
\end{aligned}\\
  & = ~
\begin{aligned} \Qcircuit @C=1em @R=.7em @! R {\prepareC
    \chi & \poloFantasmaCn{\rA} \qw &\qw}
\end{aligned}
\end{split}
\end{equation}       
having used the invariance of $\chi$. But this means that $\Psi$ and
$\Phi$ have the same marginal on system $\rA$, and, therefore,
$\K{\Psi}_{\rA \tilde \rA} = ( \tI_\rA\otimes \tU) \K{\Phi}_{\rA
  \tilde \rA}$ for some suitable $\tU \in \grp G_{\tilde \rA}$. Using
Lemma \ref{lem:revchanconj}, we can also transfer $\tU$ on system
$\rA$, getting $\K\Psi_{\rA \tilde \rA} = (\tU^\tau \otimes \tI_{\tilde \rA})\K\Phi_{\rA \tilde \rA}$. Using $\tA_i = p_i \tU_i$ we then get
\begin{equation}\label{soycansado}
\begin{split}
  p_i~ 
\begin{aligned} 
\Qcircuit @C=1em @R=.7em @! R {\multiprepareC{1} {\Phi} &\qw\poloFantasmaCn{\rA} & \gate {\tU_i} & \qw\poloFantasmaCn \rA &\qw  \\
    \pureghost{\Psi} &\qw \poloFantasmaCn {\tilde \rA} &\qw &\qw & \qw}
\end{aligned}
  & =~
\begin{aligned}
 \Qcircuit @C=1em @R=.7em @! R {\multiprepareC{1} {\Phi} &\qw\poloFantasmaCn{\rA} & \gate {\tA_i} & \qw\poloFantasmaCn \rA &\qw  \\
    \pureghost{\Psi} &\qw  \poloFantasmaCn {\tilde \rA} &\qw & \qw &\qw}
\end{aligned}\\   
 &  =~ 
\begin{aligned}
\Qcircuit @C=1em @R=.7em @! R { \multiprepareC{1}{\Phi} &\qw \poloFantasmaCn{\rA} & \gate{\tU^\tau} & \qw \poloFantasmaCn \rA & \qw  \\
    \pureghost{\Psi} &\qw \poloFantasmaCn{\tilde \rA} & \multimeasureD{1}{M_i} && \\
    \multiprepareC{1}{\Phi} &  \qw  \poloFantasmaCn{\rA}  &\ghost{M_i}&& \\
\pureghost\Phi & \qw \poloFantasmaCn{\tilde \rA} &\qw & \qw &\qw }
\end{aligned} 
\end{split}
\end{equation}       
By the states-transformations isomorphism, this means that each
$M_i$ is atomic (indeed, the corresponding state is pure). Applying the deterministic effect on system $\rA$, the above equation also implies  
\begin{equation}
 p_i ~ 
\begin{aligned} \Qcircuit @C=1em @R=.7em @! R {\multiprepareC{1} {\Phi} &\qw\poloFantasmaCn{\rA} & \measureD e  \\
    \pureghost{\Psi} &\qw \poloFantasmaCn {\tilde \rA} & \qw}
\end{aligned}  
  ~=~
\begin{aligned}
 \Qcircuit @C=1em @R=.7em @! R {\prepareC \chi &\qw \poloFantasmaCn{\tilde \rA} & \multimeasureD{1}{M_i} \\
    \multiprepareC{1}{\Phi} &  \qw  \poloFantasmaCn{\rA}  &\ghost{M_i} \\
    \pureghost\Phi & \qw \poloFantasmaCn{\tilde \rA} &\qw } 
\end{aligned}
\end{equation}
which amounts to saying $\B{M_i}_{\tilde \rA \rA} \K\chi_{\tilde \rA}
= p_i \B e_\rA$, because $\Phi$ is dynamically faithful.  Moreover,
summing over the outcomes in Eq. (\ref{soycansado}) we obtain $(\sum_i p_i \tU_i) \K{\Phi}_{\rA
  \tilde \rA} = \K\chi_\rA \K\chi_{\tilde \rA} = \tT \K\Phi_{\rA
  \tilde \rA}$. Again, since $\Phi$ is dynamically faithful, this
implies $\sum_i p_i \tU_i = \tT$. \qed
\medskip

In a theory with local discriminability one has also the following result: 
\begin{corollary}
  Let $\K{\Psi}_{\rA \tilde \rA}$, $\{\tR_i\}_{i \in \rX}$,
  $\{p_i\}_{i \in \rX}$, and $\{M_i\}_{i \in \rX}$ be the state, the
  recovery channels, the probabilities, and the observation-test in a
  deterministic teleportation protocol, as in Theorem
  \ref{theo:uniquetele}.  In a theory with local discriminability the
  number of outcomes satisfies the bound $|\rX|\ge \dim \Stset_\Reals
  (\rA)$. The bound is achieved if and only if $p_i = 1/\dim
  \Stset_\Reals (\rA)$ for every $i$, and the states $\K{\Psi_i}_{\rA
    \tilde \rA} := (\tR_i \otimes \tI_{\tilde \rA})\K \Psi_{\rA \tilde
    \rA}, \quad {i \in \rX}$ are perfectly distinguishable with the
  observation-test $\{M_i\}$, \emph{i.e.}
\begin{equation} 
\SC  {M_i} {\Psi_j}_{\rA\tilde \rA} = \delta_{ij}
\end{equation}
\end{corollary}
\Proof From Eq. (\ref{probtele}) we have $p_i \le 1/\dim \Stset_\Reals
(\rA)$ for every $i$, and, therefore $1 \le |\rX|/\dim \Stset_\Reals
(\rA)$. Clearly, the bound is achieved if and only if $p_i = 1/\dim
\Stset_\Reals (\rA)$ for every $i$. In this case, it can be seen from
the proof of Eq. (\ref{probtele}) that one has $\SC {M_i}
{\Psi_i}_{\rA \tilde \rA} = 1$. Since $\{M_i\}_{i \in \rX}$ is an
observation-test, and the probabilities of all outcomes must sum up to
unit, this implies $\SC {M_j} {\Psi_i}_{\rA \tilde \rA} =
\delta_{ji}$. \qed
\medskip

The above Corollary shows that if teleportation has the minimum
possible number of outcomes $|\rX| = \dim \Stset_\Reals (\rA)$, then
\emph{dense coding} is possible: By acting locally on one side of the
state $\Psi$ one can produce $\dim \Stset_\Reals (\rA)$ perfectly
distinguishable states. This number exceeds the maximum number of
perfectly distinguishable states available in system $\rA$, which must be strictly smaller than $\dim \Stset_\Reals (\rA)$ due to
Corollary \ref{cor:maxdiscr}. However, we didn't prove here the
existence of such a teleportation scheme with $|\rX| = \dim
\Stset_\Reals (\rA)$. This issue, which is closely related to the
topic of discrimination in theories with purification, will be
addressed in a future work.

\bigskip

\section{Conclusions and perspectives on future work}

In this paper we investigated causal probabilistic theories with
purification, and derived a surprising wealth of features that are
characteristic of quantum theory without resorting to the
framework of Hilbert spaces or C*-algebras.  Among theories with local
discriminability, quantum theory appears as the only known one that 
satisfies the purification principle. The absence of a counterexample
and the amount of quantum features derived suggest that 
quantum theory could be the only causal theory with purification
and local discriminability. However, at the moment we do not have a
derivation of quantum theory from the purification principle, and
the question whether there are other theories satisfying the above
postulates remains open.

Any answer to this question would lead to an interesting scenario: If
quantum theory is the only causal theory with purification and
local discriminability, then the machinery of Hilbert spaces is a
quite redundant way to prove theorems that in fact can be
derived directly from basic physical notions.  What is
more, the general proofs of most theorems are simpler and more
intuitive than the original quantum proofs.  On the other hand, if
quantum theory is not the only theory satisfying our postulates,
the existence of more general theories, that share with quantum
mechanics the basic structure highlighted in this paper, is also a
very fascinating perspective. Moreover, abandoning the standard
quantum formalism would be interesting especially in view of a
possible reconciliation with general relativity. In this direction,
particularly appealing is the possibility of dropping causality from
our requirements, and of working with non-unique deterministic effects. The study of non-causal theories with purification is expected to provide new insights toward a formulation of quantum gravity. Such
an approach would be related to the
informational approaches of Hardy \cite{causaloid} and Lloyd
\cite{lloyd}.  The study of theories with purification in the non-causal setting will be addressed in a forthcoming paper.

Another direction of further research is the generalization of the
notion of subsystem.  On the one hand, introducing \emph{classical systems} in the
theory and clarifying how they can be viewed as subsystems of the
non-classical ones is expected to provide an additional structure that will eventually contribute to the full derivation of quantum mecanics. On the other hand, under suitable assumptions, a face
of the convex set of states of a system can be considered as the set
of states of some subsystem. Following this observation, we plan to
consider information-theoretic tasks like state
compression in theories with purification, by analyzing the mechanism that leads the state
$\rho^{\otimes N}$ to approach a face corresponding to the state space
of $M < N$ systems.

\acknowledgements
We thank the anonymous referees for many suggestions that contributed to improve the original manuscript.  
GC is grateful to R. Spekkens, B. Coecke, R. Colbeck, S. Facchini, A.  Bisio, H. Himai, and A. Doering for useful discussions and suggestions. GMD is grateful
to L. Hardy for useful discussions. This work is supported by the Italian Ministry of Education through grant PRIN 2008.   Research at Perimeter Institute for Theoretical Physics is supported in part by the Government of Canada through NSERC and by the Province of Ontario through MRI.


\begin{thebibliography}{99}
\bibitem{nota:rule} In other words, we need to put a probabilistic
  structure on top of the formal language of circuits. This rule would
  be given by a function $\pi$ from $\Trnset (\rI,\rI)$ to the
  interval $[0,1]$, enjoying the properties $\sum_{i \in \rX} \pi
  (p_i)=1$, and $\pi (p_i \otimes q_j) = \pi (p_i \circ q_j) = \pi
  (p_i) \pi (q_j)$.  In this presentation, however, we will omit the
  function $\pi$, and we will directly identify the event $p_i$ with
  its probability $\pi (p_i)$.
\bibitem{nota:linearextension} Note that the extension by linearity
  $\hat \tC_k \left( \sum_{i} c_i \rho_i \right) := \sum_i c_i \tC_k \K{
    \rho_i}_\rA$ is well-defined: Indeed, one has $\sum_i c_i \rho_i
  =0$ if and only if $\sum_i c_i \SC a {\rho_i}_\rA =0 $ for every
  effect $a \in \Stset (\rA)$. Now, if $b$ is an arbitrary effect in
  $\Cntset (\rB)$, then $\B b_\rB \tC_k $ is an effect in $\Cntset
  (\rA)$, and, therefore, one must have $\sum_{i} c_i \B b_\rB \tC
  \K{\rho_i} = 0$. Since $b$ is arbitrary, this implies $\sum_i c_i
  \tC_k \K{\rho_i} = 0$, thus proving that the linear extension is
  well-defined.
\bibitem{ranextr}  If we toss $N$ times a coin with arbitrary bias $q_0$, all the  $\binom N  k $  sequences with $k$ zeros will have the same probability $q_0^k  (1-q_0)^{N-k}$.   This means that, conditionally to the fact that there are $k$ zeros, we have  $\binom  N  k$  equiprobable strings, and, therefore, the probability of the first $m_k$ strings with $k$ zeros is $p(m_k|k) = m_k/ \binom N  k$.     For a given $\epsilon >0$, if the number  $\binom N  k$  is sufficiently large, choosing a suitable $m_k$ we can approximate any given probability $p\in [0,1]$ with $p(m_k|k)$, in such a way that $|p-p(m_k|k)| \le \epsilon/2$.   Moreover, by choosing a sufficiently large $N$ one can guarantee that the set $S_{ok}$ of values of $k$ that allow for this approximation has total probability $p_{ok} = \sum_{k\in S_{ok}}  p_k = \sum_{k \in S_{ok}} \binom N  k  q_0^k (1-q_0)^{N-k} \ge 1- \epsilon/2$.   If for every $k$ we associate the outcome $0$ to the first $m_k$ strings, and the outcome $1$ to the remaining ones, this procedure defines a new coin with bias $p_N=\sum_{k} p_k  p(m_k|k) $ satisfying the bound $|p- p_N|  \le \epsilon$.
\bibitem{nota:choijam} The general version Choi-Jamio\l kowski
  isomorphism considered in this paper is an isomorphism between
  transformations and bipartite states.  This is not the same as the
  mathematical ``CJ isomorphism'' of Ref. \cite{maurolast}, which was
  instead the isomorphism between the cone of transformations
  $\Trnset_+ (\rA, \rA)$ and the cone of positive bilinear forms on
  the complex vector space $\Cntset_{\mathbb C} (\rA) = \Span_{\mathbb
    C} \{\Cntset(\rA)\}$.
\bibitem{limprod}  Indeed, it is immediate to see that if two sequences of channels $\{\tC_m\}$ and $\{\tD_n\}$ converge to two channels $\tC$ and $\tD$, respectively, then the sequence $\{\tC_n\tD_n\}$ converges to $\tC\tD$:  $|\!|  \tC_n  \tD_n- \tC\tD |\!|_{\rA,\rA}  \le |\!| (\tC_n - \tC) \tD_n |\!|_{\rA,\rA} +  |\!| \tC  (\tD_n-\tD)  |\!|_{\rA,\rA} \le  |\!| \tC_n - \tC |\!|_{\rA,\rA} +  |\!| \tD_n-\tD  |\!|_{\rA,\rA} \overset{n\to \infty}\longrightarrow 0$, having used the triangle inequality and the monotonicity property of Lemma \ref{lem:monotonicitytransformations}.
\bibitem{notgood} For bipartite channels the term
  ``semicausal'' was used to mean ``allowing for signalling in one direction at most'', while the term "causal'' was used to mean
  "not allowing for signalling (in any direction)''.  For an $N$-partite channel, however,
  there are $N!$ possible orderings of the input systems, and a causal
  $N$-partite channel was then defined as a channel that is ``allowing for the propagation of signals along one of these possible orderings''. Of course, for $N=2$
  the two nomenclatures are conflicting.
\end{thebibliography}
\end{document}